\documentclass[epsf]{book}
\input{epsf.tex}
\usepackage{fancyhdr}  
\usepackage{wasysym}

 \renewcommand{\chaptermark}[1]{\markboth{\thechapter.\ #1}{}}  
 \renewcommand{\sectionmark}[1]{\markright{\thesection.\ #1}}          
\def\={\!\!&=&\!\!}
\def\bu{\bar{u}}  \def\bphi{\bar{\phi}}
 \def\Na{{\bf N}} \def\V{{\cal V}}
\newcommand{\nc}{\newcommand}
\nc{\alert}[1]{#1}
\nc{\al}{\alpha}
\nc{\ga}{\gamma}   \nc{\Ga}{\Gamma}
\nc{\De}{\Delta}
\nc{\ald}{{\dot \al}}
\nc{\betad}{{\dot \beta}}
\nc{\gd}{{\dot \gamma}}
\nc{\sigmad}{{\dot \sigma}}
\nc{\mud}{{\dot \mu}}
\nc{\aldd}{{\ddot \al}}
\nc{\betadd}{{\ddot \beta}}
\nc{\gdd}{{\ddot \gamma}}
\nc{\sigmadd}{{\ddot \sigma}}
\nc{\mudd}{{\ddot \mu}}
\nc{\la}{\lambda}   \nc{\La}{\Lambda}
\nc{\var}{\varphi}  \nc{\hn}{h^\vee}
\nc{\pa}{\partial}  \nc{\hf}{\frac{1}{2}}         
\nc{\binomial}[2]{\left (\begin{array}{c} {#1}\\ {#2} \end{array}
\right )}
\nc{\ben}{\begin{equation}}
\nc{\een}{\end{equation}}
\nc{\bea}{\begin{eqnarray}}
\nc{\eea}{\end{eqnarray}}
\nc{\braket}[1]{\langle\,{#1}\rangle}

\nc{\C}{\mbox{\hspace{1.24mm}\rule{0.2mm}{2.5mm}\hspace{-2.7mm} C}}
\nc{\Nat}{\mbox{\hspace{.04mm}\rule{0.2mm}{2.8mm}\hspace{-1.5mm} N}}
\nc{\spa}{\hspace{1 cm},\hspace{1 cm}}
\nc{\vs}{\vspace}
\nc{\NP}[1]{Nucl.\ Phys.\ {\bf #1}}
\nc{\PL}[1]{Phys.\ Lett.\ {\bf #1}}
\nc{\CMP}[1]{Commun.\ Math.\ Phys.\ {\bf #1}}
\nc{\LMP}[1]{Lett.\ Math.\ Phys.\ {\bf #1}}
\nc{\TMP}[1]{Theor.\ Math.\ Phys.\  {\bf #1}}
\nc{\PR}[1]{Phys.\ Rev.\ {\bf #1}}
\nc{\PRL}[1]{Phys.\ Rev.\ Lett.\ {\bf #1}}
\nc{\PTP}[1]{Prog.\ Theor.\ Phys.\ {\bf #1}}
\nc{\PTPS}[1]{Prog.\ Theor.\ Phys.\ Suppl.\ {\bf #1}}
\nc{\MPL}[1]{Mod.\ Phys.\ Lett.\ {\bf #1}}
\nc{\IJMP}[1]{Int.\ Jour.\ Mod.\ Phys.\ {\bf #1}}
\nc{\IM}[1]{Invent.\ Math.\ {\bf #1}}
\nc{\SJNP}[1]{Sov. J. Nucl. Phys.\ {\bf #1}}

      \def\Tr{{\rm Tr}}

\def\ni{\noindent}
\def\wt{\widetilde}      
      \def\ol{\overline}
\def\nn{\nonumber \\}

\def\bra#1{\,\left\langle\, #1 \,\right\vert}
\def\ket#1{\,\left\vert\, #1 \,\right\rangle}
\def\com#1#2{ \left[\, #1 \,,\, #2 \,\right] }

\def\to{\rightarrow}      \def\too{\longrightarrow}

\def\calr{{\cal R}}
\def\V{{\cal V}}        
\def\U{{\cal U}}
      \def\H{{\cal H}}            
\def\A{{\cal A}}      \def\W{{\cal W}}      \def\F{{\cal F}}      
      \def\LL{{\cal L}}           
                  
\def\N{{\cal N}}

\def\Z{{\bf Z}}
\def\R{{\bf R}}
\def\C{{\bf C}}

\def\bz{{\bar z}}      \def\bw{{\bar w}}   \def\bh{{\bar h}}

\begin{document}
\pagenumbering{roman}     
\setcounter{page}{1}
\thispagestyle{empty}

\vspace*{20mm}

\begin{center}
{\LARGE
Two-Point Functions and Boundary States in Boundary Logarithmic
 Conformal Field Theories
}

\vspace{20mm}

{\Large
{\bf Yukitaka Ishimoto}

\vspace{5mm}
{\normalsize
Balliol College
}

\vspace{10mm}

{\large 
Theoretical Physics, Department of Physics, \\
University of Oxford, \\
1 Keble Road, Oxford OX1 3NP, U.K.}
}
\end{center}

\vspace{20mm}
\vspace{20mm}

\begin{center}
{\large 
Thesis submitted for the Degree of Doctor of Philosophy \\
in the University of Oxford\\[20pt]
- Michaelmas Term 2003 - 
}
\end{center}

\newpage
\baselineskip=14pt
\parskip=0pt
\thispagestyle{empty}
\vspace*{10mm}
\begin{center}{\Large\bf Abstract}\end{center}
{
Amongst conformal field theories,
there exist logarithmic conformal field theories 
such as $c_{p,1}$ models, various WZNW models, and a large variety of statistical models. 
It is well known that these theories generally contain
a Jordan cell structure, which is a reducible but
indecomposable representation. 
Our main aim in this thesis is to address the results and prospects of boundary
logarithmic conformal field theories: theories with boundaries that
contain the above Jordan cell structure.

In this thesis, 
we briefly review conformal field theory and the appearance of
logarithmic conformal field theories in the literature in the
chronological order. Thereafter, we introduce the conventions and
basic facts of logarithmic conformal field theory, and 
sketch an essential note on boundary conformal field theory.
We have investigated $c_{p,q}$ boundary theory in search of logarithmic
theories and have found logarithmic solutions of two-point functions in the context
of the Coulomb gas picture. Other two-point functions have also been studied in the free boson construction of BCFT with $SU(2)_k$ symmetry.
In addition, we have analyzed and obtained the boundary Ishibashi state for a $rank$-$2$ Jordan cell structure \cite{ishimoto1}.
We have also examined the (generalised) Ishibashi state construction and the symplectic fermion construction at $c=-2$ for boundary states in the context of the $c=-2$ triplet model \cite{kaw, bredthauer}.
It is also presented how the differences between two constructions should be interpreted, resolved and extended beyond each case. 
Some discussions on possible applications are given in the final chapter. 
}

\baselineskip=17pt
\parskip=5pt
\tableofcontents

\newpage
\thispagestyle{empty}
\vspace*{10mm}
\begin{center}
{\Large {\bf Acknowledgments}}
\end{center}

I am very grateful for all the support and help I had received not only in Oxford but also in many other places throughout my D.Phil course.

First and foremost, I am very grateful to my supervisors Prof Ian I. Kogan and Dr John F. Wheater for their guidance and stimulating discussions and suggestions with their great insight and enthusiasm.
I would like to thank Dr P. Austing, Dr R. Read, Dr A. Nichols for carefully reading my manuscript and encouraging me a lot.
I would also like to thank Mr A. Bredthauer for his discussions and comments.
I would also like to thank my office-mates and many others in Theoretical Physics, Oxford for their help, stimulating discussions, and sharing ideas.
I would like to thank researchers whom I have met in various conferences, from whom I have been inspired, encouraged and learnt a great deal.

I am very grateful to Dr J. W. Hodby, secretaries and many others in my Balliol College who have provided an intellectual environment and an enjoyable time in the college,
especially, to the people in the Holywell Manor who encouraged and discouraged me to write up this thesis.
I would also like to thank 
many of my friends whom I have met in Oxford and Kyoto
for their help and cheer.

I am grateful to New Century Scholarship Scheme and Balliol College, Oxford for their financial support.

Finally, special thanks to my family for their support.\\[20pt]

\newpage
\thispagestyle{empty}
\vspace*{70mm}
\begin{center}
{\it ``Once a riddle is solved, it creates ten new ones.''\\
Osamu Tezuka}
\end{center}

\newpage
\pagenumbering{arabic}
\setcounter{page}{1}
\chapter{Introduction}
\label{ch:intro}

Nearly two decades after the celebrated paper by Belavin, Polyakov and Zamolodchikov \cite{BPZ}, 
it is now clear that two-dimensional conformal field theory (CFT) 
is an essential mathematical tool and background to explore various models and theories in physics. 
Among them, there are two main branches of physics where CFT lies deeply. 
One is particle physics and the other is condensed matter physics.
CFT was extended to logarithmic conformal field theory nine years later \cite{gura}, but let us first focus on brief descriptions of CFT applications.

In particle physics, symmetry has played a great role in quantum field theory. 
It revealed the electroweak theory as a simplest example of unification, and QCD as a theory for quarks. 
They together gave rise to the successful standard model, leaving quantum gravity as a last requirement towards the final unification -- the ultimate theory of everything (TOE). 
For the consistency of quantum gravity, string theory was reintroduced in the early 1980s, when the basic formulation of CFT was established in \cite{BPZ}. 
Strings and their interactions form a two-dimensional surface called
the `world sheet' and replace many complicated graviton loops in
quantum field theory with simpler surfaces. 
It is two-dimensional conformal symmetry that emerges in the world-sheet physics and is a part of reparametrisation symmetry which in turn requires the whole theory to be in the specific dimensions, 
26 (10) for the bosonic (fermionic) case.
Here, CFT shows up as a necessity for string theories, towards the TOE.

On the other hand, in condensed matter physics, consider the $(2+1)$-dimensional electron systems which have a phase transition at some temperature. Before \cite{BPZ}, it was known that a certain class of such systems has some strange behaviours at its critical point. Namely, order parameters of the system obey power laws with constant exponents near the critical point as if scaling symmetry is enhanced. By a symmetry argument, which we will briefly mention in the following section, it turns out that CFT describes the system at criticality, giving rise to a correct set of those critical exponents of the order parameters \cite{BPZ}. It was also shown that, in some statistical models, the corresponding CFT gives not only the exponents but, in principle, all physics of the systems. 
The Ising model is an explicit example of such statistical models.
CFT arises as a requirement that any underlying non-conformal theory of such systems must flow into the CFT as it approaches the critical point.

Many contributions have been made to the conformal zoo of theories \cite{CFT,ketov}.
Starting from the Kac determinant and table \cite{kac}, the unitary minimal models have been found and developed in particle theories and statistical models \cite[3-8]{BPZ}. 
Some other mathematical frameworks have also been invented such as representation theories of various CFTs and partition functions on a torus in terms of character functions. We should also count here the affine Lie algebras, the Sugawara construction and the related Wess-Zumino-Novikov-Witten models (WZNW models) [3-13]. 
Free field realisations of the minimal models were given, including
the Dotsenko-Fateev construction \cite{Dotsenko:1984nm,Dotsenko:1984ad}, conformal ghost systems
and other fermionic systems for superstring theories
\cite{polchinskibook,GSWbook}. Superconformal field theories and their
free field representations have also been developed in relation to strings \cite{GKO,GKO1}. 
Especially, boundary conformal field theory (boundary CFT or BCFT) was nicely introduced by Cardy in the context of open strings and statistical models with finite size effects \cite{Cardy:bb}. We will mention this later. 

There is much more research on this subject. 
However, as stated above, most of the literature is restricted to the unitary CFTs with some extra symmetries.
Vast areas of other possibilities have not been greatly discussed or focussed on, 
especially those away from the unitary cages of the tamed zoo.
It is often the case with experiments, that
physical events happen and are detected in the absence of unitarity or outside the conformal region. 
Thus, we need another tool to explore the frontier of unitary CFT and beyond for a breakthrough in the subject.
In fact, there is a class of the theories called `logarithmic
conformal field theory (LCFT)' which is believed to lie widely at the border
of unitary and non-unitary conformal theories. 
It should be clarified here that what we are going to discuss in this thesis 
is those LCFTs with one or more boundaries. The Coulomb gas construction of $c_{p,q}$ models, a $c_{2,1}=-2$ model in particular, and the free boson realisation of the $SU(2)_k$ WZNW model will be given in the presence of a boundary. The results on their two-point functions will be shown explicitly. We also investigate boundary states in LCFTs, especially at $c=-2$, and study a few different constructions such as of (generalised) Ishibashi states and of coherent states [22-24]. 

It was first found and discussed by Knizhnik that a four-point
function of CFT has a logarithmic singularity in orbifold models \cite{Knizhnik:xp}.
The same sort of singularity was also discussed by Rozansky and Saleur in the $GL(1,1)$ WZNW model \cite{roz}.
Later, Gurarie revealed that, given such logarithms, 
logarithmic fields appear in the theory as a degenerate pair of fields
\cite{gura}. Thus discovered theory was named LCFT, since known CFTs
didn't contain such fields.
The main feature and a heuristic definition of LCFT is that there
appears a set of `logarithmic' operators which form a reducible but indecomposable representation of the $L_0$ operator, the zero mode of the Virasoro algebra \cite{gab3}.
In \cite{roh1}, this structure was generally defined as the {\it
Jordan cell structure}, and the pair of `logarithmic' operators in \cite{gura} emerged as a $rank$-2 Jordan cell.
One of the pair is purely logarithmic, giving a logarithmic singularity in its correlator, and the other gives a primary state of zero norm \cite{CKT}.
In principle, 
some minimal models of CFT may possess such fields,
although in many cases they are non-unitary or their central charges
are irregular. 
Nevertheless, this class of theories is worth studying
for new physics.
Note that we can always ignore their non-unitary nature as if they
were a subsystem. 
Thus far, many studies have been devoted to this subject and have found
the same sort of logarithmic behaviours in various models.
For example, 
the gravitationally dressed CFT and 
WZNW models at different levels or on different 
groups \cite[29-42]{roz}. 
$c_{p,1}$ and non-minimal $c_{p,q}$ models, including
the $c_{2,1}=-2$ 
model \cite[21-24, 27, 28, 30, 43-58]{gura}. 
$c=0$ models [59-62]. 
Critical polymers and percolation \cite{salu,kau3,card7,perco1,perco2},
quantum Hall effect,
quenched disorder and 
localization in planar systems [65-70]. 
2D-magneto-hydrodynamic and ordinary
turbulence [71-76]. 
LCFT in general [77-85]. 
In string theory, D-brane recoil, target-space symmetries and $AdS/CFT$ correspondence have been
studied and discussed with respect to LCFTs in the literature [86-106].

On the other hand, there is another class of theories called BCFT first considered by Cardy in \cite{Cardy:bb,card1,Cardy:tv} (see also [109-112]). 
BCFT is CFT with one or more boundaries. 
It was shown by Cardy that 
many tools developed in ordinary CFT can be imported into BCFT
by the use of his `mirror method', 
hence $n$-point functions become manageable \cite{Cardy:bb}.
As is often the case with experiments, when critical systems are
restricted to finite size, the corresponding theories turn out to be BCFTs.
Therefore, this class of theories is 
as essential as CFT in both particle physics and condensed matter physics.
In fact in string theory, theories of open strings are defined on an
infinite strip with two boundaries in its simplest cases.
A periodicity along its boundaries induces a dual `closed string' picture defined on the same geometry. 
Modular invariance leads to a one-to-one correspondence between the boundary conditions in the first picture and the boundary states in the other.
It was found in \cite{card1} that these boundary states are spanned by
boundary {Ishibashi} states \cite{ishibashi1}, by which the
Verlinde formula in \cite{verl1} is proven to
hold for unitary minimal models of boundary CFT. 
Note, from the point of view of brane solutions in string theory, the boundary states prescribe the dynamics between closed-string states on both branes.

Although the study of LCFT is still in progress, much progress has been made in both areas, LCFT and BCFT. 
Despite this, only a small number of papers have contributed to LCFT
with boundaries (boundary LCFT) and the effects of the presence of the boundaries [22-24, 50, 55, 56, 71, 83]. 
This is partially because there is a problem of reducible but indecomposable
representations which cannot be applied to boundary
CFT in a straightforward way.
The first systematic attempt to formulate a boundary LCFT was made by
Kogan and Wheater in \cite{KW}. Several important problems were 
discussed, including the boundary operators, their correlation functions, the structure of boundary states in LCFT, using a model at $c=-2$ as an example, and a possibility towards the Verlinde formula.

In 2001, our previous paper showed that a `pure' $rank$-2 Jordan
cell has only one {Ishibashi} state in the cell. It propagates from one boundary to another, behaving as a ghost state \cite{ishimoto1}.
From a basic mathematical framework developed by Rohsiepe \cite{roh1}, the structure of initial and final states of Jordan cells were reintroduced and a brief proof for the above statement was given in a purely algebraic manner.
The `pure' Jordan cell means that the Virasoro representations built on the cell do not contain any subrepresentations of lowest weight states nor become subrepresentations of any other representation. 
From the calculation on the cell \cite{ishimoto1, YI2}, we state that there
is only one Ishibashi state in each `pure' $rank$-2 Jordan cell
 and that it is the state built on the primary state of zero norm.

Soon, along the same lines, another computation was given in a symplectic fermion system at $c=-2$ by Kawai and Wheater \cite{kaw}. 
In this fermionic ghost system appearing in \cite{salu}, they constructed the coherent states which satisfy Cardy's equation for boundary states, and found another set of boundary states which differs from our set. 
Logarithmic terms occur in their character functions, which was absent in our previous result.
However, recent results found by Bredthauer and Flohr showed that
one may have generalised Ishibashi states for boundary states in the $c=-2$
theory, and that they might give the same logarithms in their character
functions \cite{bredthauer}. A following paper by Bredthauer revealed
that with a choice of (generalised) Ishibashi states one can see the isomorphic
relation between their results and other results on fermions in terms
of characters \cite{bredthauer2}. 

A question arises, namely what is the difference between the boundary 
states in \cite{ishimoto1}, those in \cite{kaw,kawai1}, and those in \cite{bredthauer, bredthauer2}. Our answer for the first two results is that they deal with the same model where our definition didn't count their generalised Ishibashi states but give a smaller set of their solutions. Then, the question is transfered to the difference between two constructions, namely, between the Ishibashi state construction and the coherent state, or the symplectic fermion, construction. By examining each construction, we come to an answer that these two cases should be called different models because they are constructed in mathematically and physically distinct ways. We will show these and additional results in chapter \ref{ch:boundary states}.

Apart from the boundary states, we would like to consider free field realisations in boundary logarithmic theories. While boundary states would give much information w.r.t. the boundary, the actual correlation functions need to be calculated or confirmed that they satisfy a certain differential equations. 
From such results of correlation functions, one can see the relations between their normalisation factors and those of boundary operators defined only on the boundary.
In chapter \ref{ch:free} we will examine a few cases of free boson constructions on the upper half-plane and single out necessary conditions for boundary LCFT in this context. 
Some of the boundary two-point functions will also be shown, which are in complete agreement with \cite{KW}.

The thesis is organised as follows.

In this chapter, 
we shall give a pedagogical introduction to the celebrated world
of CFTs, chiefly in two dimensions.
All referred facts and equations are well-known to those with experience in this subject.
In chapter \ref{ch:emergence}, we start reviewing the emergence of logarithms and a brief history of logarithmic CFTs, roughly in chronological order. 
In the following sections of the same chapter, we briefly review the definitions and formal
properties of LCFTs. This includes the definition of Jordan cell
structure, a review of sections of \cite{roh1} and a brief introduction to the models at $c=-2$, which we will focus on later.
In chapter \ref{ch:BCFT}, we discuss BCFTs that were introduced by Cardy in his successive papers \cite{Cardy:bb,card1,Cardy:tv,card6}.

Chapters \ref{ch:free} and \ref{ch:boundary states} contain the main results of this thesis.
In chapter \ref{ch:free}, we examine the Coulomb gas construction of general $c_{p,q}$ models and describe the $c=-2$ case in particular, in the presence of a boundary. Using the same techniques, the free boson realisation of the $SU(2)_k$ WZNW model with a boundary is studied. We also present boundary two-point functions of these models. 
In chapter \ref{ch:boundary states}, we prove the existence of the boundary Ishibashi state for
the $rank$-$2$ Jordan cell structure and show its explicit form \cite{ishimoto1}.
In the following sections, the other two constructions are reviewed and confirmed \cite{kaw,bredthauer}. Then, we describe and clarify the differences between the constructions, referring our results and \cite{bredthauer2}. Some additional results to compare will also be presented.

In chapter \ref{ch:conclusion}, we present and summarise the results on chapters \ref{ch:free} and \ref{ch:boundary states}. We give a remark on the Verlinde formula for LCFT and possible applications to string theory. 
We also discuss the generalisation of the above results and possibilities of boundary LCFTs and prospects of LCFTs to close.

The appendices contain: the first appendix is the
explanation of radial quantisation mainly for the introduction to CFT. 
The second appendix is a proof of relations between hypergeometrical functions appearing in chapter \ref{ch:free}.

\newpage
\section{Conformal Symmetry}
\label{sec:CFT1}

Nearly two decades ago, conformal field theory was formulated by Belavin,
Polyakov, and Zamolodchikov as ``the massless, two-dimensional,
interacting field theories,'' which are invariant under ``an infinite
dimensional group of conformal (analytic) transformations.''
By then, it was also known as the maximal kinetic extension of relativistic invariance. 
The connection to the context of second-order phase transitions and 
its historical point of view are also described briefly in \cite{BPZ}.

Their breakthrough revealed how rich a structure two-dimensional field
theories may possess and it boosted the establishment of integrable
models and string theory in its early stages. String theory needed a
well-established theory of the world sheet, and integrable models were built up partially by the emergence of unitary minimal models, which we will describe in the following section.

In this section, we briefly give an introduction and basic facts on
conformal invariance and the resulting CFT, based on
\cite{BPZ,CFT,ketov}. 
To begin with, in order to realise how CFT looks, we shall sketch what the conformal symmetry is, and the effects on field theories.

Consider a smooth $d$-dimensional space characterised by a symmetric
metric $g_{\mu \nu}(x)$, where $x$ is a $d$-dimensional vector, including a
time-like coordinate $x^0$.
An invariant line element is given by $ds^2(x) = g_{\mu \nu}(x) dx^\mu dx^\nu$.
Under a finite version of general coordinate transformation, $x^\mu \too x^{\prime \mu}$, the metric is transformed as
\bea
 \label{def:generalcoordinate}
   g_{\mu \nu}(x) \too {g_{\mu \nu}}^\prime (x^\prime)
   = \frac{\pa x^\lambda}{\pa x^{\prime \mu}} \frac{\pa x^\kappa}{\pa x^{\prime \nu}} g_{\lambda \kappa}(x) ,  
\eea
while the line element is invariant.

The conformal transformations are a subset of the general coordinate
transformations which preserve angles between any two vectors
in question. 
Provided that there exists a proper definition of norms, 
the angle of two vectors $v_1^\mu, v_2^\mu$ is given by:
\[ \frac{g_{\mu \nu}(x) v_1^\mu v_2^\nu }{\|v_1^\mu\| \|v_2^\nu\|} , \]
where $x$ is a point of the space where two vectors reside. 
Under the general coordinate transformation shown above, the invariance of the angle restricts the transformed metric to the following form:
\[ g^\prime_{\mu \nu} (x) = \Omega(x) g_{\mu \nu}(x) ,\]
where $\Omega(x)$ is a function of $x$. This is highly restrictive and actually leads to the whole richness of conformal field theory which we will focus on in most of our thesis.

Introducing the infinitesimal transformation of (\ref{def:generalcoordinate}) as $x^\mu \too x^\mu + \epsilon^\mu(x)$, 
one can easily see the details of this transformation as conditions on $\epsilon^\mu(x)$:
\bea
 \label{def:conformaltf}
  \pa_\mu \epsilon_\nu + \pa_\nu \epsilon_\mu
   = \frac2d \left(\pa_\lambda \epsilon^\lambda \right) \delta_{\mu \nu} ,
\eea
and its corollary is
\bea
 \label{def:conformaltfcol}
  \left\{ \delta_{\mu \nu} \Box + (d-2) \pa_\mu \pa_\nu \right\}
(\pa_\lambda \epsilon^\lambda) = 0. 
\eea
Here, we suppress the variable $x$ of $\epsilon_\mu (x)$ and it is assumed that the space is locally isomorphic to $\R^d$
and, therefore, so is the metric as $g_{\mu \nu}\sim\delta_{\mu
\nu}$. We will assume this geometry in most of our thesis, unless
otherwise stated.

From the eq. (\ref{def:conformaltfcol}), it can be easily seen that the
algebraic structure dramatically changes from $d=2$ to $d>2$. The
function $\epsilon^\mu$ is
at most quadratic in $x$ when $d>2$, and it
determines the finite number of generators of the conformal group in
question, while the two-dimensional case exhibits its infinite
dimensionality. 
For example, the conformal transformations of four-dimensional space
are given by fifteen generators: four translations, six rotations, one
dilatation, and four special conformal transformations.

From the group theoretical point of view, if we assume $d$-dimensional
Minkowski space instead of an Euclidean 
one, the Poincare symmetry of the space is the noncompact group
$SO(d-1,1)$. On the other hand, the conformal group 
of the space becomes a larger noncompact
group $SO(d,2)$, one of whose subgroups is the Poincare
group. In fact, it is $SO(3,1)\subset SO(4,2)$ for four dimensions and
$SO(1,1)\subset SO(2,2)$ for two
dimensions. 

Let us focus on two dimensions in what follows. We will consider
conformal field theory only in two dimensions in the rest of our
thesis. However, note that many of notions and techniques 
in two dimensions are
applicable to higher dimensions.

With the corollary (\ref{def:conformaltfcol}) in two
dimensions and a change of variables $z=x_0 + i x_1$, $\bz=z^*$,
we can extract the relevant conditions for $\epsilon_z (z,\bz)$ and
$\epsilon_\bz (z,\bz)$,
\[ \bar \pa_\bz \epsilon_z (z,\bz) = \pa_z \epsilon_\bz (z, \bz) = 0 , \]
where $\pa_z \equiv\frac{\pa}{\pa z},\, \bar \pa_\bz\equiv
\frac{\pa}{\pa\bz}$. We also use the notations $\pa\equiv\pa_z,\, \bar
\pa\equiv\bar \pa_\bz$ for convenience.
Clearly, this is a chiral condition of the function and indicates that
the two-dimensional conformal transformations can be represented 
by holomorphic and anti-holomorphic functions,
$\epsilon (z) \equiv \epsilon_z (z, \bz),\,\, \bar \epsilon (\bz)\equiv\epsilon_\bz (z,\bz)$.

All generators of the holomorphic part, $\{ l_{m\in\Z}\}$, can be
expressed by $l_m \equiv - z^{m+1} \frac{d}{d z}$.
Thus, the corresponding local (but classical) conformal algebra is infinite
dimensional with two sectors: the holomorphic and anti-holomorphic sectors. 
The commutation relations of the holomorphic sector are shown below.
\bea
 \label{def:conformal alg}
 \com{l_m}{l_n} = \left( m-n \right) l_{m+n} .
\eea
The anti-holomorphic sector is given similarly 
while being independent of its holomorphic counter part. 
Because of its independence, or equivalently decoupling, 
we will neglect the anti-holomorphic sector,
unless otherwise stated.

Despite its outstanding feature of infinite dimensionality, 
it should be mentioned that only a finite set of the generators are globally well-defined. 
More precisely, 
the global symmetry is $SL(2,\C)$ and is known as 
the complex M$\ddot{o}$bius transformations. 
These transformations, as a general element of the group, are given by 
$z \too z^\prime = \frac{a z + b}{c z + d}$, 
and are parameterized by four constants, $a, b, c, d \in\C$, 
with the constraint, $ad-bc=1$. 
In the presence of this global symmetry, we can assign two quantum numbers $h$
and $\bh$ to states of the theory, as eigenvalues of the Cartan
subalgebras of two mutually commutating holomorphic and
anti-holomorphic global symmetries.
Thus, it is expected that the whole local algebra may generate highest weight representations as usual in quantum field theory, which turn out to be lowest weight representations of the algebra in the following section.

\section{A Framework of CFT}
\label{sec:CFT2}

From the field theoretic point of view, let us
suppose that there is an action and thus an energy-momentum
tensor $T_{\mu \nu}(z,\bz)\equiv - \frac{4\pi}{\sqrt{g}} \frac{\delta S(z,\bz)}{\delta g^{\mu \nu}}$. The invariant line element is $d s^2 = d
z d \bz$. 

The energy-momentum tensor must satisfy the following conservation law: 
$\pa^\mu T_{\mu \nu}=0$. 
In addition, 
conformal invariance imposes tracelessness of the tensor: $T^\mu_{~\,\mu} = 0$, 
which is equivalent to the conservation law of the dilatational current.
In our complex coordinates, this yields $\pa_\bz T_{z z}= \pa_z
T_{\bz\bz}=T_{z\bz}=T_{\bz z}=0$ and a decomposition of $T_{\mu \nu}$ by $T(z)\equiv T_{z z}(z)$ and 
$\bar T (\bz)\equiv T_{\bz \bz}(\bz)$ for holomorphic and anti-holomorphic sectors, respectively.

The energy-momentum tensor is a field of conformal dimension two\footnote{
This is because the metric $g^{\mu \nu}$is a second-rank tensor while
the action is a scalar under coordinate transformations. 
However,
its finite conformal transformation is not expressed only by the form
 (\ref{def:primary field}) but with an extra Schwartzian derivative term. See
\cite{CFT,ketov}.
}, having
the following mode expansion
\bea
 \label{def:EM mode}
   T(z) = \sum_{n\in\Z} L_n z^{-n-2}.
\eea
Equivalently, 
\bea
 \label{def:Virasoro generators}
   L_n \equiv \oint \frac{d z}{2 \pi i} z^{n+1} T(z),
\eea
each of which constitutes a generator of the Virasoro algebra---quantum conformal algebra. 
The closed contour is arbitrary around the point $z=0$ in the above case.
\footnote{
Its shape is irrelevant, thanks to Cauchy's theorem.
}

The definition of the Virasoro algebra is given 
by the following operator product expansion (OPE) of the $T(z)$.
\bea
 \label{def:EM OPE}
  T(z) T(w) \sim \frac{c/2}{(z-w)^4} + \frac{2 T(w)}{(z-w)^2} + \frac{\pa_w T(w)}{z-w}, 
\eea
where a constant $c$ is called the `central charge' or `conformal anomaly'.\footnote{
One may take a different central charge $\bar c$ for the anti-holomorphic sector, but we assume $c=\bar c$. To derive the relations (\ref{def:Virasoro}) from (\ref{def:EM OPE}), see also the radial quantisation on p.\pageref{sec:radial quantisation}.
}
With eq. (\ref{def:Virasoro generators}), the commutation relations\footnote{
\label{def:operator ordering}
Given a field $A(z)$ of conformal dimension $h$ and $B(w)$,
the commutation relation between two fields can be computed by the following way 
\bea
 \label{def:ordering}
   \com{A_n}{B(w)} \equiv \oint_w d z z^{n+h-1} A(z)B(w),
\eea
where $A_n$ is a $n$-th mode of $A(z)$. For more details, see the Schwinger's time-splitting technique, for example in \cite{ketov}.
}
of the Virasoro algebra are given by
\bea
 \label{def:Virasoro}
  \com{L_m}{L_n} = \left( m-n \right) L_{m+n} + \frac{c}{12} m(m^2-1) \delta_{m+n,0} . 
\eea
The presence of the charge, $c$, is the most prominent difference between
these relations and those of (\ref{def:conformal alg}), by which the
main characteristics of the theory are also determined [See the next section.]. 
Note that the Virasoro algebra (\ref{def:Virasoro}) can be recognized as a central extension of the two-dimensional conformal algebra (\ref{def:conformal alg}), which is uniquely determined by the Jacobi identity up to constant shifts of the generators.

In addition to the fields which define the action and the
energy-momentum tensors, there are so called primary fields, or simply primaries. A definition of a primary field, labeled by two numbers $(h, \bh)$, is given by 
\bea
 \label{def:primary field}
  \Phi = \Phi_{h,\bh} (z,\bz) (d z)^h (d \bz)^\bh ,
\eea
where $\Phi$ is invariant under conformal transformations. 
$\Phi_{h,\bh}(z,\bz)$ is a non-chiral primary field, 
and $h$ $(\bh)$ is the conformal dimension of the (anti-)holomorphic sector. 
Without much loss of generality, we may assume that all primaries can
be decomposed into two sectors as $\Phi_{h, \bh}(z,\bz) = \Phi_h(z)
\otimes \Phi_\bh(\bz)$. However in some cases, one should discard this
assumption. We will discuss this point later in the appropriate
section. 

Generally speaking, a conformally invariant theory possesses more than one primary. 
Under a general conformal transformation, $z \to z+ \epsilon(z)$, their common transformation property is
\bea
 \label{def:tf o primary}
   \delta_\epsilon \Phi_h(z) = \left( h \pa\epsilon(z) + \epsilon(z) \pa
\right) \Phi_h (z).
\eea
One may regard the above equation as the definition of primaries
instead of eq. (\ref{def:primary field}). 

In terms of the OPE, one may define this $\delta_\epsilon$ by
\bea
 \label{def:delta T}
   \delta_\epsilon = \oint \frac{d z}{2 \pi i} \epsilon(z) T(z).
\eea
Setting the following OPE between a primary and the energy-momentum tensor,
the expression (\ref{def:tf o primary}) can be derived from a contour integral.
\bea
 \label{def:primary OPE}
  T(z) \Phi_{h}(w) \sim \frac{h \Phi_{h}(w)}{(z-w)^2} + \frac{\pa_w \Phi_{h}(w)}{z-w}.
\eea
By substituting $\epsilon(z)=z$ into eq. (\ref{def:tf o primary}),
or taking the zero mode of $T(z)$ in (\ref{def:primary OPE}),  
one can easily check that $h$ $(\bh)$ is in fact the eigenvalue of 
the $L_0$ $(\bar L_0)$ operator at $z=0$.\footnote{
In the context of statistical models, primaries are order parameters and these eigenvalues correspond to critical exponents of them.
}

By acting with the modes of $T(z)$ on a primary successively, there arise a family of fields: 
\bea
 \label{def:secondary}
   \{ L_{n_1}^{k_1} L_{n_2}^{k_2} \cdots L_{n_p}^{k_p} \Phi_h(z) \} ~~~{\rm for~} n_1, \cdots, n_p, k_1, \cdots, k_p \in \Z . 
\eea
These fields are called descendants or secondary fields, and the sum
$N$ of $\{n_i\}$ is called the level of the field, since it gives the difference of conformal dimensions between the primary field and its descendant.
In general, the elements of (\ref{def:secondary}) are independent of
each other and constitute a family of fields as an infinite dimensional
representation of the Virasoro algebra.
The family is called a conformal family $[\Phi_h]$, which is here a Verma module of the Virasoro algebra.
The full algebra consists of two chiral ones.

Now that we have \alert{the} basic fields and operators, 
let us turn to the state space and its properties.

Given a field $A(z,\bz)$, an IN state and an OUT state can be defined by taking the following limits of an $SL(2,C)$ invariant vacuum\footnote{
The vacuum also satisfies 
\bea
 \label{def:vacuum}
   L_{n\geq -1} \ket{vac} = 0,
\eea
since otherwise the energy-momentum tensor becomes singular at the origin.
If $n\geq 0$ in the above condition, it is nothing but the lowest
weight state condition on the vacuum. 
These are simply  manifestations of the conformal invariance of the vacuum.
}
acted on by the field,
\bea
 \label{def:in-out beakets}
  \ket{IN} &=& \lim_{z,\bz\to 0} A(z,\bz) \ket{vac},
\nn
  \bra{OUT} &=& \lim_{w,\bw\to 0} \bra{vac} A\left( \frac1w, \frac1\bw \right) w^{-2h} \bw^{-2 \bar h} .
\eea
For the reason why $z=0$ ($z=\infty$) is taken as the initial (final) point, see the radial quantisation in appendix A.

Likewise, primary states in $\ket{IN}$ can be defined by primary fields as
\bea
 \label{def:lws}
   \ket{h} \equiv \lim_{z \to 0} \Phi_h (z) \ket{vac}
\eea
Though it may be confusing, it is customary to call both fields and states `primaries'. 
The assumption of the decomposition is inherited from the fields by the states: a full
primary is a tensor product of two chiral primaries, $\ket{h,\bh} =
\ket{h}\otimes\ket{\bh}$, and so are their descendants. 

One can easily check that the primaries satisfy 
\bea
 \label{def:unitarity for L}
   \bra{h} L_{n<0} =0 \,,\quad L_{n>0} \ket{h} = 0 ,
\nn
   L_0 \left( L_{n<0} \ket{h} \right) = (h-n) \left( L_{n<0} \ket{h}\right),
\eea
where $n\in\Z$.
Therefore, primaries are indeed lowest weight states (LWS), since
$h-n>h$ in the above expression.\footnote{
These LWSs are sometimes called
highest weight states, depending on the convention. The meaning is the same.
}
Together with their descendants, the states comprise lowest weight
Verma modules $\{\V_{h}^{Vir}(c)\}$ of the Virasoro algebra. 

To summarise, a conformally invariant
theory can be defined by a symmetry algebra\linebreak
$\A \supset Vir$ with a central charge $c$, its Verma modules $\{\V_{h}^\A(c)\}$ for a set of conformal dimensions $\{h\}$, and a closure of the given modules {\it i.e.} the operator algebra hypothesis. The hypothesis is a strong version of the Wilson OPE:
\bea
 \label{def:operator alg}
   \Phi_i (z) \Phi_j (w) = \sum_{k\in\{\V_h^\A (c)\}} C_{ij}^{~ k}(z,w) \Phi_k (\frac{z+w}{2})
\eea
for any pair of fields in the given modules, provided that the whole
set of the modules are complete. If the modules are all irreducible,
it is known that the relation (\ref{def:operator alg}) can be
represented by a restricted set of the fields. In addition, the set
only consists of primaries in some special cases, where the suffices $i,j$ completely fix the possible $k$'s. 
We will mention this in the following section.

Even if we define the theory by the above set, exact forms of physical
quantities do {\it not} seem to be obvious without taking any explicit
field representations. However the partition function and, therefore,
$N$-point functions have infinitely many constraints due to the
conformal invariance. 
When the $N$-point function $G^{(N)}$ transforms to ${G^\prime}^{(N)}$ under the infinitesimal conformal transformations, the conformal invariance requires
\bea
 \label{def:conf inv on Npt}
   {G^\prime}^{(N)} - G^{(N)} = \delta_\epsilon G^{(N)} = 0.
\eea
Rewriting this using (\ref{def:delta T}) and (\ref{def:primary OPE}), we obtain the Ward identity
\bea
 \label{def:Ward}
   \braket{T(z) \Phi_1(z_1) \cdots  \Phi_N(z_N)}
 = \sum_{j=1}^{N} \left[ \frac{h_j}{(z-z_j)^2} + \frac{1}{z-z_j}
\frac{\pa}{\pa z_j} \right] \braket{ \Phi_1(z_1) \cdots  \Phi_N(z_N)}.
\eea
Likewise, the following identities hold for the $(N+1)$-point functions.
\bea
 \label{def:diff op}
   \braket{\Phi_1(z_1) \Phi_2(z_2) \cdots \Phi_N (z_N) \left( L_{-n} \Phi_{N+1}(z_{N+1})\right) }
\nn
  = \hat \LL_{-n} \braket{\Phi_1(z_1) \Phi_2(z_2) \cdots \Phi_N (z_N) \Phi_{N+1}(z_{N+1})} , 
\eea
where 
\bea
 \label{def:diff op1}
   \hat \LL_{-n} \equiv - \sum_{j=1}^N \left[ \frac{(1-n) h_j}{(z_j - z_{N+1})^n} + \frac{1}{(z_j - z_{N+1})^{n-1}} \frac{\pa}{\pa z_j} \right] . 
\eea
Replacing $\Phi_{N+1}(z_{N+1})$ by the identity operator $I(z_{N+1}=0)$ \footnote{
This is equivalent to the $sl(2,\C)$ invariant vacuum at the point $z_{N+1}$.
}, 
they reduce to the identities of global conformal invariance for generic 
$N$-point functions ($|n|\leq 1$):
\bea
 \label{def:diff op2}
  0 = \LL_{-n} \braket{\Phi_1(z_1) \Phi_2(z_2) \cdots \Phi_N (z_N)} , 
\quad
   \LL_{-n} \equiv - \sum_{j=1}^N \left[ \frac{(1-n) h_j}{z_j^n} + \frac{1}{z_j^{n-1}} \frac{\pa}{\pa z_j} \right] .
\eea
These differential operators indeed restrict the possible forms of the functions either drastically or exactly.\footnote{
In fact, two-point and three-point functions are determined only by the conformal invariance in many cases.
}
We will see the latter case in the next section.

All of the properties we have introduced above hold in any theory as long
as it is conformally invariant in the sense we showed.
We haven't introduced many other basic notions of CFT such as the fusion
product, conformal block and duality (including the crossing symmetry) on the blocks, monodromy invariance, partition function on a torus and its modular property, and so on.
As for field representations, we will briefly review them and study the Coulomb gas construction in chapter \ref{ch:free}.
For more details of CFT, refer to \cite{BPZ,CFT,ketov} and the references therein.

\section{The Unitary Minimal Models}
\label{sec:minimal}

As one can easily imagine from the previous section, there are a wide
variety of CFTs, depending on the central charge $c$ and the choice of
the modules $\{\V_{h}^\A(c)\}$. Among them, there are the unitary {\it minimal models} with a finite set of primaries, which are exactly solvable.

It is known that a conformal theory may possess a particular type of degenerate Verma module for a certain set of $c$ and $h$. This type of degeneracy can be removed from the modules, leaving only irreducible modules in the theory. 
Minimal models are the conformal theories of $c<1$ with a finite set of those truncated irreducible Verma modules, and were first discovered and classified in \cite{BPZ}.

This class also belongs to Rational Conformal Field Theory (RCFT), which is CFT with a finite number of primaries.
The reason of the word `minimal' is because 
it is RCFT, has no multiplicity in spectrum and no additional symmetry, according to \cite{ketov}.
In this section, we introduce these minimal models, and give the main outcomes of them.

Suppose that, amongst a conformal family of a primary $\ket{h}$, there is a so-called {\it null state} $\ket{\chi}$, or {\it singular vector}, 
which 
satisfies the conditions
\bea
 \label{def:null vector}
   L_0 \ket{\chi} = (h + N) \ket{\chi}, \,\, L_{n>0} \ket{\chi} =0 ,
\nn
  \braket{ any~ state \vert\chi} = 0 ,
\eea
where $N$ is a positive integer.
When the theory is degenerate, the degenerate descendants satisfy the
first equation of the first line in (\ref{def:null vector}). By
imposing the second line, one can finally get rid of the state and its
descendants. For this condition to be conformally invariant, the
second equation of the first line is required to be satisfied. Then
the null state turns out to be a lowest weight state (LWS) since the second equation is nothing but the LWS condition.
If one subtracts all of these degenerate fields from the theory, the conformal families become irreducible.

As the null state is a descendant of $\ket{h}$,
one may consider a linear space of the descendants at level $N$ and a matrix of their scalar products
\bea
 \label{def:matrix o descendants}
   \hat M_{\{m\},\{n\}}^{(N)}(c,h) \equiv \bra{h}\left( L_{m_p} \cdots L_{m_1} \right) \left( L_{-n_1} \cdots L_{-n_q} \right) \ket{h},
\eea
where $\sum_{i=1}^p m_i=\sum_{i=1}^q n_i= N$.

The determinant of this matrix is called the Kac determinant (\cite{kac} and
see also references in \cite{ketov}),
\begin{eqnarray}
 \label{def:Kac det}
   \det M^{(N)} (c, h)= \prod_{k=1}^N \prod_{rs=k} \left( h-h(r,s) \right)^{P(n-k)} ,
\end{eqnarray}
where
\begin{eqnarray}
 \label{def:h(m,n)}
   h(r,s) &=& \frac{1}{48} \Biggl\{ (13-c)(r^2+s^2)
+\sqrt{(1-c)(25-c)}\,(r^2 - s^2) 
\nn
&&\qquad  {}-24rs-2(1-c) \Biggr\} \quad for~ r,s\in\Z^+,
\end{eqnarray}
and $P(N)$ is the number of linearly independent descendants at level $N$.
The above two equations were found in a heuristic way.
The expression in (\ref{def:h(m,n)}) can be rewritten with three numbers $\alpha_0, \alpha_\pm$ by
\bea
 \label{def:h n charges}
&&   h(r,s) = -\frac12\, \alpha_0^2 + \frac18 \left( r \alpha_+ + s \alpha_- \right)^2,
\nn
&&  c=1-12 \alpha_0^2,\quad \alpha_+ + \alpha_- =2\alpha_0,\quad 
 \alpha_+ \alpha_- =-2,
\eea
while $\alpha_\pm$ can be expressed using only $c$
$$
\alpha_\pm = \frac{\sqrt{1-c} \pm \sqrt{25-c}}{\sqrt{12}}.
$$
These three numbers will appear again as `charges' in chapter \ref{ch:free} for the Coulomb gas construction.\footnote{
The normalisation of the charges varies in the literature. So, we fix it as above for later convenience.
}

Obviously, the existence of the null state means a zero of the Kac
determinant (\ref{def:Kac det}) and the reverse holds: zeros of the Kac determinant mean
degeneracies and null states of zero norm. 
If $h=h(r,s)$ then there is a null state at level $N=rs$ and the Verma module becomes irreducible.

Moreover, the determinant and the dimension $h(r,s)$ tells us that, when
$c>25$, the reduced theory is non-unitary with negative dimensional fields.
When $1<c<25$, the conformal dimensions are generally complex.
When the ratio of $\alpha_\pm$ is rational, a constant $h=h(r,s)$
contains an infinite number of $(r,s)$ pairs and, therefore, the module
contains an infinite number of null states. The quotient modules lead to
the operator algebra, and this closes within a finite set of conformal families.

The minimal models are special cases of rational CFT with the
rational number $\alpha_+/\alpha_-$ such that
\bea
 \label{def:minimal}
&& \frac{\alpha_+}{\alpha_-} = - \frac{p}{q},\quad c= 1 - \frac{6 (p-q)^2}{pq},
   \nn
&& h(r,s)= \frac{1}{4 pq} \left\{ (rq-sp)^2 - (p-q)^2 \right\},
\eea
for $p,q\in\Z^+$ and $coprime$.
Only when $q=p+1$ and $p\geq 2$, unitary representations are not
excluded, in which case 
$$
c=1-\frac{6}{p(p+1)}, \quad 
h_{r,s} = \frac{1}{4p(p+1)} \left\{ \left( r(p+1)-sp \right)^2 -1 \right\} .
$$

\begin{figure}[h]
\vspace*{1.5cm}
\hspace*{5.0cm}
\epsfysize=7cm
\epsffile{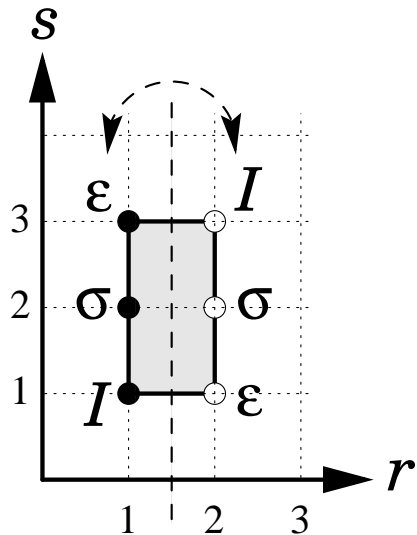}
 \caption{The Kac table for the Ising model ($c=1/2$).}
      \label{fig:Kac}
\end{figure}

Since $h(r,s)=h(p-r,q-s)$, we can restrict $r$ and $s$ to the rectangular region
\bea
 \label{def:rs}
  0<r<p,\, 0<s<q, p<q. 
\eea
These dots in the rectangle are illustrated in the Kac table (Fig.\ref{fig:Kac}). 
An outstanding feature of
the minimal models is represented by these dots: the discrete set of
conformal dimensions and families $\{(h(r,s),\Phi_{r,s})\}$ within the rectangle
closes within themselves, provided that the mirror
symmetry $\Phi_{r,s}=\Phi_{p-r,q-s}$ is imposed on the fields. Furthermore, all primaries listed in the table have null vectors at level $r\times s$, because of which any multi-point function become a solution of a certain linear differential equation. The derivation of the equation will be shown shortly after the `fusion rule'.
Note that this rectangle condition can be interpreted as a unitarity condition
on the theory, because it actually eliminates all fields of negative
conformal dimension.

The operator algebra of the minimal models is known to be completely represented by the `fusion rules' of the primaries:
\bea
 \label{def:fusion}
   [\Phi_{r,s}] \times [\Phi_{r^\prime, s^\prime}]
   = \sum_{j=r-r^\prime+1}^{r+r^\prime-1} \sum_{k= s-s^\prime+1}^{s+s^\prime-1} [\Phi_{j,k}].
\eea
For the unitary minimal models, the indices $j$ and $k$ are bounded in
 the rectangular region, and the fusion rule coefficients $\{C_{ij}^{~k}\}$ are fixed.

It is worth pointing out here that the null state in (\ref{def:null vector}) can be
expressed by a linear combination of the descendants at level $N$. 
\bea
 \label{def:null vector1}
   \ket{\chi} = \left( \sum_{\{n_i\}} b_{\{n_i\}} L_{-n_1} \cdots L_{-n_k} \right) \Phi_h(0) \ket{vac},
\eea
where $\sum_{i=1}^{k} n_i =N$ and $b_{\{n_i\}}$ denotes its coefficient. The summation is over all the descendants at level $N$.
When the null state is inserted at the point $z\not=0$ in a correlation
function, the function must vanish while the Virasoro generators are
reinterpreted as those defined at the point $z$.
Recall that every generator can be expressed by an appropriate
differential operator shown in (\ref{def:diff op1}), thus
the vanishing correlator must satisfy a nontrivial differential equation.
For example, the linear differential equation for a four-point function with a null state at level two of $\Phi_{1,2}(z)$ is: 
\bea
 \label{def:diff eq o 4pt}
   \left\{ \frac{3}{2(2h+1)}\frac{\pa^2}{\pa z^2} - \sum_{i=1}^3 \frac{h_j}{(z-z_j)^2} - \sum_{i=1}^3 \frac{1}{z-z_j} \frac{\pa}{\pa z_j} \right\} &&
\nn
 {}\times \braket{\Phi_1(z_1) \Phi_2(z_2) \Phi_3(z_3) \Phi_{1,2}(z)} &=& 0.
\eea
In general, solutions of eq. (\ref{def:diff eq o 4pt}) are given by 
some hypergeometric functions of the type ${}_2\F_1 (a,b;c; \xi)$. $\xi$ is an anharmonic ratio of four variables. 
With this sort of equation and solutions, the unitary minimal models
become exactly solvable. 

From the structure of the degeneracies, the partition function on a
torus can also be computed and turns out to be given by
the Riemann theta functions and the Dedekind eta function $\eta(\tau)$, which will appear in chapter \ref{ch:LCFT}.
It also satisfies the modular invariance of the theory, which will be discussed
further in the following chapters.

In this section, we have sketched the minimal models with two positive numbers $p$ and $q$. 
In what follows, 
we call the models `minimal series' which satisfy the condition 
(\ref{def:minimal}) and consist only of the discrete set of 
primaries $\{ h_{r,s}\equiv h(r,s), \Phi_{r,s} \}$. 

Now a question arises. If one takes different values of $p,q$, that is, $p<4$ or $q\not= p-1$, what may happen?
Some of the previous facts still hold: namely, 
the existence of linear differential equations for the correlation functions, 
the closure within the fields listed in the Kac table. 
The power of the Virasoro symmetry still remains.
When $p=3,q=2$, the central charge vanishes. If we take the 
standard way of understanding it, the theory is trivial, since the
rectangle only leaves $(r,s)=(1,1)$, 
which is the identity operator $I$.\footnote{
Note that trivial $c=0$ theory is sometimes included in the minimal models.
The rectangle doesn't shrink to a point but the only field is the trivial identity owing to the mirror symmetry.
}
Now we encounter a problem of non-unitarity, $c_{2,1}=-2$. 
The argued rectangle totally vanishes outside the unitary cage, 
and the operator algebra does not have the rationality in the RCFT sense, 
but has the quasi-rationality given by Nahm \cite{nahm}. 
Namely, when the finite (or semi-infinite) set of fields 
close under fusion, the theory is called (quasi-) rational. 
\label{def:rationality0}

\chapter{LCFT}
\label{ch:emergence}
\label{ch:LCFT}

On the basis of the beauty of CFT shown in the previous chapter, a huge number of models, theories and techniques have been studied and developed so far. However, most of them are devoted to unitary theories such as the minimal models and their supersymmetric extensions.
In 1993, LCFT emerged as a consequence of the study of a non-unitary model of $c=-2$ by Gurarie \cite{gura}. The origins of this were the same logarithms appearing in the $GL(1,1)$ WZNW model examined by Rozansky and Saleur \cite{roz}. Shortly after its first appearance in the literature \cite{gura,Knizhnik:xp,roz}, similar logarithms were found in various distinct models \cite{CKT,Caux:1996kq,KLS}. Finally, they all together led us to the newly created world of LCFT.

In this chapter, we first give a brief history of the emergence of the logarithms. Then we show a simple way of finding such a logarithm in a CFT and hence the origin of LCFT.
This phenomenon has been seen in various models as described in the introduction. 
Using them, we describe the defining features of LCFT in the end of section \ref{sec:emergence} \cite{CKT}. 
Secondly, we review mathematical definitions of Jordan cell structure
which were studied in \cite{roh1}.
Fixing their notations,
we describe Jordan lowest weight vectors, their modules and the corresponding states in bra and ket state spaces in section \ref{sec:math-def}. 
In section \ref{sec:lcft c=-2} and \ref{sec:rational characters} we address the known facts of the simplest cases -- the $c=-2$ models, and character functions of the rational models.

\section{The Emergence of Logarithms in CFT}
\label{sec:emergence}
\subsection{Logarithms in The Early Stages}
\label{sec:logarithms}

It was in 1993 that Gurarie first discovered a set of peculiar fields called `logarithmic operators' \cite{gura}. 
Namely, there appears a logarithmic singularity in a four-point
function in a $c=-2$ non-unitary model. 
In conjunction with the operator algebra hypothesis (\ref{def:operator alg}), the logarithmic term leads to the emergence of operators whose correlators contain logarithms as well. Besides, the logarithmic operators, a pair in the case of \cite{gura}, turn out to be degenerate and form a Jordan cell structure of the $L_0$ operator.

The non-trivial but straightforward evidence of such operators is a logarithmic singularity in a four-point function.
Such functions had already appeared in the literature 
before Gurarie discussed them \cite{Knizhnik:xp, roz}. 
In \cite{roz} in 1992, such singularities had been found and discussed in the $GL(1,1)$ supersymmetric WZNW model in the context of dilute polymers and percolation, where the conformal symmetry is enhanced at its criticality. 
The continuous limit of dense phase of the polymers corresponds to a model discussed by Gurarie. 
They also indicated that a spectrum of dilute polymers seems to coincide with the Fock space of the twisted $N=2$ superconformal theory, or $GL(1,1)$ WZNW model. 
In fact, they showed that the $\xi$-$\eta$ ghost system of $c=-2$ has the same partition function as the continuous limit of dense polymers. 
It should be mentioned that the symmetry of both systems is $GL(1,1)/U(1)$, a subgroup of the above mentioned $N=2$ symmetry.
This ghost system will be dealt with later in chapter \ref{ch:boundary states}.

Surprisingly, similar logarithms and structure were discovered in 1995 by Bilal and Kogan in a totally different model and formulation -- a gravitationally dressed CFT \cite{BK}. 
In the first reference of \cite{BK},
they studied correlation functions in 2d CFT coupled to induced gravity in the light-cone gauge. 
In the light-cone gauge, gravitationally dressed $N$-point functions are realised by adding an extra term $E(h_{++}) \equiv \int d^2 x h_{++}(x) T_{--}(x)$ to its action.
From the gravitational Ward identity of an $N$-point
fermion correlation function, they derived a generic differential equation for fermion four-point functions.
Introducing the anzatz of the solution and its factorization, it reduces to a known degenerate hypergeometric function in the vicinity of $\xi=1$, where $\xi$ is an anharmonic ratio.
Then, it was shown that one of two independent solutions has a logarithmic singularity at $\xi=1$.
The degeneracy of the states causes the emergence of logarithms in the correlators without spoiling the conformal invariance, and the operators therefore turned out to be logarithmic operators similar to those in the $c=-2$ model.
The logarithmic degeneracy in the Liouville theory was also suggested in \cite{BK}.

In the following year, again logarithms were found by Caux, Kogan and Tsvelik in a model of fermions in $(2+1)$ dimensions \cite{CKT}. 
In the context of replica theory in condensed matter physics, 
they set up $N$ copies of fermions with a non-Abelian gauge potential at criticality. 
They imposed a time-independence condition on the model and took only slow parts of the fermions into account.
$N$ pairs of the fermions form a composite field $Q$, an $r  \times r$ matrix, with which the theory reduces to an $SU(r)$ WZNW model plus other terms. 
At criticality, diagonal disorder in chirality was also chosen. 
From the disorder and the deduced Knizhnik-Zamolodchikov equation, they obtained the four-point function of $Q$'s and found a logarithmic singularity in eq. (29) of \cite{CKT}.
Then, it was found that the deduced OPE contains logarithmic pairs similar to those in \cite{gura}. They also found a generator of a continuous symmetry and inferred the extended symmetry in WZNW models of $SU(2)$, which was later confirmed in \cite{Caux:1996kq}.

Although we have explained LCFT roughly in the above four examples,
this new phenomenon takes place in various models listed in the introduction.
They are not at all confined to one topic but range from string theory in particle theory to quantum Hall effects in condensed matter theory.

\subsection{A Minimal Model Revisited}
\label{sec:minimal2}

Let us consider a four-point function in the $c_{2,1}=-2$ model in \cite{gura}. 

As the suffix shows, this model occurs in the minimal series with $p=2$ and $q=1$ in (\ref{def:minimal}). Therefore, there are primaries which satisfy the Kac formula (\ref{def:h(m,n)}). 
When a four-point function of the theory contains any of those fields, the function must satisfy a certain linear differential equation as eq. (\ref{def:diff eq o 4pt}). 

Let $\mu(z)$ be a primary $\Phi_{1,2}(z)$ field of conformal dimension $h_{1,2}=-1/8$, and solve the differential equation for the four-point function of four
$\mu(z)$'s. Note that the negative conformal dimension $h_{1,2}$ means that the theory is non-unitary. Besides, the two-point function
of two $\mu$'s is clearly divergent at infinity. 

Substituting the dimension $-1/8$ into all $h, h_j$ in
eq. (\ref{def:diff eq o 4pt}), one can easily obtain a second-order
differential equation for the four-point function.
From the dimensional analysis and the global conformal invariance of the four-point function, one can factorise the function as
\bea
 \label{def:harmonic decom}
   \braket{\mu(z_1) \mu(z_2) \mu(z_3) \mu(z_4) }
   = z_{13}^{\frac14} z_{24}^{\frac14} 
    \xi^{\frac14} \left(1-\xi \right)^{\frac14} G^{(4)}(\xi),
\eea
where the explicit form of the anharmonic ratio $\xi$ is given by 
$\xi=\frac{z_{12}z_{34}}{z_{13}z_{24}}$ and $z_{ij}= z_i - z_j$. We
use this convention from now on.
From the same conformal invariance, the differential equation
reduces to a second-order ordinary differential equation 
\bea
 \label{def:diff eq o c=2}
   \left[ \xi(1-\xi) \frac{d^2}{d \xi^2} + \left(1- 2\xi
\right) \frac{d}{d \xi} -\frac14 \right] G^{(4)}(\xi) = 0 .
\eea
A general solution of this equation is a linear combination of two
independent hypergeometric functions, $\F(1/2,1/2;1;\xi)$ and $\F(1/2,1/2;1;1-\xi)$,
while an integral expression of the function is given by
\bea
 \label{def:f21 integral}
   \F(1/2,1/2;1;x) = \int_0^1 \frac{d z}{\sqrt{z(1-z)(1-xz)}}.
\eea
Thus, if we define $G(\xi)\equiv\F(1/2,1/2;1;\xi)$, the function
$G^{(4)}$ reads
\bea
 \label{sol:diff o c=2}
   G^{(4)}(\xi) = A \, G(\xi) + B \, G(1-\xi),
\eea
where $A$ and $B$ are $c$-number coefficients.

The explicit form of the general solution seemed to be
troublesome because the second term in eq. (\ref{sol:diff o c=2}) has a
logarithmic singularity at $\xi=0$, whereas the first one is
regular, having an ordinary Taylor expansion, at that point. For this
simple reason, 
the second term had not been taken into account and the logarithm had
hardly appeared in the literature. Together with the non-unitarity, the
model had been paid little attention for a long time before Gurarie further investigated this theory.

In fact, even if one excludes the second term from the theory, the
first term also possesses a logarithmic singularity at
$\xi=1$. Besides, the full theory requires the second term 
in order to be single-valued in the complex plane and the correct
combination of the holomorphic and anti-holomorphic functions is 
\begin{eqnarray}
 \label{def:pair of F}
   G(x) \ol G(1-x) + G(1-x) \ol G(x) .
\end{eqnarray}
Hence, the full theory needs both solutions.

Here, the so-called logarithmic CFT arises:
Any four-point function is regarded as a pair of pairs. Since any two fields can be expressed by a sum of single fields (the operator algebra hypothesis or the fusion rule), if a four-point function has a logarithmic singularity in itself, there must be some two-point functions that exhibit the same singularity i.e.
\bea
 \label{def:mu mu general}
   \mu(z) \mu(0) \sim  \sum_{\{h\}} z^\frac14 \left( C_h(0) z^h \log z + D_h(0) z^h \right) ,
\eea
where $C_h(0)$ is not necessarily present for all $h\in\{h\}$.
It turns out that
\begin{eqnarray}
 \label{def:mu mu c-2}
   \mu(z) \mu(0) \sim z^\frac14 \left( C(0) \log z + D(0) \right) ,
\end{eqnarray}
where a logarithmic pair $\{C(z),D(z)\}$ of conformal dimension zero appear in the r.h.s.
These logarithmic pairs are known to satisfy the relations
\bea
 \label{def:gurarie degeneracy}
   L_0 \ket{D} = \ket{C}, \quad L_0 \ket{C} = 0.
\eea
Formally, the conformal dimension of $\ket{D}$ is zero, while $\ket{D}$ is {\it not} a primary state in an ordinary sense, since it doesn't fit in eq. (\ref{def:primary field}). 
Neither does the state $\ket{D}$ transform as $T(z)$, but it mixes with the other primary state $\ket{C}$, building a complicated conformal family under the transformations. Therefore, we may conclude that this is a newly found operator in representations of the Virasoro algebra. Details on the representation will be given in the following sections. Note that the state $\ket{D}$ has a degree of freedom on redefining itself by adding the state $\ket{C}$, though it can be fixed by a normalisation of the two-point functions of the pair:
\begin{eqnarray}
 \label{def:2pt CD}
   \braket{C(z)C(0)}&=& 0\,,\; \braket{D(z)D(0)}= \log z + \lambda,
   \nn
   \braket{C(z)D(0)}&=& \braket{D(z)C(0)} = 1.
\end{eqnarray}
The normalisation factors of the second line and of the log term must be equivalent for the function $G^{(4)}(\xi)$, for example, $1$ in (\ref{def:2pt CD}) for this case. A constant $\lambda$ fixes the redefinition degree of freedom of $D(z)$ and, therefore, its corresponding state.
By taking the common parts of LCFTs found later, the general forms of them are illustrated in the following subsection.

\subsection{The Common Features from the Logarithms}
\label{sec:otherLCFT}

As we saw in the previous sections, the logarithmic fields naturally appear as a consequence of the logarithms in correlation functions and the operator algebra hypothesis that is necessary in all CFTs. 
The reality condition on the full correlator functions is also natural in the context of CFT. 
In some models of LCFT, even non-unitarity is {\it not} necessary for the logarithmic fields (cf. \cite{KLS}).
Therefore, we may conclude that the critical difference between LCFT and ordinary CFT is the existence of those logarithmic fields. 
We will illustrate the basic properties of them below.

The general form of the pair of logarithmic fields $C(z)$ and $D(z)$ can be written down as
\begin{eqnarray}
 \label{def:CD}
  T(z) C(w) &\sim& \frac{h_C \,C(w)}{(z-w)^2} + \frac{\partial_w C(w)}{z-w} , \nn
  T(z) D(w) &\sim& \frac{h_C \,D(w) + C(w)}{(z-w)^2} + \frac{\partial_w D(w)}{z-w} ,
\end{eqnarray}
where $h_C$ is the conformal dimension of both fields. 
This forms a $rank$-2 Jordan cell. The formal definition of the Jordan cell in CFT will be given in the next chapter.

Their correlation functions are given by \cite{gura,CKT}
\begin{eqnarray}
 \label{def:rank2JS}
  \braket{C(z) C(w)} &\sim& 0 , \nn
  \braket{C(z) D(w)} &\sim& \frac{\al}{(z-w)^{2 h_C }} , \nn
  \braket{D(z) D(w)} &\sim& \frac{1}{(z-w)^{2 h_C }}\left(-2 \al \ln (z-w) + \al^{\prime} \right) .
\end{eqnarray}
Accordingly, a pair of initial states $\ket{C}$ and $\ket{D}$ form a $rank$-2
Jordan cell,
\begin{eqnarray}
 \label{def:rank2}
  L_0 \ket{C} &=& h_C \ket{C} ,\nn
  L_0 \ket{D} &=& h_C \ket{D} + \ket{C} . 
\end{eqnarray}
Their Verma modules are obtained from the above states by successively acting with the generators of the chiral algebra on them.

\section{Mathematical Definitions}
\label{sec:math-def}

\subsection{Formal Properties of CFT Revisited}

The commutation relations of the Virasoro algebra $\LL$ are given in (\ref{def:Virasoro}). 
For general use, we replace the central charge $c$ by the capital letter $C$, in order to specify that this is a central element of the symmetry. Although one can consider a non-diagonal $C$ elsewhere, 
we assume this is diagonal $C=c I$ with the identity operator $I$ in what follows.

Given the whole set of chiral primaries and its conformal dimensions, the corresponding conformal families $[\Phi_h]$ are generated as the Verma modules $\{V_h(c)\}$ of the symmetry algebra $\LL$. With them and their operator algebra (\ref{def:operator alg}), one can define a CFT of the central charge $C$.
This is the basic formulation described in chapter \ref{ch:intro}.
Roughly speaking, all the information of the theory is encrypted in the set of the modules $\{V_h(c)\}$, so let us first concentrate on these formal properties.

From the mathematical point of view, these modules can be built schematically as follows: Let $\U$ be the universal enveloping algebra of $\LL$, and $\U_n$ be an $n$-th level of $\U$. 
In addition, let $\U^{\pm}\subset\U$ and $\U^0\subset\U$ be the enveloping algebras of the subalgebras of $\LL^\pm \equiv \{L_n, n \!{\tiny \begin{array}{l} >\\[-3pt]< \end{array}} \!0 \}$ and $\LL^0 \equiv \{ L_0, C \}$ respectively.\footnote{
The $\U$ can be expressed by $\U=\U_- \U_0 \U_+$.
The decomposition can be naturally extended to
any chiral algebras that are graded in the same way.
}
The lowest weight module (LWM) $V$ of $\LL$ is defined as a module which contains a subspace $V_0$ such that $dim~ V_0=1$ and $V= \U^- V_0$. The vector $v\in V_0$ is called the lowest weight vector (LWV), or lowest weight state (LWS) in Hilbert state space. In other words, the LWV is a primary, satisfying the conditions
\bea
 \label{def:conf dim}
   L_0 \, v &=& h \, v,\quad C\, v \,=\, c\, v, \quad h, c\in\C,
\nn 
   \U^+ v &=& 0 , 
\eea
similar to (\ref{def:unitarity for L}),
where $h$ is the lowest weight (conformal dimension) of $v$.
A Verma module is defined as a LWM, $V$, which fulfills the statement: 
{\it There is a unique $\LL$-homomorphism $V \to W$ and $v \to w$ for any LWV $w\in W$ and LWM $W$.}\footnote{
This Verma module is unique up to $\LL$-isomorphism.
}
An important property is that a LWM becomes irreducible when it is a quotient of $V$ by its maximal proper submodule. 
As for the minimal models, such submodules are of null vectors and subtracted from the Verma modules. Their LWMs thus become irreducible.

Even in LCFT,
many similarities are expected to hold, however the lowest weight space $V_0$ is known to be degenerate $i.e.$ $dim V_0 >1$. Furthermore, one may consider the degeneracy of more than two operators for LCFT as a formal introduction of Jordan cell structure.

\subsection{Jordan Cell Structure}
\label{sec:jordan cell}

For a reducible but indecomposable representation, we demand the $L_0$ operator to be decomposed into a diagonal part and a non-diagonal nilpotent part, 
$L_0 \equiv L_0^d + L_0^{nil}$ such that
\bea
 \label{def:L^n}
   \com{L_0^d}{L_n} = n L_n, \quad \com{L_0^{nil}}{L_n} = 0.
\eea
This is compatible with the commutators (\ref{def:Virasoro}).

Clearly, the case shown in (\ref{def:rank2}) satisfies the following property
\bea
 \label{eq:L^n CD}
   L_0^{nil} \ket{D} &=& \ket{C} , \quad L_0^{nil} \ket{C} \,=\, 0 ,
\nn
   L_0^d \ket{D} &=& h \ket{D}, \quad L_0^d \ket{C} \,=\, h \ket{C}.
\eea
Therefore, $L_0^{nil}$ is in fact nilpotent as $(L_0^{nil})^2=0$ on the above pair of states.
From this, one may consider the $k$-th nilpotency of $L_0^{nil}$ such that $(L_0^{nil})^k = 0$,
and thus $rank$-$k$ Jordan cell structure. 

The Jordan lowest weight module (JLWM) is defined by Rohsiepe as an ${\cal L}$-module $\V$, satisfying 
\bea
 \label{def:JLWM}
   &(0)& C v^{(i)} = c v^{(i)} , \nn
   &(1)& L_0 v^{(i)} = h v^{(i)} + v^{(i-1)} ,
    \quad   L_0 v^{(0)} = h v^{(0)} \quad (i>0 {~~and~~} h,c \in{\C} ), \nn
   &(2)& v^{(i)} \in \V_0 \qquad ( 0 \leq i \leq k-1 ),  \nn
   &(3)& \V = \U . v^{(k-1)} , 
\eea
where $\V_0 \equiv \{v\in\V|{}^\forall v^{\prime} ; {\cal U}^+ v = 0,
v\not=\U^- v^{\prime} \}$ is the lowest weight space, $h$ is the lowest weight, $\{ v^{(i)} \}$ are
linearly independent Jordan lowest weight vectors (JLWVs). The integer $k$
is called the $rank$ of the JLWM, and the above mentioned nilpotency $(L_0^{nil})^k = 0$ is realised on this module. When $v^{(i)}, v^{(j)} ~s.t.~  i>j$, the former is called higher than the latter, and the reverse is called lower.
Note however that, in $rank$-2 cases, $v^{(0)}$ is called the lower JLWV and $v^{(1)}$ is the upper JLWV in convention. The Verma module $\V^{(0)}$ built on the lower JLWV is called the lower JLWM, and $\V^{(1)}\equiv\V/\V^{(0)}$ is called the upper JLWM.

This Jordan cell structure can easily be illustrated by a matrix representation of $L_0$ on $\V_0$ as:
\begin{eqnarray}
 \label{def:matirx L0}
   L_0 = \left( 
   \begin{array}{cccccc}
     h&      1&        &        &       &       \\[-5pt]
      &      h&       1&        & \mbox{\strut\rlap{\smash{\Huge$O$}}\quad}&       \\[-5pt]
      &       &\:\ddots&  \ddots&       &       \\[-5pt]
      &       &        &\:\ddots&      1&       \\[-5pt]
      &       &        &        &      h&      1\\[-5pt]
  \mbox{\strut\rlap{\smash{\Huge$O$}}\quad}&       &       &       &       &      h
   \end{array}
   \right),   
\end{eqnarray}
where $v^{(0)}$ is expressed by $(1, 0,\cdots,0)^{T}$.
Note that all vectors can be generated from $v^{(k-1)}$, so that $\V=\{\U\,v^{(k-1)}\}$.

Likewise, one can define the action of $\LL$ on its dual space $\V^*$ by setting 
\begin{eqnarray}
 \label{def:*}
   \left( L_{i_1}^{n_1} \cdots L_{i_p}^{n_p} 
 \right)^\dag
   &=& 
  L_{-i_p}^{n_p} \cdots L_{-i_1}^{n_1}, \nn
   \left( \LL \phi \right) (w) &=& \phi (\LL^\dag w) {\rm ~~for~\phi\in
   \V^* ,~} w \in \V .
\end{eqnarray}
The action of $\LL^\dag$ on $\V$ induces the same conformal structure in
$\V^*$, and then $\V^*$ appears as a JLWM with lowest weight vectors $\{ v^{(i) *} \}$, which satisfy eq. (\ref{def:JLWM}). 
In particular, the LWVs satisfy
\begin{eqnarray}
\label{def:v*}
  L_0 v^{(i) *} = h v^{(i) *} + v^{(i-1) *}
\end{eqnarray}
One can further define the dual JLWM, $\V^\dag \subset
\V^*$, which is naturally induced by a map:
\bea
 \label{def:dual}
  \V=\{u.v\} &\to& \V^\dag=\{u.v^*\},
 \nn
  u.v &\to& u.v^* , 
\eea
where $u\in\U$, and $v,v^*$ are the lowest weight vectors of $\V$ and $\V^\dag$, respectively.

In fact, as the form (\ref{def:*}), the Shapovalov bilinear form $\braket{ |\, }$ can be defined by\footnote{One may relax this orthogonality condition for eq. (\ref{def:*JLW}). For example in rank-$2$ cases, the sufficient condition for eq. (\ref{def:*JLW}) is that $\braket{v^{(j)}|v^{(i)}}$ be an upper triangular matrix.}:
\bea
 \label{def:shapo}
   {}^\forall v^{(i)} \in \V, ~{}^\exists v^{(j) \dag}\in\V^\dag ~;  \quad
   \braket{v^{(j)}|v^{(i)}} \,=\, \delta_{ij}, 
\eea
which satisfies $\braket{v_i |L_n^{\dag}|v_j} = \braket{v_i | L_{-n} |v_j }$. 
From eqs. (\ref{def:v*}) and (\ref{def:shapo}), one finds the relation between $*$ and $\dag$. 
The $*$ transformation is an isomorphism
which acts on $\V$ as:
\begin{eqnarray}
 \label{def:*JLW}
   * :\quad v^{(i)} &\too& v^{(i) *} = v^{(k-1-i) \dag} .
\nn
      \ket{v^{(i)}} &\too& \bra{v^{(k-1-i)}}
\end{eqnarray}
It does {\it not} map $v^{(i)}$ to $v^{(i) \dag}$, but to $v^{(k-1-i) \dag}$.
Then $\bra{v^{(i)}} \rightarrow \ket{v^{(k-1-i)}}$ finds $*^2=1$.
In $rank$-2 cases, $v^{(0) *}= v^{(1) \dag}, v^{(1) *}= v^{(0) \dag}$, and these can be explicitly written with $C, D$ by:
\begin{eqnarray}
 \label{def:rank2 map}
   \bra{D} \stackrel{*}{\longleftrightarrow} \ket{C},\;  
   \bra{C} \stackrel{*}{\longleftrightarrow} \ket{D}.
\end{eqnarray}
Note that $\bra{C}L_0=h_C \bra{C} + \bra{D}$ and that $\bra{C}=\bra{D^*}\equiv \lim_{z\to OUT} \bra{0} D(z) z^{2h_C}$, where the limit is taken to the point where the $sl(2,\R)$-invariant $OUT$ vacuum is defined.

On this $\V$, we further introduce our decomposition of the Virasoro generators \cite{ishimoto1}, namely
$L_n = L_n^d + L_n^{nil}$ such that,
\bea
 \label{def:d+n}
    L_n^d &:& \ket{v^{(i)}, N} \rightarrow \ket{v^{(i)}, N-n} , \nn
    L_n^{nil} &:& \ket{v^{(i)}, N} \rightarrow \ket{v^{(i-1)}, N-n} \quad
    for ~i\not= 0 , \nn
   &&  \ket{v^{0}, N} \rightarrow  0 ,
\eea
where $\ket{v^{(i)}, N}$ denotes the orthogonal basis of
$\ket{v^{(i)}}$-descendants at level $N$ for short.\footnote{
The algebraic structure of this decomposition cannot always be algebraically unique, depending on the monomials of the modes of $\LL$ in $\ket{v^{(i)},N}$.
We simply regard this as a symbol of the mapping to the other conformal family $\ket{v^{(i-1)},N-n}$, and there is no ambiguity in this sense.
} 
We will use these notations in chapter \ref{ch:boundary states}.

\subsection{More on Cells \& Modules}

So far, we only deal with holomorphic sectors while the anti-holomorphic counterparts are neglected as its trivial copies. 
However, one can see from eq. (\ref{def:mu mu general}) that the naive chiral form may explicitly violate the reality condition on the full correlation functions, so that the theory should require some non-chiral form. 
For example, a $rank$-3 case should have the following non-chiral fusion,
\bea
 \label{def:rank3}
   \Phi_h(z,\bz) \Phi_{h^\prime}(0,0) &\sim& z^{-h-h^\prime+h_C}
  \nn
&&\times \left( C_0(0,0) (\log |z|^2)^2 + C_1(0,0) \log |z|^2 + C_2(0,0) \right) + \cdots.
\eea
Nevertheless, it is useful to consider the relevant chiral modules in $rank$-2 cases, as the building block of such non-chiral theories. 
We concentrate on $rank$-2 cases in the rest of this thesis.\footnote{
In non-chiral $rank$-2 cases, the reality puts a strong condition on the states as
\bea
 \label{def:local LCFT}
   (L_0 - \ol L_0) \ket{phys} = 0,
\eea
which denies the naive chiral decomposition \cite{gab3}. We will mention this condition again in chapter \ref{ch:conclusion}.
}

Soon after \cite{gura}, it was similarly found that there exists a class of LCFTs when $c=c_{p,1}$.\footnote{
The $c_{p,1}$ models --- $c_{p,1}= 1 - 6 \frac{ (p-1)^2}{p}$, $h_{1,2k+1} = k^2 p + k p - k$.
The fundamental region of the minimal models is empty for this type 
as $ \{r,s|1\leq r<p, 1\leq s<q\} = \phi $. 
}
The analogy is trivially on the existence of null fields and their related differential equations. Namely, the submodules which make JLWMs `minimal' exist in a similar way to null fields in the unitary minimal models.

These submodules are defined as maximal preserving submodules by the following property: 
The quotient of a JLWM $\V_h(c)$ by its maximal preserving submodule $J$ gives the same nilpotency length (N-length($\V_h(c)$)$=$N-length($\V_h(c)/J$)) 
and there is no such submodule $K$ satisfying $J\subset K$. Note that, roughly speaking, the nilpotency length is the $rank$ of JLWM. For the precise definition of the nilpotency length, see \cite{roh1}.

The decomposition $\V_h(c)=\V_h^{(0)}(c) + \V_h^{(1)}(c)$ induces the decomposition of $J$ such that
\bea
 \label{def:cp1 preserve}
   J^{(0)} &\subset& \V_h^{(0)}(c),
\nn
   J^{(1)} &\subset& \V_h^{(1)}(c),
\eea
and in the $c_{p,1}$ models it turns out that
\bea
 \label{def:cp1 submodule}
   J^{(0)} = \V_{h_k}(c),\quad
   J^{(1)} = \V_{h_m}(c)
\eea
where $h_k$ ($h_m$) is the conformal dimension of $k$($m$)-th singular vector in $\V_h^{(0)}(c)$, and $k<m$.\footnote{
The existence of the submodule is given in the theorem 6.1 in \cite{roh1}.
}

From these facts, one may compute the character functions for the partition function on a torus. We will review these later in the case of $c=-2$.

For later convenience, we also introduce the $\W$-algebra $\W(2,3^3)$ and its modules. This algebra belongs to {\it triplet algebras} found and discussed in \cite{kau1}. It consists of $\{L_{n\in\Z}\}$ and $\{W_{n\in\Z}^{a}\}$, where $a \in \{0,\pm\}$ and $\com{L_0}{W_0^a}=0$. The definition of this algebra at $c=-2$ is given in eq. (\ref{def:W2333}) in the following section \cite{kau1}.

A generalised lowest weight module (GLWM) $M$ of $\W(2,3^3)$ is defined by
\bea
 \label{def:GLWM}
  &(1)& \forall v\in M_0,\, \U_\W^+ v =0,
  \nn
  &(2)& M_0 {\rm ~is~ irreducible~} \W_0{\rm -module},
  \nn
  &(3)& M = \U_\W \, M_0,
\eea
where a linear subspace $M_0\subset M$ is the lowest weight space of $M$, and $\{v\in M_0\}$ are the lowest weight vectors, as in JLWM. Its lowest weight $h$ is given by $L_0^d v = h\,v$ for $v\in M_0$.
When $dim M_0=1$, $M$ is called a singlet module, and when $dim M_0=2$ it is called a doublet module. By replacing LWM in the definition of JLWM by GLWM, we obtain a generalised Jordan lowest weight module (GJLWM). All other definitions can be generalised in the same way.

\section{The $c=-2$ Models}
\label{sec:lcft c=-2}

In this section we focus on models at $c=-2$ and show their mathematical complexity. 

We first remind the reader that, when we remove the unitarity constraint on $(p,q)$ in the minimal models,
the $c_{p,1}$ models appear as non-unitary theories, 
where the rectangular regions vanish and so do the restrictions on $(r,s)$.
Instead, due
to the relations $h_{r,s}=h_{-r,-s}=h_{r+1,s+p}$, the fundamental region for $(r,s)$
is stretched to a semi-infinite rectangular region $1\leq r ,\, 1\leq s\leq p$.

In such models, the central charge $c=-2$ has an entry as $(p=2,q=1)$. 
As we saw in section \ref{sec:minimal2}, from the four-point function of $\mu(z)\sim \Phi_{1,2}(z)$ and its null field condition, there appear a pair of degenerate states $C(z)$ and $D(z)$, which form a reducible but indecomposable representation. 
We label these by $\Omega(z)\equiv C(z)$ and $\omega(z)\equiv D(z)$ in accordance with the notation in the literature. 
The common conformal dimension of $\Omega(z)$ and $\omega(z)$ is zero and the state $\ket{\Omega}$ is the $sl(2,\R)$-invariant vacuum and has zero-norm as in (\ref{def:2pt CD}, \ref{def:rank2JS}). 
Since $L_0 \ket{\omega} = \ket{\Omega}$, $\omega(z)$ can generate all descendants of $\Omega(z)$ by the actions of $\{L_{-n\leq0}\}$. Thus, from the definition (\ref{def:JLWM}), they form a $rank$-2 JLWM 
$\V_0(c=-2)$.

It has already been shown in the literature that there are several types of model at $c=-2$
corresponding to those algebraic structures. 
One is the `normal' $c=-2$ model whose symmetry is the Virasoro symmetry
with one dimension-3 operator \cite{gura}. The other is the $c=-2$ `triplet' model with a triplet $\W$-symmetry \cite{gab3,kau1}. Another model is a `local' theory not only with the triplet $\W$-symmetry but with a non-chiral constraint, which changes the algebraic structure of the states \cite{gab5}.

In the `normal' case, there occurs no singlet nor doublet structure due to the absence of such an extended symmetry. There is also no finite restriction nor truncation w.r.t. the conformal grid mentioned above. Therefore, this model is not {\it rational} in terms of the closure of the fusion rules.
In addition, by a modular transformation, their character functions are transformed to another set of functions which in turn yield infinitely many 
characters beyond the trivial set of characters 
\cite{flohr1}. We will mention this again in chapter \ref{ch:boundary states} in the context of boundary LCFT.

The $c=-2$ `triplet' model is an extension of the `normal' model with three dimension-3 operators.
The operators constitute a so-called $\W$-algebra $\W(2,3^3)$
including the Virasoro algebra. With these fields and some appropriate 
null conditions, corresponding characters of those representations 
close under modular transformations and there is a possibility to build a Verlinde formula in a slightly modified way.
The difference mentioned here is related to the concept of {\it ``rationality''}
in LCFTs.
In fact, because of the closure of the extended symmetry, the only possible set of LWSs are restricted to $\{-1/8, 0, 3/8, 1\}$ for ${\cal W}(2,3^3)$ at $c=-2$.

On the other hand, there is a non-chiral LCFT
which has been constructed as it strictly satisfies the ``locality'' condition (\ref{def:local LCFT}) \cite{gab5}. 
This `local' logarithmic theory has a bigger and more complicated indecomposable representation than that of the triplet model. 
Unless some sort of simplification occurs in the theory, it gives a 
highly nontrivial and complicated partition function which might be 
unable to produce a Verlinde formula. We are not going to discuss this theory further here. 

In the following subsections, we describe the lowest weight spaces and the character functions of the first two models.

\subsection{Lowest Weight Spaces}

\begin{figure}[h]
\hspace{1cm}
\hspace*{4cm}
\epsffile{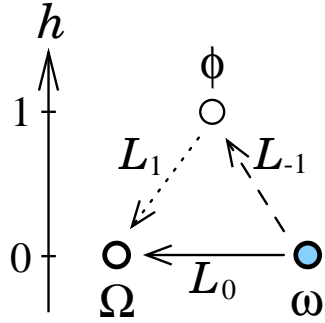}
 \caption{$\Omega$ and $\omega$ are degenerate logarithmic operators while $\phi$ is the subrepresentation of $\omega$, as $\phi\equiv L_{-1}\omega$. Hence, $L_1 \phi = 2 \Omega$.}
      \label{fig:normal_c-2}
\end{figure}

We start with the normal $c=-2$ model. Most of the arguments are applicable to the other cases as well.
Remarkably, $\omega(z)$ is not $sl(2,\R)$-invariant and does have a subrepresentation $\phi(z)$ at level $1$. 
This precisely agrees with eq. (\ref{def:cp1 submodule}). 
The null vector of $\ket{\Omega}$ is at level $1$, while that of $\ket{\omega}$ is at level 3. 
Subtracting the submodules built on the null vectors\footnote{
For more information on null vectors, please consult \cite{flohr3,flohr5,flohr6}.
} 
from $\V_0(-2)$, we obtain the minimal JLWM $\bar \V_0(-2)$ such that $\V_\Omega \equiv \bar \V_0^{(0)}(-2) \not\sim \bar \V_0^{(1)}(-2)$.
Although the upper and lower submodules are not isomorphic to each other, if we define $\V_\omega$ by the quotient of $\bar \V_0^{(1)}(-2)$ with the minimal submodule $\bar \V_1$ on $\phi(z)$, one can obtain an isomorphism
$$
\V_\omega \sim \V_\Omega.    
$$
The relation between these lowest weight spaces and the subrepresentation can be drawn illustratively as in Fig.\ref{fig:normal_c-2}.

The character function of the minimal JLWM $\bar \V_0(-2)$ is given by:
\bea
 \label{def:normal character}
   \chi_{\bar \V_0(-2)}(q) &=& \chi_{\bar \V_0^{(0)}(-2)}(q) + \chi_{\bar \V_0^{(1)}(-2)}(q)
   \nn
   &=& \chi_{\V_\Omega}(q) + \chi_{\V_\omega}(q) + \chi_{\bar \V_1}(q).
   \nn
   &=& 2 \chi_{\bar \V_0^{(0)}(-2)}(q) + \chi_{\bar \V_1}(q),
\eea
where the character function $\chi_{\V}(q)$ of $\V$ is defined by the
following {\it trace} over the module: 
$\chi_{\V}(q) \equiv \Tr_{\V}\; q^{L_0-\frac{c}{24}}$.
$q$ is the moduli parameter. 
The functions for this case can be given similarly to those in the unitary minimal models.

On the other hand, the $c=-2$ triplet model possesses the extended algebra $\W(2,3^3)$ at $c=-2$, 
whose commutation relations are given in \cite{kau1} as below:
\begin{eqnarray}
 \label{def:W2333}
   \com{L_m}{L_n} &=& (m-n) L_{m+n} - \frac16 m ( m^2-1 ) \delta_{m,-n} \,,\nn
   \com{L_m}{W_n^a} &=& (2m -n ) W_{m+n}^a \,,\nn
   \com{W_m^a}{W_n^b} &=& g^{ab} \Bigg( 2 (m-n) \Lambda_{m+n} 
   + \frac{1}{20} (m-n) ( 2 m^2 +2 n^2 -mn -8 ) L_{m+n} \nn
   &&\qquad - \frac{1}{120} m(m^2 - 1) (m^2 - 4) \delta_{m+n} \Bigg) \nn
   && + f^{ab}_c \left( \frac{5}{14} ( 2 m^2 + 2 n^2 - 3mn -4) W_{m+n}^c 
   + \frac{12}{5} V_{m+n}^c \right) ,  
\end{eqnarray}
where $m,n\in\Z$, and $a, b \in\{0,\pm\}$ are the $su(2)$ indices. 
$g^{ab}$ and $f^{ab}_c$ are the $su(2)$ metric and structure constants, respectively. $\Lambda$ and $V$ are quasi-primary normal ordered fields such that
\begin{eqnarray}
 \label{def:composite}
   \Lambda(z) = : T T :(z) - \frac{3}{10}\pa^2 T(z), \quad
   V^a(z) = : T W^a :(z) -\frac{3}{14} \pa^2 W^a(z) .
\end{eqnarray}
For the closure of this algebra, that is, the Jacobi identity, one has
to impose the null field conditions on physical states, which restrict the possible set of primaries to be $\{ -1/8,0,3/8,1 \}$ \cite{gab3}.
Under such circumstances, $\{ W_0^a \}$ becomes an $su(2)$ global symmetry, mutually commuting with the $L_0$ operator. Hence, all primaries must be irreducible under this $su(2)$ and each representation yields a GLWM defined in (\ref{def:GLWM}).
$\V_{-1/8}$ and $\V_{0}$ are singlet modules while $\V_{3/8}$ and $\V_{1}$ are doublet.

\begin{figure}[h]
\hspace{1cm}
\hspace*{2cm}
\epsffile{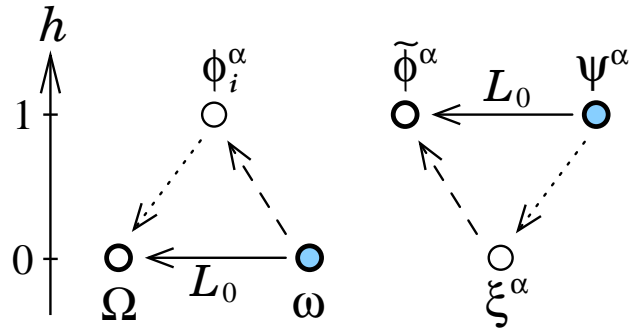}
 \caption{$\Omega$ and $\omega$ remain unchanged while $\phi_i^\al$ with $i=1,2$ are two doublet representations generated by $\{L_{-1}, W^a_{-1}\}$. $\phi_i^\al$ are also mapped onto $\Omega$ by positive modes of $\W(2,3^3)$.}
      \label{fig:triplet_c-2}
\end{figure}

Comparing this model to the normal one, the subrepresentation of $\omega(z)$ clearly differs, because under the triplet algebra a primary of $h=1$ must be a doublet $\phi^\al(z)$.
Moreover, though we have not mentioned it, there is another Jordan cell at $h=1$ of states whose lower vector is a descendant of a field $\xi$ of $h=0$.
By the same argument, all of them must be doublet under the $su(2)$.
The relations are illustrated in Fig.\ref{fig:triplet_c-2}.

The $\W$-character of the GJLWM at $h=0$ is also given by the first line of eq. (\ref{def:normal character}). The remarkable difference is that the number of the $h=1$ fields is doubled in the GJLWM case. Then, this difference drastically cancels cumbersome logarithms in the $\W$-characters.
The details of the $\W$-characters will be given in the following section.
Note that the characters of the normal model and of the triplet model are all different even when the lowest weight spaces coincide, due to the presence of $\{W_n^a\}$.

\section{Characters of the Rational Models}
\label{sec:rational characters}

As the last preliminary knowledge of LCFT that we shall need, we draw attention to the rationality of the theory again. 
The modular transformations appearing in boundary LCFT have matrix representations on the space of character functions which, for the matrices to be finite, should also be finite and closed {\it i.e.} rational under such transformations.

\subsection{The $c=-2$ Characters}

As is mentioned in the previous section, the $c=-2$ triplet model is suggested to be a rational theory having the given finite set of modules. Its closure under their fusion products was confirmed by Kausch and Gaberdiel up to level six by computer analysis \cite{gab3},
on which one may state that the model is rational. Actually, certain combinations of characters give the rationality under modular transformations.

The $\W(2,3^3)$-characters at $c=-2$ are given in \cite{kau2} by:
\begin{eqnarray}
 \label{def:Wchara}
   \chi_{\V_0} ( q )
   &=& \frac{1}{2\,\eta(q)} \left( \Theta_{1,2}(q) + \pa\Theta_{1,2}(q) \right) =  \frac{1}{2\,\eta(q)} \Theta_{1,2}(q) +\frac12 \eta^2(q), \nn
   \chi_{\V_1} ( q )
   &=& \frac{1}{2\,\eta(q)} \left( \Theta_{1,2}(q) - \pa\Theta_{1,2}(q) \right) = \frac{1}{2\,\eta(q)} \Theta_{1,2}(q) -\frac12 \eta^2(q), \nn
   \chi_{\V_{-1/8}} (q) 
   &=& \frac{1}{\eta(q)} \Theta_{0,2}(q) ,\quad
   \chi_{\V_{3/8}} (q) 
   \,=\, \frac{1}{\eta(q)} \Theta_{2,2}(q) ,
\end{eqnarray}
where $q=e^{2\pi i \tau}$, $\eta(q)\equiv q^{1/24} \prod_{n=0}^{\infty}
\left(1-q^n\right)$ is the Dedekind eta function, 
$\Theta_{l,k}(q)\equiv\sum_{n\in\Z} q^{(2kn + l)^2/4k}$ 
is a Riemann theta function,
and $\pa\Theta_{l,k}(q)\equiv\sum_{n\in\Z}
\left(2kn+l\right) q^{\left( 2kn+l \right)^2 / 4k}$. 
The first character is of the $sl(2,\R)$-invariant vacuum representation and of the Jordan cell.
The set of characters (\ref{def:Wchara}) do not close under a modular
transformation $q\to \wt q= e^{2\pi i \left( -1/\tau\right)}$, but generate a new function of $\tau$, $\Delta\Theta_{1,2}/\eta\equiv
i\tau \pa\Theta_{1,2}/\eta$:
\begin{eqnarray}
\label{mod W-chara}
 \chi_{\V_0} (\wt q) &=& \frac14 \chi_{\V_{-1/8}}(q) - \frac14 \chi_{\V_{3/8}}(q) - \frac{\Delta\Theta_{1,2}(q)}{2 \eta(q)} , 
\nn
\chi_{\V_1} (\wt q) &=& \frac14 \chi_{\V_{-1/8}}(q) - \frac14 \chi_{\V_{3/8}}(q) + \frac{\Delta\Theta_{1,2}(q)}{2 \eta(q)} , 
\nn
\chi_{\V_{-1/8}}(\wt q)
 &=& \frac12 \left(\chi_{\V_{-1/8}}(q) + \frac{2\Theta_{1,2}(q)}{\eta(q)} + \chi_{\V_{3/8}}(q) \right) ,\nn
\chi_{\V_{3/8}}(\wt q)
 &=& \frac12 \left(\chi_{\V_{-1/8}}(q) - \frac{2\Theta_{1,2}(q)}{\eta(q)} + \chi_{\V_{3/8}}(q) \right) .
\end{eqnarray}

For the modular invariance, we introduce the $rank$-2 GJLWMs $\calr_0$ and $\calr_1$, and define $\chi_{\calr_0} = \chi_{\calr_1}$.
From eq. (\ref{def:normal character}) and Fig.\ref{fig:triplet_c-2}, one obtains
\bea
 \label{def:GJLWM0}
  \chi_{\calr_0} (q) &=& 2 \chi_{\V_0}(q) + 2 \chi_{\V_1}(q)
  \,=\, \frac{2\,\Theta_{1,2}(q)}{\,\eta(q)},
\eea
with which 
$\{ \chi_{\V_{-1/8}}, \chi_{\V_{3/8}} \}$ form a modular invariant set.
The $\calr_0$ contains JLWVs at $h=0$ and the $\calr_1$ contains different JLWVs at $h=1$.
This was first proposed in \cite{gab3}, 
and its $S$-matrices are given in \cite{gab3,roh1}.

\subsection{Characters for the $c_{p,1}$ Models}

According to \cite{gab3}, the only rational LCFT which had been studied, 
is the triplet algebra at $c=-2$. 
Rohsiepe suggested in \cite{roh1} that similar results to the $c=-2$ case will hold for the whole series of triplet algebras ${\cal W}(2,(2p-1)^3)$ at $c=c_{p,1}$. This is yet to be proven. 

Below are the details and proposals on the $c_{p,1}$ models.
Some characters of $c_{p,1}$ models were first shown in \cite{flohr1}:
\begin{eqnarray}
 \label{def:allchara}
   \chi_{V_0} ( q )
   &=& \frac{1}{p\,\eta(q)} \left( \Theta_{p-1,p}(q) + \pa\Theta_{p-1,p}(q) \right) , \nn
   \chi_{V_{k,1}} (q) 
   &=& \frac{k}{p \eta(q)} \Theta_{p-k,p}(q) + \frac{1}{p\,\eta(q)}\pa\Theta_{p-k,p}(q) , \nn
   \chi_{V_1} ( q )
   &=& \frac{p-1}{p\,\eta(q)} \Theta_{p-1,p}(q) - \frac{1}{p\,\eta(q)}\pa\Theta_{p-1,p}(q) , \nn
   \chi_{V_{p-k,2}} ( q )
   &=& \frac{p-k}{p\,\eta(q)} \Theta_{p-k,p}(q) - \frac{1}{p\,\eta(q)}\pa\Theta_{p-k,p}(q) ,\nn 
   \chi_{V_{p,1}} (q) 
   &=& \frac{1}{\eta(q)} \Theta_{0,p}(q) ,\quad
   \chi_{V_{p,2}} (q) 
   \,=\, \frac{1}{\eta(q)} \Theta_{p,p}(q) ,
\end{eqnarray}
where $k$ takes a positive integer value from 2 to $p-1$, 
The vacuum representation $\chi_{V_0}(q)$ is calculated as follows,
\begin{eqnarray}
 \label{def:Wvac}
   \chi_{V_0} (q) &=& \sum_{k\geq0} (2k +1) \chi_{1,2k+1}^{Vir} (q) \nn
&& {\rm where~} \chi_{1,2k+1}^{Vir} (q) \,=\, \frac{1}{\eta(q)} 
 \left( q^{h_{1,2k+1}} - q^{h_{1,-2k-1}} \right).
\end{eqnarray}
This is same as $\chi_{\V_0}$ in eq. (\ref{def:Wchara}).
In eq. (\ref{def:allchara}), the first character is of the vacuum
representation and forms an indecomposable representation together with
$\chi_{V_1}$, the character of the representation $V_1$, whose lowest
weight state is at $h=1$. 

A finite set of characters which close under \alert{modular transformations}
can be given similarly by the following set:
\begin{eqnarray}
 \label{def:allchara1}
   \chi_{R_i} ( q )
   &=& 2 \left( \chi_{V_{k,1}} (q) + \chi_{V_{p-k,2}} (q) \right) 
  \,=\, \frac{2}{\eta(q)} \Theta_{p-i,p}(q) \qquad {\rm for~} i=1,...,p-1 , \nn
   \chi_{V_{(p,I)}} (q) 
   &=& \frac{1}{\eta(q)} \Theta_{(I-1)p,p}(q) \qquad {\rm for ~} I=1,2 ,
\end{eqnarray}
where $V_{1,1}=V_0, V_{p-1,2}= V_1$, $R_i$ may correspond to 
GJLWMs under the triplet $\W$-algebra.
The more precise construction should follow the detailed investigation
on null field conditions of the triplet algebra and their fusion
products as in \cite{gab4}. Both are beyond our scope at present.

\chapter{Boundary CFT}
\label{ch:BCFT}

The early eighties has been called the first ``golden age'' of string
theory with advances supported greatly by the celebrated paper \cite{BPZ}.
Soon after, theorists began to search for the
proper formulation of the open string sector of the theory. 
The open string here means a one-dimensional connected object 
with two end-points, which is embedded smoothly in spacetime.
Therefore, the corresponding mathematical
framework is CFT with boundaries (boundary CFT or BCFT). 
In 1986, Cardy introduced his `method of images' to the boundary CFT 
analogous to that of electromagnetism. 
From the following development, we are now able to treat
almost all boundary CFTs as chiral CFTs without boundary. 
Boundary conditions must be examined carefully, though.

In successive papers, Cardy also introduced the duality concept of
string theory into cylindrical geometry and derived a prominent
BCFT version of the Verlinde formula. 
The Verlinde formula is the formula which relates the $S$-matrix of 
the modular group to the fusion rule matrices. 
In addition, it also plays a key role on the relation between CFT
and quantum algebra. This formula is known to hold for the unitary
minimal models. 

The emergence of such theories has been extraordinary beneficial 
in the sense that we explained in the introduction.
We will give an introduction to this class of theories and prepare for establishing the boundary LCFT.

This chapter consists of two sections: A brief introduction of BCFT on
an upper half-plane, and another introduction for the theory on a
cylinder of a finite length. 
We briefly review Cardy's work and related techniques and add several
remarks to guide the reader through the rest of this thesis.

\section{CFT on an Upper Half-Plane}
\label{sec:CFT plane}

\begin{figure}[h]
\hspace{1cm}
\hspace*{4cm}
\epsffile{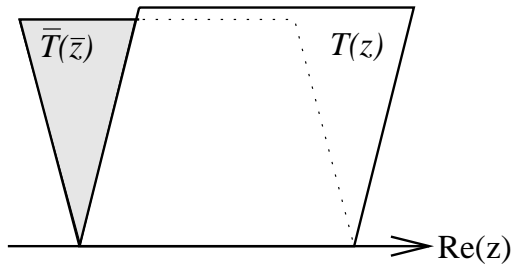}
 \caption{Holomorphic and anti-holomorphic sectors defined on a UHP
 are equivalent to one sector defined on two UHPs but connected to
 each other at the boundary.}
      \label{fig:2UHP}
\end{figure}

Boundary CFT is defined on a two-dimensional surface with
one or more boundaries and the prototype of its geometry is an upper
half-plane (UHP). On a UHP, the theory is decomposed into and described by holomorphic and anti-holomorphic sectors, 
however they are not totally independent but coupled to
each other by the boundary conditions.
In this case the boundary is the real axis (Fig.\ref{fig:2UHP}). 

\ni
The
boundary itself must be invariant under conformal transformations.
The allowed infinitesimal conformal coordinate transformations are $z\to z+ \epsilon(z)$ such that
\bea
 \label{def:conformally invariant bc}
  \epsilon(z) \in \R ~{\rm if~ z\in\R} .
\eea
This makes it a real analytic function $\epsilon(\bz)=\ol{\epsilon(z)}$.
On operators, the conformal transformations are expressed, as in eq. (\ref{def:delta T}), by a contour integral:
\bea
 \label{def:BCFTtr}
   \delta_\epsilon = \oint dz\, \epsilon(z) T(z) - \oint d\bz\, \epsilon(\bz) \ol T(\bz) , 
\eea
where the closed contours surround all the relevant operators on the UHPs respectively. This actually produces the conformal Ward identity on correlation functions as we have shown in eq. (\ref{def:Ward}). For the identity to be valid, the above integral on the boundary should vanish and it gives the following condition
\bea
 \label{def:bc_T}
   T(x)=\ol T(x), \ x\in\R ,
\eea
for this case. This is equivalent to $T_{tx}=0$ in the Cartesian coordinate, meaning no energy flows across the boundary. 
In (\ref{def:BCFTtr}), one can take the contours so that two infinite semi-circles are tied up at the boundary without contours on the axis and, if we map the anti-holomorphic sector to the LHP, the theory can be interpreted as a chiral theory defined on the whole plane.

$N$-point functions in BCFT are re-interpreted as $2N$-point functions on the whole plane in a chiral theory and therefore the differential equations for four-point functions are also naturally defined for the minimal models. 

For example, as we have shown in section \ref{sec:minimal2} for $c=-2$, the four-point function of four $\mu(z)$'s obeys the differential equation (\ref{def:diff eq o c=2}). With the anzatz (\ref{def:harmonic decom}), we have obtained the general solution (\ref{sol:diff o c=2}). The general result for our case is the one replacing $z_3,z_4$ with $\bz_2, \bz_1$.

On the other hand, a one-point function of the field $\Phi_\Delta$ with conformal dimension $\Delta$ can be drawn as
\bea
 \label{def:2pt anonymous}
   \braket{\Phi_\Delta(z,\bz)}_{boundary}
   = \braket{\Phi_\Delta(z) \Phi_\Delta(\bz)}
   = (z-\bz)^{-2 \Delta},
\eea
on the basis of the operator algebra (\ref{def:operator alg}). Since $(z+\bz)\in \R$, the operators appearing in the $r.h.s.$ of (\ref{def:operator alg}) become special having no anti-holomorphic counterparts. This type of operator is called a `boundary operator' and is important in the construction of BCFT.
For, from eq. (\ref{def:mu mu c-2}), one can express a $\mu(z,\bz)$ field in the $c=-2$ models as
\bea
 \label{def:mu mu bcft}
   \mu(z,\bz) \sim (z-\bz)^\frac14 \left( \Omega(x) \log (z-\bz) + \omega(x) \right) ,
\eea
where $x= (z+\bz)/2 \ i.e.$ $x$ is real.
As in the bulk, the OPEs between boundary logarithmic operators are given similarly by substituting $z=x_1$, $w=x_2$ and $h=0$ 
into eq. (\ref{def:rank2JS}) \cite{KW}:
\bea
 \label{def:bop CD}
   \braket{C(x_1) C(x_2)} &\sim& 0,
   \nn
   \braket{C(x_1) D(x_2)} &\sim& \al (x_1 - x_2)^{-2\Delta_D} ,
   \nn
   \braket{D(x_1) D(x_2)} &\sim& (x_1 - x_2)^{-2\Delta_D} \left( \alpha \log(x_1-x_2) + \al^\prime \right).
\eea
Here $\Omega(x)$ and $\omega(x)$ are lower and upper JLWV respectively, and $x_1,x_2\in\R$.
However, the coefficients $\al, \al^\prime$ are related to $A,B$ in the solution (\ref{sol:diff o c=2}) and all of them depend on the boundary conditions.
In fact, if $B=0$, all the logarithms must vanish near the boundary
   and result in $\al=0$ \cite{KW}. In chapter \ref{ch:free}, we will look into the relations between two-point functions and the boundary operators. 
In the following section, we move to the dual picture where the boundary conditions and boundary states are strongly correlated.
Note that it is not straightforward to interpret the boundary conditions even in the context of dual pictures, 
whereas $free$ and $fixed$ $(+-)$ boundary conditions are known and easy to be interpreted in the Ising model ($c=1/2$).

\section{CFT on a Cylinder}
\label{sec:CFT cyl}

In the following, we use the word `theory of (open or closed) strings' for explanations. Precise speaking, they are merely a CFT with two different `time' directions defined on the same geometry.

\begin{figure}[h]
\vspace*{1.5cm}
\hspace*{4cm}
\epsfxsize=7cm
\epsfysize=3cm
\epsffile{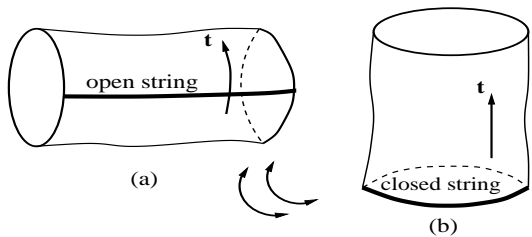}
 \caption{The geometry for open-closed string duality.}
      \label{fig:open-closed}
\end{figure}

Imagine that we have a one-loop diagram of an open string, starting from one
place, going around and back to the same place, 
as is shown in Fig.\ref{fig:open-closed}(a). 
Naturally, the geometry of this diagram is equal to a tube with 
two one-dimensional ends, circles. 
Here, a physical object doesn't rely on the direction of
the string's move which, in the above case, 
we interpret as the closed direction. 
Even if one chooses the closed string viewpoint, 
going straight from one end to the other as in Fig.\ref{fig:open-closed}(b),
the theory must result in the same physics observation.
However, we still need those two distinctive pictures to facilitate our calculations.

In the open string picture, both ends have boundary
conditions while, in the closed string picture, 
an initial state travels from one end to another and 
reaches the final point in the time direction.
The final state is not arbitrary at all but 
a state giving non-vanishing partition function with the initial state.
These initial and final states are called boundary states.
As they have the same physics, these two different physical pictures on
boundaries are associated with each other. Let us move to this.

Consider an infinite strip of width $L$ on which theories of
open strings can lie. 
By a conformal map,
$
w = \frac{L}{\pi} \ln z 
$, 
a theory on a $z$ upper half-plane 
is mapped onto a $w$ infinite
strip, where time $t$ goes
 along two parallel edges. 
A pair of conformally invariant boundary conditions are put onto these two
edges, labeled by $\al, \beta$, and the Hamiltonian of this
system is given by $(\pi/L)H_{\al \beta}$ with a generator of
$t$-translations $H_{\al \beta}$. 
The eigenstates of $H_{\al \beta}$
fall into irreducible representations of the chiral algebra and the partition
function becomes a linear combination of the functions 
of these representations. 
By imposing a periodicity $T$ along $t$, 
we get the partition function of the open string picture:
\bea
 \label{def:Zopen}
 Z_{\al \beta}^{open} (q)
= \Tr\, q^{H_{\al \beta}}
= \sum_{i} n_{\al \beta}^{\;\; i} \, \chi_i (q) , 
\eea
where $q\equiv e^{2\pi i \tau}, \tau\equiv i T/2L$ and $n_{\al \beta}^{\;\; i}$
is the number of times which a representation $i$ occurs in the presence
of boundary conditions $(\al \beta)$. $\chi_i (q)$ denotes a character
function of the representation $i$.

On the other hand,
as a consequence of the periodicity,
the dual description of the theory appears by the change of $t$
direction to one across the strip. 
The boundary conditions are converted into the
boundary states on the boundaries of the \alert{cylindrical geometry.}
This cylindrical geometry, an annulus on the $\zeta$-plane with the radial quantisation, 
is obtained from the strip by a map,
$w = i \frac{T}{2 \pi} \ln\zeta$, 
and the conformal invariance of the boundary conditions amounts to
the following conditions on the 
boundary states $\{\ket{B}\}$:
\bea
 \label{def:bc0}
   ( A_n - (-1)^s \ol{A}_{-n} ) \ket{B} &=& 0 ,
\eea
where $A_n$ ($\,{\ol{A}_n}\,$) denotes an $n$-th mode of
the (anti-) holomorphic sector of the chiral algebra $\A$, and $s$ is a dimension
of the operator. 
Among solutions of eq. (\ref{def:bc0}), {Ishibashi} states \cite{ishibashi1} are known to
form a basis of boundary-state space, at least for the minimal RCFTs, and 
express the partition function of the closed string picture in a more
convenient way as below.
\bea
 \label{def:Zclose}
   Z_{\al \beta}^{closed} (\wt q) 
= \bra{\wt\al} \wt q^{\frac12 \left( L_0 + \ol{L}_0 - \frac{c}{12}
\right)} \ket{\wt\beta} 
= \sum_{i} \braket{\wt\al|i}\braket{i|\wt\beta} \chi_{i}(\wt q)
= \sum_{i} (M_{\al})_{\beta}^{\;\; i} \chi_{i}(\wt q),
\eea
where $\wt q\equiv e^{2\pi i \wt\tau}, \wt\tau\equiv - 1/\tau$.
Note that $\{\ket{i}\}$ denote Ishibashi states.

Equating the above two partition functions, we obtain 
the modified version of Cardy's equation:
\bea
 \label{def:modified Cardy eq}
   Z_{\al \beta}^{closed} (\wt q) &=& Z_{\al \beta}^{open} (q) ,\nn
(M_{\al})_{\beta}^{\;\; j} &=& (n_\al)_{\beta}^{\;\; i} S_i^{\;\; j} .
\eea
where $S_i^{\;\; j}$ is a matrix of the modular transformation $S$
realised on the characters. 
Modular properties 
of the characters lead to the relations between boundary states and 
$n_{\al \beta}^{\;\; i}$, by which  
the boundary states in terms of Ishibashi states
and $n_{\al \beta}^{\;\; i}$ are equated.

In the unitary minimal models,
modular properties completely determine the form of states and the $n_{\al\beta}^{\;\; i}$, and lead to the Verlinde formula of boundary CFT.
However, the condition for the existence of a Verlinde formula 
is that $n_{\al \beta}^{\;\; i}=\delta_\beta^i$ for some $\al$, in other
words, the theory should have a unique `vacuum' and the $n_{\al \beta}^{\;\; i}$ are
identical to the fusion rule coefficients $\N_{\al \beta}^{\;\; i}$. 
The diagonality of Ishibashi states is essential in this construction 
but is absent in LCFT. 
This is because LCFT possesses reducible but indecomposable representations which are obviously {\it not} diagonal. 

However, in the final form of eq. (\ref{def:modified Cardy eq}), the diagonality is hidden,
and a more important point is whether we can find the matrix $M_\al$.
Once the basis of the boundary states is obtained, we can equate the partition functions as in the first line of (\ref{def:modified Cardy eq}) and find possible $M_\alpha$. For such a reason, we will investigate the basis of the boundary states in detail in chapter \ref{ch:boundary states}.
We will consider Ishibashi states first and then coherent boundary states.
Note that it would be natural to obtain Ishibashi states for the Verlinde formula as it was first found from this type of states.

Before we turn to the boundary states, we will go further into the theories on the upper half-plane and their correlators.
In the following chapter, we will construct free boson realisations
of $c=-2$ and of $SU(2)_k$ with a boundary, and calculate their two-point functions.

\chapter{Free Field Realisations of Boundary LCFT}
\label{ch:free}

Among free field representations, the Coulomb gas picture (CG), or
Dotsenko-Fateev construction, is known to represent a wide variety
of conformal field theories (CFTs) in two dimensions \cite{Dotsenko:1984nm,Dotsenko:1984ad}.
Besides, free boson realisations are available and useful in various
models such as the $SU(2)_k$ WZNW models via bosonisation \cite{Gerasimov:fi}.
However, the boundary conformal field theory (BCFT) has not been much studied
in CG picture \cite{schulze,kawaic1,kawaic2} and in the free boson realisations, especially for logarithmic conformal field theories (LCFTs).

By the method described in chapter \ref{ch:BCFT}, BCFT on the upper half-plane can be realised by a chiral theory on the whole complex plane. In this chapter, we start with the above free boson realisations on the upper half-plane and examine some models by the method in search of LCFTs. We also calculate their boundary two-point functions with logarithms and their relations with boundary operators, and confirm the results in \cite{KW}.

\section{The Coulomb Gas Picture of LCFT with Boundary}
\label{sec:CG}

In this section, we consider the Coulomb gas picture of the $c_{p,q}$ models with a boundary, for LCFTs in particular. 
We will later focus on a $c_{1,2}=-2$ case with the Neumann boundary condition.

\subsection{The $c_{p,q}$ Models with Boundary}
\label{sec:pq}

The action functional of a free boson $\Phi$ is given by:
\begin{eqnarray}
\label{def:action pq-model}
 S_{CG}[\Phi] = \frac1{8\pi} \int_\Sigma d^2 z \sqrt{g} 
 \left[ g^{\mu\nu} \pa_\mu \Phi(z,\bz) \pa_\nu \Phi(z,\bz) + 2\,i \alpha_0 R^{(2)} \Phi(z,\bz) \right].
\end{eqnarray}
$\alpha_0$ is the background charge $\alpha_0 = (p-q)/\sqrt{2pq}$,
and the central charge is $c_{p,q}=1-12\alpha_0^2=1-\frac{6(p-q)^2}{pq}$ with
$p,q\not=0$. 
$R^{(2)}$ is the two-dimensional scalar curvature and $(z,\bz)$
are the complex coordinates of the 2d Riemann surface $\Sigma$.
$\Sigma$ has a boundary and is equipped with the globally well-defined metric $g_{\mu\nu}(z,\bz)$.
A prototype of such geometries is the upper half-plane ($UHP$), thus 
let us consider $\Sigma=UHP$ and the flat metric $g_{\mu\nu}=\left(\begin{array}{cc}
                               0 & 1/2 \\
                               1/2 & 0 \\
                                    \end{array} \right)$.
Here the background charge is concentrated at infinity, 
and the 1d boundary $\pa\Sigma$ is merely the real axis.\footnote{
We use the conventions $d^2 z = dx dy$ and 
$\delta^2 (z)=\delta(x)\,\delta(y)$ so that 
$\int d^2 z\; \delta^2 (z) = 1$.
}
When $p=q+1, |p|>2, |q|>2$ and both are coprime integers,
the theory becomes the well-known unitary minimal model \cite{BPZ,Dotsenko:1984nm,Dotsenko:1984ad},
but we do not specify $p$ and $q$ here for a more general case.
A restriction will come up later in (\ref{assum:pq}).

The first variation of the action in $\Phi$ gives the equation of motion
in the bulk and the boundary conditions: 
\begin{eqnarray}
\label{def:EOM pq-model}
& \pa \bar\pa \Phi(z,\bz) = 0 \,,&
\nn
& \delta \Phi(z,\bz) = 0 \;\Bigr|_{\pa\Sigma} &{\rm ~~~(Dirichlet~boundary~condition)}, 
\nn
& \left(\pa - \bar\pa\right) \Phi(z,\bz) = 0 \;\Bigr|_{\pa\Sigma} &{\rm ~~~(Neumann~boundary~condition)} .
\end{eqnarray}
$\pa$ and $\bar\pa$ denote $\frac{\pa}{\pa z}$ and $\frac{\pa}{\pa\bz}$,
respectively.
The contribution from the second term in (\ref{def:action pq-model}) vanishes here due to the flat metric, \alert{except at infinity}.
The first equation of (\ref{def:EOM pq-model}) implies that
$\Phi(z,\bz)$ is analytic.
It is therefore
customary to decompose $\Phi$ as $\Phi(z,\bz)= \phi(z) + \bphi(\bz)$
with their propagators:
\begin{eqnarray}
 \braket{\phi\phi} = -\ln (z-w)\,,\,\,  \braket{\bphi\bphi} = -\ln (\bz-\bw) .
\end{eqnarray}
For simplicity, the variables ($z,\bz,w,\bw$) are often omitted, while $\braket{\phi\phi}$ stands for $\braket{\phi(z)\phi(w)}$.
Although the two boundary conditions are listed in eq. (\ref{def:EOM
pq-model}), 
we only deal with the Neumann boundary condition ($N$ for short) in what
follows. 
The imposition of the boundary condition ($N$) adds up two more propagators to
the above:
\begin{eqnarray}
\label{def:propagator in N}
 \braket{\phi\bphi} = - \ln (z-\bw) \,,\;\, \braket{\bphi\phi} = - \ln (\bz-w) .
\end{eqnarray}
Note that the Dirichlet boundary condition in CG was discussed in \cite{schulze}.

On the other hand, the variation of the action in $g^{zz}$ leads to the stress tensor
$T(z)$ (the energy-momentum tensor):
\begin{eqnarray}
 T(z) = - \frac{4\pi}{\sqrt{g}} \frac{\delta S[\phi]}{\delta g^{zz}} = - \frac12 :\left(\pa\phi(z)\right)^2: + i \alpha_0 \pa^2 \phi(z) ,
\end{eqnarray}
while its anti-holomorphic counterpart is given similarly.
The normal ordering \linebreak ($: operators :$) is introduced to the first term to
subtract its divergence.

By the ``method of images'' initiated by Cardy \cite{Cardy:bb}, the
anti-holomorphic sector of the theory can be mapped onto the lower
half-plane ($LHP$) by $\bphi(\bz) = \Omega \phi(z^*)$ with the parity 
$\Omega=1$ ($\Omega=-1$ for the Dirichlet case). In the given geometry,
$\bz$ is equal to $z^*$, but we keep \alert{$z^*$} for a while for convenience. 
Thereby, all the propagators
$\braket{\phi\phi}$, $\braket{\bphi\bphi}$, $\braket{\phi\bphi}$, $\braket{\bphi\phi}$,
are simplified to $\braket{\phi\phi}$ with corresponding arguments, 
and the theory becomes effectively chiral. Indeed, Cardy's equation on
the boundary:
\begin{eqnarray}
 T(z)=\ol T(\bz) \Bigr|_{\pa\Sigma}
\end{eqnarray}
is trivially satisfied, and the conformal transformations of the primary
fields are generated by the single tensor field $T(z)$ being analytically
continued to $z\in\C$.

The chiral primary Kac fields $\Phi_{r,s}(z)$ of the $c_{p,q}$ 
model have a pair of integer indices $r,s$.
Their conformal dimensions obey the Kac formula \cite{kac}: $h_{r,s}=
-\frac12 \alpha_0^2 + \frac18 \left( r \alpha_+ + s \alpha_- \right)^2$ 
with $\alpha_\pm \equiv \alpha_0 \pm \sqrt{\alpha_0^2 + 2}$.
It is assumed that the corresponding non-chiral primaries are symmetric products
of the chiral primaries:\footnote{
The ``symmetric'' condition is often called the ``spinless'' condition since it is equivalent
to $\Phi_{r,s}(z,\bz) \in Ker(L_0-\ol L_0)$.
}
\begin{eqnarray}
 \Phi_{r,s}(z,\bz) = \Phi_{r,s}(z) \otimes \bar\Phi_{r,s}(\bz).
\end{eqnarray}
Accordingly, 
by the method of images, 
the non-chiral primary becomes a simple product of two identical chiral
fields: $\Phi_{r,s}(z,\bz) = \Phi_{r,s}(z)\,\Phi_{r,s}(z^*)$.
\alert{Then the} problem of a boundary $N$-point function is reduced 
to that of a chiral $2N$-point function. 
For example, the boundary two-point function is illustrated as below:
\begin{eqnarray}
\label{def:2pt-4pt}
 \braket{\Phi_{r,s}(z,\bz)\Phi_{r^\prime,s^\prime}(w,\bw)}_N 
= \braket{\Phi_{r,s}(z)\Phi_{r^\prime,s^\prime}(w)\Phi_{r^\prime,s^\prime}(w^*)\Phi_{r,s}(z^*)}_0 \,\,.
\end{eqnarray}
$\braket{\cdots}_0$ denotes the vacuum expectation value in the mapped
chiral theory. Its charge condition in CG will appear shortly.

In CG, chiral fields are realised by the vertex operators:
\begin{eqnarray}
\label{def:vop}
 V_\alpha \equiv :e^{i \alpha \phi}: {\rm ~~~and~~~} \wt V_\alpha \equiv V_{2 \alpha_0 -\alpha}, 
\end{eqnarray}
with their common conformal dimension $h_\alpha=\frac12 \alpha\left(\alpha - 2 \alpha_0\right)$.
The former operator is called 'normal' and the latter is called
'conjugate' or 'dual'. 
We often denote the conjugate charge $2\alpha_0-\alpha$ by $\wt \alpha$.
As in the expression (\ref{def:vop}), the vertex operator realisation provides two ways
of representing the chiral primaries:
\begin{eqnarray}
\label{def:V_{r,s}}
 \Phi_{r,s}(z) \sim V_{r,s}(z) {\rm ~~or~~} \wt V_{r,s}(z) .
\end{eqnarray}
$V_{r,s}$ denotes $V_{\alpha_{r,s}}$ for short, with the charge  
$\alpha_{r,s}=\frac{1-r}2 \alpha_+ + \frac{1-s}2 \alpha_-$.

In addition to the above primaries, there are assumed to exist
marginal operators $J_\pm$ called the ``screening currents''.
These vertex operators of dimension 
$(h,\bar h) = \{(1,0), (0,1)\}$
are invariant up to total derivative under the conformal transformations, 
and thus they don't disturb conformal properties of correlation
functions when inserted. 
The $(1,0)$ screening current gives rise to the following conserved operator
under the conformal transformations:
\begin{eqnarray}
\label{def:scop BCFT}
 Q_\pm \equiv \oint_{C} d\zeta\, J_\pm (\zeta), \; J_\pm \equiv V_{\alpha_\pm},
\end{eqnarray}
where the contour $C$ is on $\Sigma=UHP$. 
The other currents are given similarly.

The vacuum expectation value of an arbitrary product of
these vertex operators may result in a trivial value,
unless the charges in 
$\braket{V_{\alpha_1}(z_1) \cdots V_{\alpha_N}(z_N)\, Q_+^{\,n_+} Q_-^{\,n_-} \ol Q_+^{\,\bar n_+} \ol Q_-^{\,\bar n_-}}_0$
satisfy the following neutrality condition (charge asymmetry condition
in \cite{Dotsenko:1984nm}): 
\begin{eqnarray}
\label{def:charge cond pq}
  \sum_{i=1}^N \alpha_i + \sum_{\epsilon=+,-} 
  \left(n_\epsilon + \bar n_\epsilon\right) \alpha_\epsilon
 = 2 \alpha_0 \;\;. 
\end{eqnarray}
This condition is as seen from integration of the constant mode
 $\phi_0$.\footnote{
This can easily be viewed in the path integral expression.
}

As an simple example, a boundary one-point function is illustrated
blow:
\begin{eqnarray}
\label{eq:1pt pq}
 \braket{\Phi_{r,s}(z,\bz)}_N = \braket{\Phi_{r,s}(z)\Phi_{r,s}(z^*)}_0
 \sim \braket{V_{r,s}(z) \wt V_{r,s}(z^*)}
 = (z-z^*)^{- 2 h_{r,s}} \,.
\end{eqnarray}
First, the method of images is used. Then the fields are replaced by 
the equivalent product of vertex operators which leads to the
final result trivially. 
This simplicity comes from the fact that the above pair of 
vertex operators trivially satisfy the condition 
(\ref{def:charge cond pq}) as $\alpha_{r,s}+\wt \alpha_{r,s} = 2\alpha_0$.
This also shows that $p=q$ is a special point where $\alpha_0$ becomes
zero and the screening charges are not required, as $\braket{V_{\alpha_1}\wt V_{\alpha_1}\cdots V_{\alpha_N}\wt V_{\alpha_N}}$ satisfies the condition (\ref{def:charge cond pq}).
Note that the expression (\ref{eq:1pt pq}) is exact up to a normalisation
constant, which is related to boundary states \cite{Cardy:tv}. 
We will not deal with them here.

\subsection{Boundary Two-Point Functions of $\Phi_{1,2}$ and $\Phi_{r,s}$ Fields with Logarithms}

Suppose $p\not= q$.
Consider a boundary two-point function with the Neumann
boundary condition:
\begin{eqnarray}
\label{2pt pq}
 \braket{\Phi_{r,s}(z_1,\bz_1) \Phi_{1,2}(z_2,\bz_2)}_N \,.
\end{eqnarray}
From eq. (\ref{def:2pt-4pt}), this can be expressed in a chiral and
symmetric form by placing the symmetric charges in the $LHP$.\footnote{
The word ``symmetric'' for CFT correlators was first introduced to four-point correlation functions of the type $\braket{ABBA}$ \cite{Dotsenko:1984nm}.
}
With the relation (\ref{def:V_{r,s}}) and the condition 
(\ref{def:charge cond pq}), there are two choices in its vertex operator
realisation. 
The general form of the two-point function seems to be given by a linear combination of them:
\begin{eqnarray}
\label{eq:linear comb pq}
\braket{\Phi_{r,s}(z_1,\bz_1) \Phi_{1,2}(z_2,\bz_2)}_N
&=& \braket{\Phi_{r,s}(z_1) \Phi_{1,2}(z_2) \Phi_{1,2}(z_2^*) \Phi_{r,s}(z_1^*)}_0
\nn
&=& \alpha_1^{(pq)}\, F_1(z_1, z_2) + \alpha_2^{(pq)}\, F_2(z_1, z_2) \,,
\end{eqnarray}
where $\alpha_1^{(pq)}, \alpha_2^{(pq)}$ are arbitrary constants 
and the two choices are:
\begin{eqnarray}
\label{def:2 int pq}
F_1(z_1, z_2) &\equiv& \braket{V_{r,s}(z_1) V_{1,2}(z_2) V_{1,2}(z_2^*) \wt V_{r,s}(z_1^*)\; Q_- } \,,
\nn
F_2(z_1, z_2) &\equiv& \braket{V_{r,s}(z_1) V_{1,2}(z_2) V_{1,2}(z_2^*) \wt V_{r,s}(z_1^*)\; \ol Q_- } \,.
\end{eqnarray}

The problem of $\braket{\Phi_{r,s}\Phi_{1,2}}_N$ is now 
reduced to the calculation of two integrals.
First we have:
\begin{eqnarray}
\label{eq:rs-12 vertex}
F_1 &=& \oint_{C\subset\Sigma} \!\!\!d\zeta\, 
 \braket{V_{r,s}(z_1) V_{1,2}(z_2) V_{1,2}(z_2^*) \wt V_{r,s}(z_1^*)\,J_-(\zeta) }
\nn
&=& \prod_{i<j} {z_{ij}}^{\alpha_{i} \alpha_{j}} \oint_{C\subset\Sigma} \!\!\!d\zeta\, 
 \prod_{i=1}^4 \left( z_i - \zeta \right)^{\alpha_{i} \alpha_-} ,
\end{eqnarray}
with $\alpha_2=\alpha_3=\alpha_{1,2}$, $\alpha_1=\alpha_{r,s}$,
$\alpha_4= 2 \alpha_0 - \alpha_{r,s}$,
and $z_{ij}=z_i - z_j$, $z_3=z_2^*$, $z_4=z_1^*$.
Then one can easily obtain the function $F_2$ from $F_1$ by replacing $C$ by $\bar C$ on $\Sigma^*=LHP$.
Since the exponents in the integrand are:
\begin{eqnarray}
\label{eq:exponents}
&& \alpha_1 \alpha_- = 
\frac1p \left( (1-s)q-(1-r)p \right),\,\, 
 \alpha_4 \alpha_- = \frac1p \left( (1+s)q-(1+r)p \right) , 
\nn
&&
\alpha_2 \alpha_- = \alpha_3 \alpha_- 
 = - \frac{q}p \,\,, 
\end{eqnarray}
$\zeta= z_1, z_2, z_3, z_4$ are branch points of order $p$, 
when $p,q$ are coprime integers and $(1\pm s)$ are not multiples of $p$.

\begin{figure}[h]
\vspace*{0.5cm}
\hspace*{4.8cm}
\epsfxsize=5.7cm
\epsffile{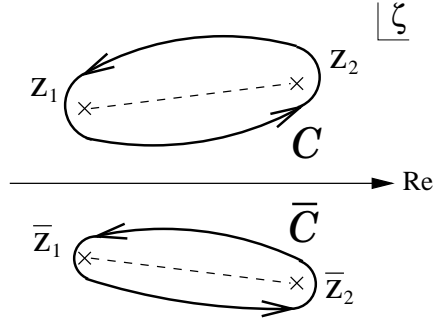}
 \caption{Non-contractable contours $C,\,\bar C$ for the $p=2$
 case. Dotted lines are branch cuts.}
      \label{fig:pochhammer}
\end{figure}

\begin{figure}[h]
\vspace*{0.5cm}
\hspace*{2.5cm}
\epsfxsize=10cm
\epsffile{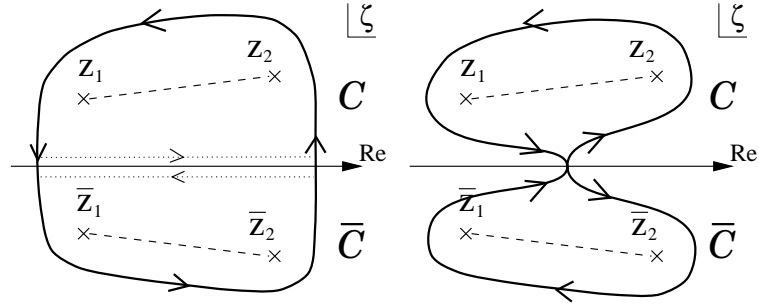}
 \caption{The contours on both half-planes. (a) Left and (b) Right
 have opposite relative signs of $\alpha_1^{(pq)}$ and $\alpha_2^{(pq)}$.}
      \label{fig:pinch}
\end{figure}

As the integration contours $C,\bar C$ are bounded on each half-plane, 
nontrivial non-contractable contours are the ones encircling
two points, or the Pochhammer-type contour between the two points, on
the same half-planes (see Fig.\ref{fig:pochhammer} for $p=2$).
By dragging one point for each contour to the
boundary, and jointing them with their contour directions unchanged, 
one can construct two types of contours on the $\zeta$-plane 
(Fig.\ref{fig:pinch} for $p=2$). 
The functional form is just of the form (\ref{eq:linear
comb pq}), a linear combination of two contour integrations.

\begin{figure}[h]
\vspace*{0.5cm}
\hspace*{2.5cm}
\epsfxsize=10cm
\epsffile{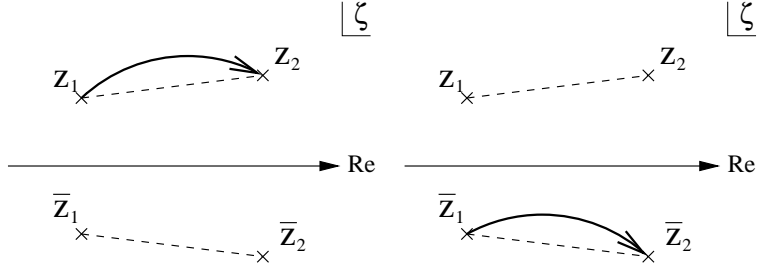}
 \caption{The line integrals from $z_1$ to $z_2$ (Left) and from $\bz_1$ to $\bz_2$ (Right).}
      \label{fig:line int}
\end{figure}

Provided that integrations of the infinitesimal arcs around 
the four points don't
contribute ($i.e.$ let the contours be the Pochhammer-type or the pair
$(r,s)$ be chosen as such\footnote{The conditions for the vanishing arc integrations
are:
$
 1 > \frac{q}p \,,\;\; r > \frac{q}p (s-1) {~and~} r< \frac{q}p (s+1) .
$
}), 
both closed contour integrals reduce to two line integrals of 
$z_1\to z_2$ and of $z_1^* \to z_2^*$ up to constant 
(Fig.\ref{fig:line int}). 
These constants can be absorbed by $\alpha_1^{(pq)}$ and $\alpha_2^{(pq)}$.
For this reduction with no poles, we assume:
\begin{eqnarray}
\label{assum:pq}
 \frac{q}{p}\not\in\Z {~~~and~~~} \frac{q\,(1\pm s)}{p}\not\in\Z .
\end{eqnarray}

The integral (\ref{eq:rs-12 vertex}), the function of $z_1, z_2, z_3,
z_4$, is projectively invariant so that one can fix three formal degrees of
freedom by a certain conformal mapping.
A standard choice of such mappings is $u=\frac{z_{34}}{z_{31}}\frac{\zeta-z_1}{\zeta-z_4}$ : $(z_1,z_2,z_3,z_4)\to (0,\xi,1,\infty)$. 
Under this mapping, the boundary real axis is
mapped onto a circle of radius $|{z_{34}}/{z_{31}}|$, and the
contour of the interval $(z_1,z_2)$ is mapped to $C_1=(0,\xi)$ in the interior of the circle. 
The integral is now transformed into:
\begin{eqnarray}
\label{eq:F1 pq}
 F_1 = 
 z_{14}^{-2 h_{r,s}} z_{23}^{-2 h_{1,2}} \xi^{\alpha_1 \alpha_2} \left(1-\xi\right)^{2\alpha_2 (\alpha_2 - \alpha_0)}
 \int_0^\xi du\, u^{\alpha_1 \alpha_-} (\xi-u)^{\alpha_{2} \alpha_-}  (1-u)^{\alpha_{2} \alpha_-} ,
\end{eqnarray}
where $\xi$ is an harmonic ratio, 
$\xi=\frac{z_{12}z_{34}}{z_{13}z_{24}}=\left|\frac{z_1-z_2}{z_1-z_2^*}\right|^2$
and $0<\xi<1$.
We omitted the phase $(-1)^{(\alpha_4+2\alpha_2)\alpha_- +1}$. 
This clearly agrees with the conformal property of the boundary two-point function in question.
The function $F_2$ is given similarly by the contour $C_2=(1,\infty)$.

The integrals in $F_i$'s are the integral representations of the first two
Kummer's solutions of the hypergeometric differential equation \cite{slater}:\footnote{
This second-order ordinary differential equation is also known as
Euler's hypergeometric differential equation.
}
\begin{eqnarray}
\label{gauss eq}
 \left[ \xi (1- \xi) \frac{d^2}{d\xi^2} 
 + \left\{c - \left(1+a+b\right)\xi \right\} \frac{d}{d\xi} -a b \right]
 F_i (\xi) = 0 \,,
\nn
\left\{
\begin{array}{rcl}
 a =& \alpha_-^2/2 &= \frac{q}{p} \\
 b =& -1 - \left(\alpha_1 - \alpha_- \right)\alpha_- 
    &= \frac{q}p \left(1+s\right) - r \\
 c =& - \left( \alpha_1 -\alpha_-/2 \right) \alpha_- 
    &= \frac{q}p\;s -r +1 
\end{array}
\right. \,,
\end{eqnarray}
\begin{eqnarray}
\label{sol:pq}
 F_1 &\sim& \frac{\Gamma(1+a-c)\Gamma(1-a)}{\Gamma(2-c)}\; \xi^{1-c}\; 
  {}_2F_1\left(1+a-c,1+b-c;2-c;\xi \right) \,, 
\nn
 F_2 &\sim& \frac{\Gamma(b)\Gamma(c-b)}{\Gamma(c)}{}_2F_1\left(a,b;c;\xi \right) \,, 
\end{eqnarray}
The prefactor of the integral in (\ref{eq:F1 pq}) is in common for both
solutions and is therefore neglected. 
As is usual in the theory of hypergeometric
functions, it is assumed that the integral in (\ref{eq:F1 pq}) is 
analytically continued from a convergent integral unless
${\rm Re}(\alpha_1\alpha_-)>-1$ and ${\rm Re}(\alpha_2\alpha_-)>-1$. 
The form of the solutions (\ref{sol:pq}) is
essentially same as $\braket{\phi_{n,m}\phi_{1,2}\phi_{1,2}\phi_{n,m}}$
in \cite{Dotsenko:1984nm}, since the calculations are symmetric under
$\alpha_+ \leftrightarrow \alpha_-$.
This observation also agrees with 
$\braket{\phi_{1,2}\phi_{1,2}}_{boundary}$ in \cite{kawaic2}.
It should be mentioned that eq. (\ref{gauss eq}) is in complete
agreement with the differential equation that the function
(\ref{eq:linear comb pq}) of the Kac fields must satisfy
\cite{BPZ,Dotsenko:1984nm,Kogan:1997fd,KW}. 
Thus, the linear combination in eq. (\ref{eq:linear comb pq}) indeed
gives the general solution.

However, 
as was discussed in \cite{gura,Kogan:1997fd}, the second-order linear differential
equation may possess a solution logarithmic at $\xi=0$, through the Frobenius
method or other similar methods, and lead to a logarithmic field of LCFT.
In such {\it logarithmic} cases, either two hypergeometric functions in the solutions (\ref{sol:pq}) coincide, or one of them becomes indefinite at $\xi=0$. 
They do not provide two independent solutions of
eq. (\ref{gauss eq}). 
The condition for such an emergence of logarithms is 
that the coefficient $c$ takes integer value \cite{slater}.
Since $r$ and $s$ are integers by definition, the condition can be translated
into that of $p,q,s$:
\begin{eqnarray}
\label{cond:log}
 s\cdot \frac{q}{p} \in \Z .
\end{eqnarray}
It should be mentioned here that this
condition is compatible with the condition (\ref{assum:pq}).

Under the conditions (\ref{assum:pq}, \ref{cond:log}), one can explicitly
observe that the two solutions (\ref{sol:pq}) become identical. 
Suppose that $c$ is a positive integer: $c=1+n,\, n\in\Na$. 
The relevant part of $F_2$ turns out to be a solution of
eq.(\ref{gauss eq}) regular at $\xi=0$. 
On the other hand, from eq. (\ref{sol:pq}) and the definition of
hypergeometric functions, one finds:
\begin{eqnarray}
\label{eq:identical}
 F_1 &\sim& \frac{\Gamma(1-a)}{\Gamma(1+b-c)}\; \xi^{-n} \sum_{k=0}^{\infty} 
    \frac{\Gamma(a-n+k)\Gamma(b-n+k)}{\Gamma(1-n+k)\, k!}\; \xi^k 
\nn
 &=& \frac{\Gamma(1-a)\Gamma(a)\Gamma(b)}{\Gamma(1+b-c)\Gamma(c)} 
     {}_2 F_1 (a,b;c;\xi), 
\nn
 &\Leftrightarrow& F_1 = \frac{\Gamma(1-a)\Gamma(a)}{\Gamma(1+b-c)\Gamma(c-b)} F_2 
\nn
 &\Leftrightarrow& F_1 = F_2 \,,
\end{eqnarray}
up to phase factor. 
The first $n$ terms of the series in the first line vanish, because of
$\frac{1}{\Gamma(-n+k)}=0$ and $a,b \not\in Z$. 
Non-integer $a$ and $b$ also ensures that none of the other gamma 
functions in (\ref{eq:identical}) are divergent and 
the above identical solution is well-defined with integer $c$. 
Consequently, one can write down the
solution for integer $c$ as:
\begin{eqnarray}
\label{sol:pq2}
 \braket{\Phi_{r,s}(z_1,\bz_1)\Phi_{1,2}(z_2,\bz_2)}_N
= \alpha \;C(z_1,z_2)\; 
{}_2 F_1\left(\frac{q}{p}, \frac{q}{p} +|n| ; 1+|n|; \xi\right) \,,\,\,
\end{eqnarray}
where $n=(\frac{q}{p} s-r)\in\Z$, $\alpha$ is an arbitrary constant.
The function $C(z_1,z_2)$ is the prefactor in eq. (\ref{eq:F1 pq}):
\begin{eqnarray}
 C(z_1,z_2) = 
(z_1-\bz_1)^{-2 h_{r,s}} (z_2-\bz_2)^{-2 h_{1,2}} 
 \xi^{\frac1{2p}\left\{(s-1)q+(1-r)p\right\}} \left(1-\xi\right)^{\frac{2q}{p} -1} ,
\end{eqnarray}
with the conformal dimensions 
$h_{r,s}=\frac12 \alpha_{r,s}(\alpha_{r,s}- 2\alpha_0)$ and 
$h_{1,2}= \frac12 \alpha_{1,2}(\alpha_{1,2}- 2\alpha_0)$.
The theory of hypergeometric functions further shows that the
above solution is logarithmic at $\xi=1$ in some cases. In 
section \ref{sec:c=-2 bop}, we will see this logarithm and discuss its meaning in a
particular case, $c_{2,1}=-2$.

The solution found in eq. (\ref{sol:pq2}) is, of course, not the general solution of eq. (\ref{gauss eq}).
The other independent solution of eq. (\ref{gauss eq}), which is logarithmic at $\xi=0$, can be obtained through various methods. 
However, given two functions (\ref{sol:pq}), the method is almost
unique, that is, 
the differentiation method of the two hypergeometric functions with appropriate prefactors \cite{salu,Hata:2000zg,slater}. 
Omitting again the function $C(z_1,z_2)$ in eq. (\ref{sol:pq}), it reads 
\begin{eqnarray}
\frac{\pa}{\pa c} 
\left(\frac{\Gamma(1+b-c)}{\Gamma(1-a)} F_1 - \frac{\Gamma(a)}{\Gamma(c-b)} F_2 \right) 
= \lim_{c\to n} \frac{\frac{\Gamma(a)}{\Gamma(1-a)}(F_1 - F_2)}{c-n} \,.
\end{eqnarray}
Under the conditions (\ref{assum:pq}, \ref{cond:log}), the gamma
functions become convergent.
Thus, in order to obtain the other solution as the $c \to n$ limit, 
the normalisation constants $\alpha_1^{(pq)}$ and $\alpha_2^{(pq)}$ must
unusually be $\alpha_1^{(pq)}= - \alpha_2^{(pq)}$ and of order $O((c-n)^{-1})$. 
Otherwise, 
the difference of two functions either vanish or remains proportional
to (\ref{sol:pq2}). 
Those normalisation constants are related to inner products of
boundary states, which are not specified yet. So, let us not expect
the constants to take such extraordinary values, but discuss another
way. We will show in the next section how the logarithmic solution
naturally emerges in the CG.

Before we move to the following section, we would like to make some remarks on
the condition (\ref{cond:log}).
The condition itself can be interpreted in two ways: 
Firstly, in terms of $s$, the condition means that it is possible for any $c_{p,q}$ model to have such a two-point function as (\ref{sol:pq2}) in the presence of boundary, if one includes the $(r,s=p)$ field in the theory. 
The result may contain a logarithm in the vicinity of $\xi=1$.
Secondly, in terms of $(p,q)$, the condition means that the theory
becomes logarithmic if $p$ is a multiple of $q$.
In this case,
$\braket{\Phi_{r,n \frac{p}{q}}\Phi_{1,2}}_{N}$ for $n\in\Z$ plays
the role, sometimes with both fields inside the conformal grid $\{\Phi_{r^\prime,s^\prime}\vert\; |r^\prime|<q, |s^\prime|<p\}$. 
Note that $c_{p,1}$ models are a part of this class 
since integer $p$ is always a multiple of $1$.
This is consistent with Flohr's conjecture and observation that the augmented conformal grid of the $c_{p,q}$ models gives the rational logarithmic CFT \cite{flohr2,flohr3,flohr6}.

\subsection{Another Operator for the Logarithmic Solutions}
\label{sec:logsol}

We now apply the CG method to show that one can naturally obtain the 
other independent solution, which is logarithmic at $\xi=0$. 
Define two screening currents analytically continued from $\Sigma$ to $\C$:
\begin{eqnarray}
\label{def:new screener}
  \wt J_\pm = V_{\alpha_\pm} (z) ~~~for~~z\in\C.
\end{eqnarray}
Recall that the conformal transformations are realised by the single
tensor field $T$, and the conformal dimension of $\wt J_\pm$ is 
$\wt h=1$ under this $T$. 
Such transformations of $\wt J_\pm$ merely give total derivatives. 
Therefore, by making a closed contour integration on the whole
 plane, one can define the analytically continued version of
the screening charges:
\begin{eqnarray}
\label{def:new charge}
 \wt Q_\pm = \oint_{\wt C} d\zeta J_\pm^c(\zeta) . 
\end{eqnarray}
Here, $\wt C$ can lie anywhere on the whole complex plane $\C$, so that
the chiral screening currents can also be realised by
these operators. 

\begin{figure}[h]
\vspace*{0.5cm}
\hspace*{4.8cm}
\epsfxsize=5.7cm
\epsffile{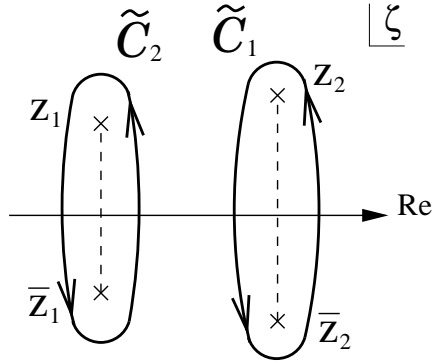}
 \caption{Two non-contractable contours for $\wt C$ for the $p=2$
 case. Subscripts of the contours are for later convenience.}
      \label{fig:c23}
\end{figure}

With these operators, one can compute a chiral four-point function:
\pagebreak
\begin{eqnarray}
 \wt F(z_1,z_2) &=& \braket{V_{r,s}(z_1) V_{1,2}(z_2) V_{1,2}(z_2^*) V_{r,s}(z_1^*) \,\wt Q_-} . 
\nn
 &=& \prod_{i<j} z_{ij}^{\alpha_i \alpha_j} \oint_{\wt C\subset \C} d\zeta \prod_{i=1}^4 \left(z_i-\zeta\right)^{\alpha_i \alpha_-} ,
\end{eqnarray}
with the exponents (\ref{eq:exponents}).
Again, $z_3$ and $z_4$ denote $z_3=z_2^*$, $z_4=z_1^*$ 
while $\bz_i=z_i^*$ in our case.
The only difference between the above and eq. (\ref{eq:rs-12 vertex}) is
 the contour $\wt C$, so we restrict $\wt C$ to be analytically
 inequivalent to $C$ and $\bar C$ in eq. (\ref{eq:rs-12 vertex}) 
for nontrivial results.
The branching structure of the integrand is not unique 
(see Fig.\ref{fig:pochhammer}).
For example, a different structure with different non-contractable
contours for the branch covering of $\C$ is shown in Fig.\ref{fig:c23} for $p=2$. 
The new contours are now encircling the points $z_2$, $\bz_2$ ($z_1$, $\bz_1$), or Pochhammer-type contours between them.
The closed contours can be reduced to the line
integrals under the condition (\ref{assum:pq}).

When the two hypergeometric functions (\ref{sol:pq}) become
identical ($i.e.$ the {\it logarithmic} cases), the two line integrals
around $\wt C_1$ and $\wt C_2$ also become identical. 
This can easily be seen by analytically changing the contour of
$(\bz_1,\bz_2)$ to that of 
$(\bz_1,z_1)+(z_1,z_2)+(z_2,\bz_2)$ and neglecting the phases: 
\begin{eqnarray}
 0 = \int_{\bz_1}^{\bz_2}  - \int_{z_1}^{z_2}  
   = \left[\int_{\bz_1}^{z_1} + \int_{z_1}^{z_2} + \int_{z_2}^{\bz_2}\right]
     - \int_{z_1}^{z_2}
   = \int_{z_2}^{\bz_2} - \int_{z_1}^{\bz_1} .
\end{eqnarray}

By the same conformal mapping as for eq. (\ref{eq:F1 pq}), one obtains
for $\wt C_1$:
\begin{eqnarray}
  \wt F_1 &=& z_{14}^{-2h_{r,s}} z_{23}^{-2h_{1,2}} 
  \xi^{\alpha_1 \alpha_2} (1-\xi)^{2\alpha_2 (\alpha_2 -\alpha_0)}
  \int_\xi^1 du\, u^{\alpha_1 \alpha_-} (u-\xi)^{\alpha_2 \alpha_-} (1-u)^{\alpha_2 \alpha_-} 
\nn
  &=& C(z_1, z_2)\, \int_0^{1-\xi} du\, u^{\alpha_2 \alpha_-} (1-\xi-u)^{\alpha_2 \alpha_-} (1-u)^{\alpha_1 \alpha_-} 
\nn
  &=& C(z_1, z_2)\, (1-\xi)^{c-a-b} \frac{\Gamma(c-b)\Gamma(1-a)}{\Gamma(1-a-b+c)}\,
  {}_2 F_1 \left( c-a, c-b; 1-a-b+c; 1-\xi\right) 
\nn
  &=& C(z_1, z_2)\, (1-\xi)^{1-\frac{2 q}{p}} \frac{\Gamma^2\left(1-\frac{q}{p}\right)}{\Gamma\left(2\left(1-\frac{q}{p}\right)\right)} \,
  {}_2 F_1 \left( 1-\frac{q}{p}+n, 1-\frac{q}{p}; 2\left(1-\frac{q}{p}\right); 1-\xi\right) .
\nn
\end{eqnarray}
Similarly, the line integral of $\wt C_2$ reduces to:
\begin{eqnarray}
 \wt F_2 &=& C(z_1, z_2)\, \frac{\Gamma(a)\Gamma(1+b-c)}{\Gamma(1+a+b-c)}\,
  {}_2 F_1 \left( b, a; 1+a+b-c; 1-\xi\right) 
\nn
 &=& C(z_1, z_2)\, \frac{\Gamma^2\left(\frac{q}{p}\right)}{\Gamma\left(\frac{2q}{p}\right)}\,
  {}_2 F_1 \left( \frac{q}{p}+n , \frac{q}{p}; \frac{2q}{p}; 1-\xi\right) ,
\end{eqnarray}
where $(a,b,c)$ are the same as in eq.(\ref{gauss eq}) and $n=\frac{q}p s-r \in\Z$.
That is to say, $\wt F_1$ and $\wt F_2$ are the sixth and
fifth Kummer's solutions of the same hypergeometric differential equation (\ref{gauss eq}),
regular at $\xi=1$.
Under the condition (\ref{assum:pq}), they are 
two independent solutions in the region of $0<\xi<1$.

There is of course a logarithmic condition analogous to the former
case (\ref{cond:log}):
\begin{eqnarray}
  2\,\frac{q}p\in\Z.
\end{eqnarray}
Because of the condition (\ref{assum:pq}), it reads:
\begin{eqnarray}
\label{cond:log2}
 \frac{q}p = \frac12 + m \,,~~ m\in\Z .
\end{eqnarray}
Since the relation between $\wt F_i$'s is the same as between
the $F_i$'s, eq. (\ref{eq:identical}) is also applicable to this case,
and we have
\begin{eqnarray}
 \wt F = C(z_1, z_2)\, 
  \frac{\Gamma^2\left(\frac12 + |m|\right)}{\Gamma(1+2|m|)}
  {}_2 F_1 \left( \frac12+|m|+n, \frac12+|m|; 1+2|m|; 1-\xi \right) .
\end{eqnarray}
The hypergeometric function in above is singular at $\xi=0$ for
$n\geq0$. 
It can be checked by Euler's transformation that the function is also singular for $n<0$.
Therefore, this is actually the second independent solution of
eq. (\ref{gauss eq}), and 
a general solution of the boundary two-point function can be given by:
\begin{eqnarray}
\label{sol:pq double}
   \braket{\Phi_{r,s}(z_1,\bz_1) \Phi_{1,2}(z_2,\bz_2)}_N
 =  \alpha F(z_1,z_2) + \wt \alpha \wt F(z_1,z_2) .
\end{eqnarray}

It should be mentioned here that, if one replaces $\wt V_{r,s}$ with
the puncture-type operator $:\phi\; \wt V_{r,s}:$ in eq. (\ref{def:2 int
pq}), the imposition of $\braket{:\phi\; V_{2\alpha_0}:} \sim 1 $ leads
to the same results appearing in this and the previous sections. 
Note again that when $p=q$ the above discussions and solutions make no
sense because screening charges are not required and the gamma functions
in the solutions (\ref{sol:pq}) diverge. 
Therefore, for instance, we cannot discuss the flow from a $c_{2,2}=1$ theory to a $c_{2,1}=-2$ theory.

\subsection{An Example at $c_{2,1}=-2$ and the Boundary Operators}
\label{sec:c=-2 bop}

Consider the $c_{2,1}$ model in CG. The central charge is $c_{2,1}=-2$.
Since the condition (\ref{cond:log}) is $s/2 \in\Z$ in this case, 
a number of $(r,s)$ pairs can satisfy the condition for logarithms.
The condition (\ref{cond:log2}) is always satisfied
by $(p,q)=(2,1)$, 
so the general solution for the boundary two-point functions is given in
eq. (\ref{sol:pq double}). 

Denoting the Kac field $\Phi_{1,2}$ of dimension $h_{1,2}=-1/8$ by $\mu$,
let us depict the relevant two-point functions with the boundary:
\begin{eqnarray}
&& \braket{\Phi_{r,s}(z_1,\bz_1) \mu(z_2,\bz_2)}_N 
\nn
&& = (z_1 - \bz_1)^{-2h_{r,s}} (z_2 - \bz_2)^{\frac14}\;
 \xi^{\frac14\left(1+2n\right)}
\nn&&\quad\times
 \left\{
 \alpha^{(r,s)}\,
 {}_2 F_1\left(\frac12, \frac12 +|n| ; 1+|n|; \xi\right)
  + \wt \alpha^{(r,s)}\, 
 {}_2 F_1\left(\frac12+n, \frac12; 1; 1-\xi\right) 
 \right\} \,,\,\,
\end{eqnarray}
where $n=s/2 -r \in\Z$. $\alpha_N$ and $\wt \alpha_N$ are constants.
Substituting $(r,s)=(1,2)$ and $(r,s)=(2,2)$, one finds:
\begin{eqnarray}
\label{sol:mumu numu}
&& \braket{\mu(z_1,\bz_1) \mu(z_2,\bz_2)}_N 
\nn
&& = (z_1 - \bz_2)^{\frac14} (z_2 - \bz_1)^{\frac14}\;
 \xi^{\frac14} (1-\xi)^\frac14
 \left\{ \alpha^\mu\, 
 {}_2 F_1\left(\frac12, \frac12; 1; \xi\right)
       + \wt \alpha^\mu\, 
 {}_2 F_1\left(\frac12, \frac12; 1; 1-\xi\right) \right\}\,,\,\,
\nn
&& \braket{\nu(z_1,\bz_1) \mu(z_2,\bz_2)}_N 
\nn
&& = (z_1 - \bz_1)^{-\frac34} (z_2 - \bz_2)^{\frac14}\;
 \xi^{-\frac14} 
 \left\{ \alpha^\nu\, 
 {}_2 F_1\left(\frac12, \frac32; 2; \xi\right)
       + \wt \alpha^\nu\, 
 {}_2 F_1\left(-\frac12, \frac12; 1; 1-\xi\right) \right\}\,,\,\,
\end{eqnarray}
where $\nu$ denotes $\Phi_{2,2}$ of $h=3/8$, $\alpha^{\mu,\nu}$ and
$\wt \alpha^{\mu,\nu}$ are arbitrary constants.
Substituting $z_i=x_i+i\,y_i$ and $x=x_2-x_1$ into (\ref{sol:mumu numu}), one
further finds that both of them exhibit the
logarithmic singularity in the vicinity of $\xi=1$ ($i.e.$ $y_i \ll |x|$):
\begin{eqnarray}
\label{munu asym}
\braket{\mu(z_1,\bz_1) \mu(z_2,\bz_2)}_N
 &\sim& \left(y_1 y_2\right)^\frac14 \ln\left(\frac{y_1 y_2}{x^2}\right), 
\nn
\braket{\nu(z_1,\bz_1) \mu(z_2,\bz_2)}_N
 &\sim& \left(y_1\right)^{-\frac34} \left(y_2\right)^\frac14 \ln\left(\frac{y_1 y_2}{x^2}\right) .
\end{eqnarray}
Comparing (\ref{sol:mumu numu}) with the general solution (10) of \cite{KW}:
\begin{eqnarray}
&& \braket{\mu(z_1,\bz_1) \mu(z_2,\bz_2)}_{boundary} 
\nn&&
 = (z_1 - \bz_2)^{\frac14} (\bz_1 - z_2)^{\frac14} \xi^{\frac14} (1-\xi)^{\frac14}
  \left\{ A\; {}_2 F_1\left(\frac12,\frac12;1;\xi\right) + B \; {}_2 F_1\left(\frac12,\frac12;1;1-\xi\right) \right\},
\nn
\end{eqnarray}
one immediately reads:
\begin{eqnarray}
 A = \alpha^\mu\,i^{\frac12} {\rm ~~~and~~~} B=\wt \alpha^\mu\,i^{\frac12}.
\end{eqnarray}

From the work by Kogan and Wheater \cite{KW}, these constants are known to
be related to two-point functions of {\it logarithmic} boundary operators.
Namely, 
in the case of logarithmic theories, 
the bulk-boundary OPE relation of dimension $h$ field is to be of the
form \cite{KW,Cardy:tv}:
\begin{eqnarray}
 \Phi_h (x,y)
 = (2y)^{\Delta_d -2h} C_{\Phi\Phi}^{d} \left( d(x)+c(x) \ln(2y) \right)
 + \sum_{i} (2y)^{\Delta_i -2h} C_{\Phi\Phi}^{i} \Psi_{i}(x), 
\end{eqnarray}
where $\Psi_i(x)$'s are {\it ordinary} boundary operators of boundary scaling
dimensions $\Delta_i$, normalised as $\braket{\Psi(0)\Psi(x)}=x^{-2\Delta_i}$.
$c(x)$ and $d(x)$ are {\it logarithmic} boundary operators of dimension
$\Delta_d$, whose two-point functions are given by \cite{KW}:
\begin{eqnarray}
\label{2pt bop}
 \braket{d(0)d(x)} &=& x^{-2\Delta_d} \left(-2 \alpha_d \ln(x)+\alpha_d^\prime\right) , 
\nn
 \braket{c(0)d(x)} &=& \braket{d(0)c(x)} = x^{-2\Delta_d} \alpha_d, 
\nn
 \braket{c(0)c(x)} &=& 0 .
\end{eqnarray}
Substituting these, $\Delta_d=0$, $z_i=x_i+i\,y_i$ with $x_1=0$, and $x=x_2$ into the forms
$\braket{\mu(z_1,\bz_1)\mu(z_2,\bz_2)}$ and $\braket{\nu(z_1,\bz_1)\mu(z_2,\bz_2)}$, then taking the limit $y_i\ll x$,
one finds the relations between $\alpha^\mu_N$, $\alpha^\nu_N$, and $\alpha_d$ from the
asymptotic behaviours (\ref{munu asym}):
\begin{eqnarray}
\label{rel:constants}
 \alpha^\mu \sim \alpha_d \left( C_{\mu\mu}^d \right)^2 \,,\;\; 
 \alpha^\nu \sim \alpha_d \left( C_{\nu\nu}^d C_{\mu\mu}^d \right) \,,\;\; 
\end{eqnarray}
up to phase.
This result holds even if there is another logarithmic pair of $\Delta_d=1$.

Hence, with the Neumann boundary condition, the decoupling of $c(x)$,
that is $\alpha_d=0$, may occur only when 
$\alpha^\mu=0$.
Note that, in such a case, the $\log$ term in eq. (\ref{2pt bop}) and
$\braket{\mu(z_1,\bz_1)\mu(z_2,\bz_2)}$ vanish simultaneously.

Lastly, it should be noted that, if $\wt \alpha^{\mu}=0$ $i.e.$ $B=0$, the
asymptotic behaviours of the two-point functions 
in the vicinity of $\xi=0$ match with the one-point function
$\braket{C(z,\bz)}=\braket{C(z)C(\bz)}=0$. 
We do not have this condition so that the theory may break the scaling
covariance as was discussed in \cite{KW}.

\section{Free Boson Realisation of the $SU(2)_k$ WZNW model with Boundary}
\label{sec:su2}

In the late 1980s, it was shown that the $SU(2)_k$
Wess-Zumino-Novikov-Witten model (WZNW model) can be
described by free field realisations. In particular, Gerasimov et al.
showed that the $SL(2)_k$ WZNW model can be realised by three free bosons \cite{Gerasimov:fi}.
We begin with the quantum equivalent 
action functional of three scalar boson fields $U$, $V$, and $\Phi$ with
a boundary:
\begin{eqnarray}
 S_{SU(2)} [U,V,\Phi] = 
 \frac{1}{8\pi} \sum_{\Phi_i=\{U,V,\Phi\}} \int_{\Sigma} d^2 z \sqrt{g}
 \left[ g^{\mu\nu} \pa_\mu \Phi_i \pa_\nu \Phi_i + 2 i \alpha_{\Phi_i} R^{(2)} \Phi_i \right] ,
\end{eqnarray}
where $\Sigma$ is set to be the upper half-plane with
$g_{\mu\nu}=\left(\begin{array}{cc}
 0& \frac12\\
 \frac12& 0
            \end{array}\right)$, and 
the background charges are 
$\alpha_{U}= \frac{i}{2},\, \alpha_{V}= \frac12,\, \alpha_{\Phi}= - \frac{1}{\sqrt2 \hat{q}}$. 
$\hat{q}$ gives the level $k$ of the affine Kac-Moody algebra (AKM algebra) as
$\hat{q}^2=k+2$ while the central charge is 
$c= 
\frac{3k}{k+2}$. For the theory to be analogous to the WZNW model, we
assume $k>-2$.
The arguments $z$ and $\bz$ are suppressed in the above.
The first variations of the action with respect to the bosons imply the
holomorphic and anti-holomorphic decompositions, $U=u+\bu$, $V=v+\bar v$,
$\Phi=\phi+\bphi$, and one can choose two different boundary conditions
for each boson. Taking only the Neumann boundary conditions into
consideration, the propagators of holomorphic fields are given by:
\begin{eqnarray}
\label{def:prop su2}
 u(z)u(w)\sim -\ln(z-w)\,,\,\, 
 v(z)v(w)\sim -\ln(z-w)\,,\,\, 
\nn
 \phi(z)\phi(w)\sim -\ln(z-w)\,.
\end{eqnarray}
Propagators of anti-holomorphic fields are given similarly, and so are
those of holomorphic and anti-holomorphic fields, while the ones
with different species such as $\braket{uv}$ vanish. 
If we introduce a notation $\phi_i = \left\{ u,v,\phi \right\}$,
eq. (\ref{def:prop su2}) takes a rather simpler form:
\begin{eqnarray}
\label{def:propagator su2k}
 \phi_i(z) \phi_j(w) \sim - \delta_{ij} \ln(z-w) . 
\end{eqnarray}
Note that all the boson propagators can be represented
by eq. (\ref{def:propagator su2k}) through Cardy's method of images.

The energy-momentum tensor can now be written in terms of $\{u,v,\phi\}$
by:
\begin{eqnarray}
   T = \sum_{\phi_i=\{u,v,\phi\}} \left( -\frac12 (\pa\phi_i)^2 + i
\alpha_{\phi_i} \pa^2 \phi_i \right) ,
\end{eqnarray}
where $\alpha_{\Phi_i}$ are relabeled by $\alpha_{\phi_i}$ for convenience.
The $SU(2)_k$ AKM currents can locally be expressed by:
\begin{eqnarray}
 J^+ &=& \frac{1}{\sqrt2}\, \pa v\, e^{-u+iv} \,,
\nn
 J^0 
  &=& \frac{i \hat{q}}{\sqrt2}\, \pa\phi + \pa u \,,
\nn
 J^- &=& \frac{1}{\sqrt2} \left[ \sqrt2 \hat{q}\, \pa\phi - i \hat{q}^2 \pa u + (1-\hat{q}^2) \pa v \right] e^{u-iv} \,.
\end{eqnarray}
They satisfy the following OPEs:
\begin{eqnarray}
  T(z) T(w) &\sim& \frac{c/2}{(z-w)^4} + \frac{2\,T(w)}{(z-w)^2} + \frac{\pa_w T(w)}{z-w} \,,
\nn
  T(z) J^a(w) &\sim& \frac{J^a(w)}{(z-w)^2} + \frac{\pa_w J^a}{z-w} {~~~for~~a=0,\pm}\,,
\nn
  J^0(z) J^\pm(w) &\sim& \frac{\pm J^\pm(w)}{z-w} \,,\;\; 
  J^\pm (z) J^\pm(w) \;\sim\; 0,
\nn
  J^0(z) J^0(w) &\sim& \frac{k/2}{(z-w)^2} \,,\;\;
  J^+(z) J^-(w) \;\sim\; \frac{k/2}{(z-w)^2} + \frac{J^0(w)}{z-w} \,.
\end{eqnarray}
We omit the terms regular as $z\to 0$, so $J^\pm(z)J^\pm(w)$ is merely
equivalent to zero up to those regular terms.
The anti-holomorphic currents such as $\ol T$ are given similarly and
are mapped to the lower half-plane, $\ol T(\bz)= T(z^*)$, by the method of images.

At this stage, one can follow the procedures of the $(p,q)$ model.
The non-chiral primaries of the $(j,m)$ highest weight representation
(hwrep) of the $SU(2)_k$ AKM algebra ($(j,\bar m)$ for the $\ol{SU(2)_k}$) are given by the chiral primaries
$\Phi_{j,m}(z)$: 
\begin{eqnarray}
\label{def:nonchiral su2}
 \Phi_{j;m,\bar m}(z,\bz) = \Phi_{j,m}(z)\,\Phi_{j,\bar m}(z^*) , 
\end{eqnarray}
whose conformal dimensions are degenerate at $h_j=\bar h_j=\frac{j(j+1)}{k+2}$.
By this mapping, their boundary two-point functions are given by chiral
four-point functions as was shown in the $c_{p,q}$ model:
\begin{eqnarray}
  \braket{\Phi_{j_1;m_1,\bar m_1}(z_1,\bz_1)\Phi_{j_2;m_2,\bar m_2}(z_2,\bz_2)}_N 
 = \braket{\Phi_{j_1,m_1}(z_1)\Phi_{j_2,m_2}(z_2)\Phi_{j_2,\bar m_2}(z_2^*)\Phi_{j_1,\bar m_1}(z_1^*)}_0 . 
\end{eqnarray} 
On the other hand, one can construct the above multiplet via the vertex
operator realisation: 
\bea
  V_{j,j-m} \equiv e^{i\alpha_j \phi} e^{(j-m) (u-iv)} ,\;
  \alpha_j = -2\alpha_0 j, 
\eea
with the conformal dimension $h_j = \frac12 \alpha_j (\alpha_j -
2\alpha_0) = \frac{j(j+1)}{k+2}$. 
Amongst all vertex operators, only the above set obey
the following OPEs of the $(j,m)$ hwrep:
\begin{eqnarray}
 J^0(z) V_{j,j-m}(w) &\sim& \frac{m\, V_{j,j-m}(w)}{z-w} \,,
\nn
 J^\pm(z) V_{j,j-m}(w) &\sim& \frac{\frac{i}{\sqrt2}(\pm j-m)\,V_{j,j-(m\pm 1)}(w)}{z-w} . 
\end{eqnarray}
Their conjugate states are defined by 
\begin{eqnarray}
 \label{def:su2dual}
  \wt V_{j,j-m} = V_{-1-j,-1-j-m} ,
\end{eqnarray}
so that the $SU(2)$-invariant function $\braket{V_{j,m} V_{j,-m}}_0$ 
should only take a nontrivial value among such chiral two-point
functions of $\{\Phi_{j,m}\}$.
Consequently, the $(j,m)$ chiral primary is realised in two ways by:
\begin{eqnarray}
 \Phi_{j,m} \sim V_{j,j-m} {\rm ~or~} \wt V_{j,j-m} .
\end{eqnarray} 
A screening charge and a screening current of dimension $(h,\bar h)=(1,0)$ is
given by\footnote{
There is a degree of freedom of having $\pa u$ in addition to $\pa v$, but
it is only equivalent up to constant factor in correlators.
}
\begin{eqnarray}
   Q_+ = \oint_{C\subset\Sigma} d\zeta\, J_+(\zeta),\;\; J_+ \equiv e^{2 i \alpha_\phi \phi} e^{-u +iv} \frac{\pa v}{i} =
 \frac{\pa v}{i} V_{-1,-1} ,
\end{eqnarray}
where $C$ is on $\Sigma$. The $(0,1)$ screening charge $\ol Q_+$ is given
similarly by the replacement of $C$ with $\ol C$ on $\Sigma^*$.
The analytically continued version of the screening charge is 
given with the closed contour $\wt C$ defined anywhere on $\C$:
\begin{eqnarray}
   \wt Q_+ = \oint_{\wt C\subset \C} d\zeta\, \wt J_+(\zeta) .
\end{eqnarray}
Note that the above screening charges are invariant under the $SU(2)_k$ transformations.

Via this realisation, one can express $N$-point functions of the
non-chiral fields,
and therefore chiral $2N$-point functions, 
by $\braket{ V_{j_1,m_1} \cdots V_{j_{2N},m_{2N}} Q_+^n}_0$ 
with the insertion of the background charges at infinity.
The charge neutrality (asymmetry) conditions read 
\bea
  \sum_i^{2N} \alpha_{j_i} + 2 n \alpha_\phi = 2 \alpha_{\phi},\, 
  \sum_i^{2N} m_i -n = -1 .
\eea
The first condition of the above can be rewritten by $\sum_i j_i -n = -1$.
Accordingly, the two-point function of $\tilde V_{j,j+m}$ and $V_{j,j-m}$
provides a non-vanishing two-point function, satisfying the charge conditions: 
$\braket{\tilde V_{j,j+m}(z) V_{j,j-m}(0)}_0 = z^{-2 h_j} $. This leads
to the definition of the conjugate, or dual, operators (\ref{def:su2dual}), 
and one-point functions: 
$\braket{\Phi_{j;m,m^\prime}(z,\bz)}_N \sim \frac{\delta_{m+m^\prime,\,0}}{(z-\bz)^{2h_j}}$.

\subsection{A Boundary Two-Point Function of the Doublet representations}
\label{sec:2pt su2}

Let us consider the doublet representations
$\Phi_{\pm}\equiv\Phi_{\frac12,\pm\frac12}$ of conformal dimension
$h_{1/2}$ $(\bar h_{1/2})$ $=\frac{3}{4(k+2)}$. The corresponding
chiral vertex operators are
\bea
  V_+ = V_{1/2,0},\; V_- = V_{1/2,1} , \nn
  \tilde V_+ = V_{-3/2, -2} ,\; \tilde V_- = V_{-3/2, -1} .
\eea
Accordingly, from eq. (\ref{def:nonchiral su2}), there are four distinct non-chiral fields, $\{\Phi_{++},\Phi_{+-},\Phi_{-+},\Phi_{--}\}$.
The correlation function, analogous to $\braket{g(z,\bz)g^\dag(w,\bw)}$
of the $SU(2)_k$ WZNW model \cite{WZNW}, is 
$\braket{\Phi_{m,\bar m} \Phi_{\bar m^\prime, m^\prime}}$ 
with the indices $\bar m$ ($\bar m^\prime$) for the conjugate doublet representation. 
Contractions between fundamental representations of $SU(2)_k$ and
their conjugates result only in terms proportional to 
$\delta_{m,\bar m}\delta_{m^\prime, \bar m^\prime}$
and
$\delta_{m,\bar m^\prime}\delta_{m^\prime, \bar m}$. Reflecting the
$SU(2)_k$ invariance of the functions, the relevant correlation functions are:
\begin{eqnarray}
 \left\{
 \braket{\Phi_{- -} \Phi_{+ +}} , 
 \braket{\Phi_{+ -} \Phi_{+ -}} , 
 \braket{\Phi_{+ +} \Phi_{- -}} , 
 \braket{\Phi_{- +} \Phi_{- +}} 
 \right\} . 
\end{eqnarray}
The latter two entries are the charge conjugates of the former two.

Among them, we begin with the 
boundary two-point function $\braket{\Phi_{--}\Phi_{++}}$:
\begin{eqnarray}
\label{def:2pt--++}
 \braket{\Phi_{--}(z_1, \bz_1)\Phi_{++}(z_2, \bz_2)}_N &=& \braket{\Phi_-(z_1)\Phi_+(z_2)\Phi_+(z_2^*)\Phi_-(z_1^*)}_0
\nn
 &=& \alpha_{1}^{-++-} F_1^{-++-}(z_1,z_2) 
   + \alpha_{2}^{-++-} F_2^{-++-}(z_1,z_2),
\end{eqnarray}
where $\alpha_{1}^{-++-}$ and $\alpha_{2}^{-++-}$ are constants, and the functions
$F_i^{-++-}$ are
\begin{eqnarray}
 F_1^{-++-}(z_1,z_2) 
 &\equiv& \braket{V_-(z_1) V_+(z_2) V_+(z_2^*) \wt V_-(z_1^*)\,Q_+} \,,
\nn
 F_2^{-++-}(z_1,z_2) 
 &\equiv& \braket{V_-(z_1) V_+(z_2) V_+(z_2^*) \wt V_-(z_1^*)\,\ol Q_+} \,.
\end{eqnarray}
These turn out to be contour integrals on the branch covering of $\C$:
\begin{eqnarray}
 F_1^{-++-}(z_1,z_2) = \prod_{i<j} z_{ij}^{\alpha_i\,\alpha_j} z_{14}
  \oint_{C\subset\Sigma} d\zeta\, \frac{
  \prod_{i=1}^3 \left( z_i-\zeta\right)^{-2\alpha_\phi^2}
  \left( z_4-\zeta\right)^{3\cdot 2\alpha_\phi^2}}
  {\left( z_1-\zeta\right)\left( z_4-\zeta\right) } ,
\end{eqnarray}
where $\alpha_1=\alpha_2=\alpha_3=-\alpha_\phi$, $\alpha_4=-3\alpha_\phi$, 
and $\alpha_\phi^2= 1/2\hat{q}^2= 1/2(k+2)$. $F_2^{-++-}$ is given similarly.
Quite conveniently, the integral in the above is equivalent to that in 
eq. (\ref{eq:rs-12 vertex}) with $(p,q,r,s)=(k+2,1,0,2)$. Then, one may
follow the same instructions: reduce the contour integral to the line
integral by absorbing its difference in $\alpha_i^{-++-}$'s, and conformally map
the four points to the standard ones.
For the reduction from the closed contours to the line integrals,
a condition $1/(k+2)\not\in\Z$ is necessary and sufficient. One may 
assume a sufficient condition $k>-1$, instead.

By performing the conformal map of $\zeta\to w= \frac{z_{34}}{z_{31}}\frac{\zeta-z_1}{\zeta-z_4}$, 
the function $F_1^{-++-}$ with the line integration is reduced to:
\begin{eqnarray}
 F_1^{-++-} = e^{2\alpha_\phi^2 \pi i} (z_{13}z_{24})^{-2h_{1/2}} \xi^{\alpha_\phi^2} (1-\xi)^{\alpha_\phi^2}
  \int_0^{\xi} dw\, w^{-2\alpha_\phi^2-1} (\xi-w)^{-2\alpha_\phi^2} 
  (1-w)^{-2\alpha_\phi^2} .
\end{eqnarray}
Together with an integral in $F_2^{-++-}$, one again finds that those
integrals are the solutions of the hypergeometric differential equation (\ref{gauss eq}) with $(a,b,c)=(2 \alpha_\phi^2, 6 \alpha_\phi^2, 1+4 \alpha_\phi^2)$.
\begin{eqnarray}
\label{sol:su2-1}
 F_1^{-++-} &=& e^{2\alpha_\phi^2 \pi i}
  (z_{13}z_{24})^{-2h_{1/2}} \xi^{\alpha_\phi^2} (1-\xi)^{\alpha_\phi^2} 
\nn&&\quad\times
  \frac{\Gamma(1+a-c)\Gamma(1-a)}{\Gamma(2-c)} 
  \,\xi^{1-c}\, {}_2 F_1 \left( 1+a-c,1+b-c;2-c;\xi \right) 
\nn
  &\sim& \xi^{\frac{1}{2(k+2)}} (1-\xi)^{\frac{1}{2(k+2)}} \frac{\Gamma(-\frac{1}{k+2})\Gamma(\frac{k+1}{k+2})}{\Gamma(\frac{k}{k+2})} \,\xi^{-\frac{2}{k+2}}\, {}_2 F_1 \left(-\frac{1}{k+2},\frac{1}{k+2};\frac{k}{k+2};\xi \right) \,,\,\,
\nn
 F_2^{-++-} &=& e^{6\alpha_\phi^2 \pi i}
  (z_{13}z_{24})^{-2h_{1/2}} \xi^{\alpha_\phi^2} (1-\xi)^{\alpha_\phi^2} \cdot
  \frac{\Gamma(b)\Gamma(c-b)}{\Gamma(c)} {}_2 F_1 \left( a, b; c; \xi\right)
\nn
  &\sim& \xi^{\frac{1}{2(k+2)}} (1-\xi)^{\frac{1}{2(k+2)}} \frac{\Gamma(\frac{3}{k+2})\Gamma(\frac{k+1}{k+2})}{\Gamma(\frac{k+4}{k+2})}\, {}_2 F_1 \left( \frac{1}{k+2}, \frac{3}{k+2}; \frac{k+4}{k+2}; \xi\right) \,.
\end{eqnarray}
Likewise, the two-point function $\braket{\Phi_{+-}\Phi_{+-}}_N$
can easily be found, with hypergeometric functions
obeying the hypergeometric differential equation of 
$(a,b,c)=(2\alpha_\phi^2, 6\alpha_\phi^2, 4\alpha_\phi^2)$.
For later convenience, we also list the function 
$\braket{\Phi_{+-}\Phi_{-+}}_N$ which in turn corresponds to the hypergeometric differential equation of 
$(a,b,c)=(1+2\alpha_\phi^2, 6\alpha_\phi^2, 1+4\alpha_\phi^2)$.
\begin{eqnarray}
\label{sol:su2-2}
  \braket{\Phi_{\epsilon_1 \epsilon_4}\Phi_{\epsilon_2 \epsilon_3}}_N
  &=& \sum_{i=1}^2 \alpha_{i}^{\epsilon_1 \epsilon_2 \epsilon_3 \epsilon_4}
      F_i^{\epsilon_1 \epsilon_2 \epsilon_3 \epsilon_4} (z_1,z_2) , 
\nn
  F_1^{++--} &\sim& \frac1k\, \xi^{\frac{1}{2(k+2)}} (1-\xi)^{\frac{1}{2(k+2)}} \frac{\Gamma(-\frac{1}{k+2})\Gamma(\frac{k+1}{k+2})}{\Gamma(\frac{k}{k+2})} \,\xi^{\frac{k}{k+2}}\, {}_2 F_1 \left(\frac{k+1}{k+2},\frac{k+3}{k+2};\frac{2k+2}{k+2};\xi \right) \,,\,\,
\nn
  F_2^{++--} &\sim& -2\, \xi^{\frac{1}{2(k+2)}} (1-\xi)^{\frac{1}{2(k+2)}} \frac{\Gamma(\frac{3}{k+2})\Gamma(\frac{k+1}{k+2})}{\Gamma(\frac{k+4}{k+2})}\, {}_2 F_1 \left( \frac{1}{k+2}, \frac{3}{k+2}; \frac{2}{k+2}; \xi\right) \,,
\nn
  F_1^{+-+-} &\sim& \xi^{\frac{1}{2(k+2)}} (1-\xi)^{\frac{1}{2(k+2)}} \frac{\Gamma(-\frac{1}{k+2})\Gamma(\frac{k+1}{k+2})}{\Gamma(\frac{k}{k+2})} \,\xi^{-\frac{2}{k+2}}\, {}_2 F_1 \left(\frac{k+1}{k+2},\frac{1}{k+2};\frac{k}{k+2};\xi \right) \,,\,\,
\nn
  F_2^{+-+-} &\sim& \xi^{\frac{1}{2(k+2)}} (1-\xi)^{\frac{1}{2(k+2)}} \frac{\Gamma(\frac{3}{k+2})\Gamma(\frac{k+1}{k+2})}{\Gamma(\frac{k+4}{k+2})}\, {}_2 F_1 \left( \frac{k+3}{k+2}, \frac{3}{k+2}; \frac{k+4}{k+2}; \xi\right) \,,
\end{eqnarray}
up to common phases for $F_1$'s and $F_2$'s respectively.
The solutions (\ref{sol:su2-1}) are the solutions shown in (2.3.14) of
\cite{Gerasimov:fi}. They and the first two $F_i$'s 
in (\ref{sol:su2-2}) are
equivalent to the solutions of the Knizhnik-Zamolodchikov equation (KZ-equation),
$\F_1^{(0)}(x)$, $\F_1^{(1)}(x)$, $\F_2^{(0)}(x)$, and $\F_2^{(1)}(x)$
in (4.10) of \cite{Knizhnik:nr} up to constant.
The general solutions are given by linear combinations of the
two corresponding hypergeometric functions for general $k>-1$.

Since those coefficients of the hypergeometric differential equation only differ by integers,
a contiguous relation of hypergeometric functions (1.4.5) in
SL\cite{slater} can be applied to $F_1^{+-+-}$:
\begin{eqnarray}
\label{id:Fiab}
  F(a+1) &=& \frac{c-1}{a} F(c-1) - \frac{c-a-1}{a} F
\nn
  {}_2 F_1(1+A,3A;1+2A) &=& 2 {}_2 F_1(A,3A;2A) - {}_2 F_1(A,3A;1+2A) 
\nn
  F_1^{+-+-} &=&  - F_1^{-++-} - F_1^{++--} ,
\end{eqnarray}
where $A=2\alpha_\phi^2=1/(k+2)$.
The same relation holds for $F_2$'s, and thus
$\braket{\Phi_{+-}\Phi_{-+}}_N$ is turned out to be linearly dependent
of $\braket{\Phi_{--}\Phi_{++}}_N$ and $\braket{\Phi_{+-}\Phi_{+-}}_N$
[see Appendix B]. 

The calculations for the charge conjugate configurations are not as
simple as the former cases, because corresponding chiral correlation
functions with vertex operators give sums of hypergeometric functions.
For example, for $i=1,2$:
\begin{eqnarray}
 \braket{V_+ V_- V_- \wt V_+ Q_+} &=& \braket{V_+ V_- V_+ \wt V_- Q_+} + \braket{V_+ V_+ V_- \wt V_- Q_+} 
\nn
 F_i^{+--+}(z_1,z_2) &=& F_i^{+-+-}(z_1,z_2) + F_i^{++--}(z_1,z_2) \,.
\end{eqnarray}
However, from the relation (\ref{id:Fiab}), they can be simplified as:
\begin{eqnarray}
\label{id:Fiab2}
 F_i^{+--+}(z_1,z_2) &=& 
 - F_i^{-++-}(z_1,z_2) ,
\nn
 F_i^{--++}(z_1,z_2) &=& 
 - F_i^{++--}(z_1,z_2) .
\end{eqnarray}
Therefore, by flipping the signs of $\alpha_{i}$'s, one ends up with: 
\begin{eqnarray}
 \braket{\Phi_{++}\Phi_{--}}_N \,=\, \braket{\Phi_{--}\Phi_{++}}_N \,,\;\;
 \braket{\Phi_{-+}\Phi_{-+}}_N \,=\, \braket{\Phi_{+-}\Phi_{+-}}_N \,.
\end{eqnarray}
Their explicit forms were shown in (\ref{sol:su2-1}, \ref{sol:su2-2}).

We are now ready to discuss the logarithms in the boundary two-point correlation functions. 
The condition for
logarithms is $4 \alpha_\phi^2\in\Z ~~i.e.$
\begin{eqnarray}
\label{cond:logsu2}
 \frac{2}{k+2}\in\Z. 
\end{eqnarray}
Therefore, some fractional levels $k<-1$ may satisfy the condition such as
$k=-4/3$ \cite{Gaberdiel:2001ny}. 
When $-1<k<\infty$, $k=0$ is the only choice, so let us consider the $SU(2)_{k=0}$ case in what follows. 
As was shown in the preceding section, the two independent
solutions coincide with each other when $c$ becomes integer while $a,b$
are not negative integers. 
In this logarithmic case, the general solution may be given by the
solutions regular at $\xi=1$, which in turn can be given by the contour
integrations with the analytically continued screening charges as in
Fig.\ref{fig:c23}. 
For general $k$:
\begin{eqnarray}
\label{sol:su2log2}
 \wt F_1^{-++-} &\sim& 
 \frac{\Gamma^2\left( 1-A \right)}{\Gamma\left(2-2A\right)} \,
 (1-\xi)^{1-2A} 
 {}_2 F_1 \left( 1+A, 1-A; 2-2A; 1-\xi \right) \,,
\nn
 \wt F_2^{-++-} &\sim& 
 \frac{\Gamma^2\left( A \right)}{\Gamma\left(2A\right)} \,
 {}_2 F_1 \left( 3A, A; 2A; 1-\xi \right) \,,
\nn
 \wt F_1^{++--} &\sim& 
 \frac{\Gamma\left(-A\right)\Gamma\left( 1-A \right)}{\Gamma\left(1-2A\right)} \,
 (1-\xi)^{-2A} 
 {}_2 F_1 \left( A, -A; 1-2A; 1-\xi \right) \,,
\nn
 \wt F_2^{++--} &\sim& 
 \frac{\Gamma\left( A \right)\Gamma\left( 1+A \right)}{\Gamma\left(1+2A\right)} \,
 {}_2 F_1 \left( 3A, A; 1+2A; 1-\xi \right) .
\end{eqnarray}
$i=1$ in $F_i$'s corresponds to the interval $(z_2,z_3)$ while $i=2$ is for $(z_1,z_4)$.
Their phases and the common factor of
$(z_{13}z_{24})^{-2h_{1/2}} \xi^{\alpha_\phi^2} (1-\xi)^{\alpha_\phi^2}$
appearing in $F_i^{-++-}$ are omitted.
The logarithmic condition for the above two pairs is in common and is equivalent to the condition (\ref{cond:logsu2}):
\begin{eqnarray}
 2 A = \frac{2}{k+2} \in \Z . 
\end{eqnarray}
Therefore, the logarithmic cases for $SU(2)_k$ doublets 
necessarily contain a logarithmic solution in each limit, $\xi\to0$ or
$\xi\to 1$.

When $k=0$ or $k\to 0$, one obtains identical solutions from
(\ref{sol:su2log2}) and the general solutions are then given by: 
\begin{eqnarray}
 \braket{\Phi_{--}\Phi_{++}}_N
 &=& \alpha^{-++-} F_2^{-++-} (z_1,z_2) 
 + \wt \alpha^{-++-} \wt F_2^{-++-} (z_1,z_2)
\nn
 &=& \frac{\pi}{2} \,
  |z_1-\bz_2|^{-\frac32} \,\xi^{\frac14} (1-\xi)^{\frac14}\,
\nn&&\times
  \left\{
  \alpha^{-++-}
  {}_2 F_1 \left( \frac12, \frac32; 2; \xi \right) 
  + \wt \alpha^{-++-} \cdot 2\,
  {}_2 F_1 \left( \frac12, \frac32; 1; 1-\xi \right)
  \right\} \,,
\nn
 \braket{\Phi_{+-}\Phi_{+-}}_N
 &=& \alpha^{++--} F_2^{++--} (z_1,z_2)
 + \wt \alpha^{++--} \wt F_2^{++--} (z_1,z_2)
\nn
 &=& \frac{\pi}{2} \,
  |z_1-\bz_2|^{-\frac32} \,\xi^{\frac14} (1-\xi)^{\frac14}\,
\nn&&\times
  \left\{
  \alpha^{++--} \cdot 2\,
  {}_2 F_1 \left( \frac12, \frac32; 1; \xi \right)
  + \wt \alpha^{++--}
  {}_2 F_1 \left( \frac12, \frac32; 2; 1-\xi \right)
  \right\} \,.
\end{eqnarray}
Both solutions are symmetric to each other under the change of
$\xi\leftrightarrow 1-\xi$. They are logarithmic in the vicinity of $\xi=0,1$, that is, at
both the far-from-boundary limit and the near-boundary limit. 
The above expressions agree with the general solution of the chiral
four-point function in \cite{Caux:1996kq} and
the boundary two-point functions in \cite{KW} up to constant factors. 

Substituting $z_i=x_i+i y_i$
and $x=x_1-x_2$, the asymptotic behaviours at the near-boundary limit ($\xi=1$) are:
\begin{eqnarray}
\label{asym su2}
 \braket{\Phi_{--}\Phi_{++}}_N
 &\sim& \alpha^{-++-} (2 y_1)^{-\frac34} (2 y_2)^{-\frac34} \,
  \left\{- \left( \frac{4y_1 y_2}{x^2} \right)
  \ln \left(4\,\frac{y_1 y_2}{x^2}\right)\right\} \,,
\nn
 \braket{\Phi_{+-}\Phi_{+-}}_N
 &\sim& \alpha^{++--} (2 y_1)^{-\frac34} (2 y_2)^{-\frac34} \,
  \left\{2
  - \frac12 \left(\frac{4\,y_1 y_2}{x^2}\right) \ln \left(4\,\frac{y_1 y_2}{x^2}\right)\right\} \,.
\end{eqnarray}

The bulk-boundary OPE relation is to be of the form \cite{Caux:1996kq,KW}:
\begin{eqnarray}
\label{eq:b-b OPE su2}
  \Phi_{\epsilon_1 \epsilon_2}(x,y)
  \=  (2y)^{-\frac34} \left\{ 
   C_{\epsilon_1 \epsilon_2}^I I_{\epsilon_1 \epsilon_2^\vee} + 
   2y \,C_{\epsilon_1 \epsilon_2}^d \, t_{\epsilon_1 \epsilon_2^\vee}^{i}
   \left( D^i(x) + C^i(x) \ln(2y) \right) + \cdots \right\} ,
\nn
\end{eqnarray}
where $I$ is the unit matrix, $t^i_{\epsilon_1 \epsilon_2^\vee}$ are the
Pauli matrices divided by two, and $\epsilon^\vee$ is the weight
conjugate to $\epsilon$. 
The two-point functions of the {\it logarithmic} boundary operators
$C^i$ and $D^i$ were given by \cite{Caux:1996kq}:
\begin{eqnarray}
 \braket{D^i(0) D^j(x)} &=& - \left( \beta + 2 \alpha_d^{su(2)} \ln(x) \right) \frac{\delta^{ij}}{x^2} \,,
\nn
 \braket{C^i(0) D^j(x)} &=& \frac{\alpha_d^{su(2)} \delta^{ij}}{x^2}\,,
\nn
 \braket{C^i(0) C^j(x)} &=& 0 .
\end{eqnarray}
We added the normalisation factor $\alpha_d^{su(2)}$ to the original expressions.
Substituting the expression (\ref{eq:b-b OPE su2}) and the above into
the forms of the boundary two-point functions, one finds the following
relations from the asymptotic forms (\ref{asym su2}) up to phase:
\begin{eqnarray}
 \alpha_N^{-++-} &=& 
- 
\frac12 \alpha_d^{su(2)} (C_{--}^d) (C_{++}^d) \,,
\nn
 \alpha_N^{++--} &=&  
\frac12 \left(C_{+-}^I \right)^2 
 \;=\; - \frac12 \alpha_d^{su(2)} \left(C_{+-}^d\right)^2 \,.
\end{eqnarray}
An interesting point is in the second relation that the normalisation
constant is related and proportional to the squared fusion coefficient
$\left(C_{+-}^I\right)^2$. This means that given fusion rules one can
fix $\alpha_d^{su(2)}$, and moreover, $\alpha^{-++-}$ and $\alpha^{++--}$. 
This feature isn't present in the
$\braket{\Phi_{r,s}\Phi_{1,2}}$ of the $(p,q)$ models.
Note that one can also assign different $\alpha_d^i$ instead of $\alpha_d^{su(2)}$ for different suffices
of the logarithmic pairs. In such a case, $\alpha_d^1=\alpha_d^2$ and
the different $\alpha_d^1$ and $\alpha_d^3$ only show up in the first
and second lines separately.

\chapter{Boundary States of Boundary LCFT}
\label{ch:boundary states}

  Since Kogan and Wheater commenced research on the boundary states of
  LCFTs , at $c=-2$ in particular \cite{KW}, 
  several papers have been published on the same topic.
  In 2001, the author analysed the boundary Ishibashi states
  of the Jordan cell from a mathematical physics point of view, 
  and sketched its oddity and the physical boundary
  states at $c=-2$ \cite{ishimoto1}.
  It was also shown that a conventional form of the Ishibashi states gives
  only one boundary state in each Jordan cell, which is found to be null in the character
  functional form.
  Soon after that work, the $c=-2$ symplectic fermion system 
  with boundaries were constructed by Kawai and Wheater, and the boundary
  coherent states were shown to be basis of the physical boundary states: 
  five out of six states enter into the physical boundary states \cite{kaw}.
  On the other hand, a more general survey on Cardy's equation
  (\ref{def:bc0}) was
  carried out in the case of LCFT by Bredthauer and Flohr, 
  being based, as \cite{ishimoto1}, on generalised Ishibashi states.
  Investigating them in the
  $c=-2$ triplet model, ten basis states were proposed \cite{bredthauer}.

  Since those works, a lasting question was why three cases in two
  different constructions show different numbers of basis states and
  hence different results, 
  whereas the symplectic fermion system supplies the free field
  realisation of the $c=-2$ triplet model. 
  Although it is widely known that there is no unique way
  of obtaining the basis of the boundary states, 
  the same model
  of LCFT was expected to give the same set of character functions to
  form the same partition function. 

  Our main aim in this chapter is to see differences between the above
  cases and the constructions, 
  and to resolve the question in the case of a particular LCFT, the
  $c=-2$ triplet model.
  For this purpose, we will first show and review the works on the (generalised)
  Ishibashi states in section \ref{sec:ishibashi} and the work on the
  symplectic fermions in section \ref{sec:coherent} as the coherent state
  construction. In section \ref{sec:comparison}, we will then give
  the comparison between the two constructions and clarify the differences.
  Finally, it will be shown that they are indeed distinct models.
  The conclusion of this chapter and the discussion on the Verlinde formula will be given in chapter \ref{ch:conclusion}.

  First, let us look back briefly 
  to the starting point of the boundary states and the character
  functions at $c=-2$.

\paragraph{Cardy's equation}

Cardy's equation or conditions on boundary states were given in chapter \ref{ch:BCFT} in (\ref{def:bc0}). We write down this once again for
accessibility, for the Virasoro algebra:
\begin{eqnarray}
 \label{def:bc1}
 (L_{n} - \ol L_{-n}) \ket{B} = 0 {\rm ~~~for~ n\in\Z}.
\end{eqnarray}
In the presence of extended symmetry $\W$ on the boundary, whose
conformal tensor is given by $W^a(z)$ of dimension $h$, 
we have in addition
\begin{eqnarray}
 \label{def:W cardy's eq}
 (W_{n}^a - (-1)^s \ol W_{-n}^a) \ket{B} =0 {\rm ~~~for~ n\in\Z}, 
\end{eqnarray}
where $s$ is a spin of the field $W^a(z)$. For the case of the $c=-2$
triplet model, the equation for the $\W$-generators ($h=s=3$) turns out to
be:
\begin{eqnarray}
 (W_{n}^a + \ol W_{-n}^a) \ket{B} =0 .
\end{eqnarray}
From the commutation relations of the Virasoro algebra and the triplet
algebra in chapter \ref{ch:intro} and \ref{ch:LCFT}, higher modes of
the algebras can be given by a smaller set of the lower ones:
\begin{eqnarray}
 L_n &=& \frac{1}{n\mp 2} \com{L_{n\mp 1}}{L_{\pm 1}} ~~~for~~n\not= \pm2,
\nn
 W_n^a &=& \frac{1}{2n} \com{L_n}{W_0^a} ~~~for~~ n\not=0 ~and~ a=+,-,0. 
\end{eqnarray}
Actually, $\{L_{\pm 2}, L_{\pm 1}, L_0, W_0^a\}$ and its holomorphic
counterpart generate the whole set of modes in this case. 
With the above, one can extract the finite number of necessary and sufficient
conditions of the equations (\ref{def:bc1}, \ref{def:W cardy's eq}) by induction method [see for example \cite{bredthauer}]:
\begin{eqnarray}
\label{def:minimal bc}
&& (L_{n} - \ol L_{-n}) \ket{B} = 0 { ~~~for~ n\in \{0,\pm 1,\pm 2\}},
\nn
&& (W_{0}^a + \ol W_{0}^a) \ket{B} =0 .
\end{eqnarray}
This also holds for other cases with $h\not= 1$. 
Even when $h=1$, only eq. (\ref{def:W cardy's eq}) with $n=-1$ must be added.

\paragraph{A Note On Characters at $c=-2$}
\label{sec:c=-2}

In boundary CFTs on a wrapped strip, it is required that 
the characters have 
rationality under the modular transformations.
That is, the partition function as a sum of
characters has to be transformed into a sum of the same set of 
characters under modular transformations, otherwise it violates the duality on the geometry.
However, the characters in the normal $c=-2$ model 
are not modular-transformed into the same set but generate infinitely
many characters \cite{flohr1}. 
We therefore exclude the normal model and consider only the $c=-2$ triplet model.
It should be noted that this loss of rationality always happens 
in the normal $c_{p,1}$ models.

The modular-invariant set of the $\W(2,3^3)$-characters were given in eqs. (\ref{def:Wchara}) and (\ref{def:GJLWM0})
based on the analysis of fusion rules \cite{gab3},
while the $S$-matrix is also given in \cite{gab3,roh1}.
Another set was also given in \cite{flohr1}, by tracing back from
a modular invariant partition function.

\section{Boundary Ishibashi states}
\label{sec:ishibashi}

\subsection{Ishibashi States in JLWMs}
\label{sec:ishibashiYI}

Let us first assume that a solution $\ket{B}$ of Cardy's boundary equation 
takes the form,
\begin{equation}
 \label{def:B}
  \ket{B} \equiv \sum_{\{N\}} \ket{ \al \,,\,N }\otimes\ol{\ket{ \beta \,,\,N}},
\end{equation}
on the ground that the initial (final) states are in the space of the
tensor products of the Hilbert spaces, 
\vspace{-15pt}
\begin{eqnarray}
 \label{def:hilbert space}
   \oplus_{i,j} \H_i \otimes \ol\H_j,
\end{eqnarray}
where $i$ and $j$ denote distinct Virasoro LWMs and JLWMs.
The form (\ref{def:B}) is known to yield a basis of boundary states,
called Ishibashi states, for the boundary states of non-logarithmic CFTs \cite{ishibashi1}.
Accordingly, eq. (\ref{def:W cardy's eq}) is equivalent to
\begin{eqnarray}
 \label{calc}
   \bra{j,N_1}\otimes\ol{\bra{k,N_2}}\, ( W_n - (-1)^s \ol{W}_{-n} )
   \ket{B} &=& 0 , 
\end{eqnarray}
where 
$j, k, N_1, N_2$ are arbitrary.
For the Virasoro parts (\ref{def:bc1}), one obtains:
\bea
 \label{eq:L}
   \bra{j,N_1}\otimes\ol{\bra{k,N_2}}\, ( L_n - \ol{L}_{-n} ) \ket{B}
 &=& 0 .
\eea
Now we simply find the conditions on $\alpha, \beta$ for (\ref{def:B})
to be a boundary state satisfying (\ref{calc}) and (\ref{eq:L}).

According to the decomposition of the Virasoro algebra
in section \ref{sec:jordan cell}, the $l.h.s.$ of eq. (\ref{eq:L}) can
be decomposed and simplified into two parts:
\pagebreak
\begin{eqnarray}
 \label{calc:L}
   lhs &=& \sum_{N} \Bigg\{ \braket{j,N_1|(L_n^d + L_n^{nil})|\al,N}
 \braket{ \ol{k,N_2}|\ol{\beta,N} }
\nn
&&\qquad\qquad - \,\,\braket{j,N_1|\al,N} 
 \braket{ \ol{k,N_2}|(\ol L_{-n}^d + \ol L_{-n}^{nil})|\ol{\beta,N}} \Bigg\} 
\nn
   &=& ({\rm diagonal~part}) + ({\rm non{-}diagonal~part}),
\nn
({\rm diagonal~part})
&\hspace{-1pt}\equiv&\hspace{-1pt} 
\delta_{N_1,N_2-n} \delta_{k,\beta} \delta_{j,\al}
\nn[0pt] 
&&\hspace{-1pt}
  \times  \Big\{ \bra{\al,N_1} L_n^d \ket{\al,N_1 + n} - \bra{\beta^*,N_1}
  L_{n}^d \ket{\beta^* , N_1+n} \Big\} ,
\nn
({\rm non{-}diag~part}) 
&\hspace{-1pt}\equiv&\hspace{-1pt} 
\delta_{N_1,N_2-n} \left( \delta_{j,\al} \delta_{\beta^* < k^*}
 + \delta_{j<\al} \delta_{\beta^*,k^*}
 + \delta_{j<\al} \delta_{\beta^* < k^*} \right)
\nn 
&&\hspace{-10pt}
  \times  \bra{j,N_1} \Big( L_n \ket{\al,N_2} \bra{\beta^*,N_2} -
     \ket{\al,N_1} \bra{\beta^*,N_1}L_n \Big) \ket{k^*,N_2} .
\nn
\end{eqnarray}
Here the diagonal part means the contribution from diagonal $(j,\alpha)$
and $(k,\beta)$ pairings with $\{L_n^d, \ol L_n^d\}$, while the non-diagonal one is from all others.
We introduce $\delta_{f>i}$ such that if $
f=v^{(a)}, i=v^{(b)}$ and 
$a>b$ then $\delta_{f>i}=1$ otherwise it
vanishes. 
Here, it is assumed that the above bilinear form is a Shapovalov form (\ref{def:shapo}) which is 
not necessarily simple under $*$ as in eq. (\ref{def:*JLW}) and
(\ref{def:rank2 map}). 

For our purpose, it is sufficient to consider that $\alpha$ and $\beta$ are in
JLWMs. For the diagonal
representations we simply drop the non-diagonal
part.
Since $\braket{j,N_1|L_n^{nil}|\al,N_2}\not=0$
if and only if $h_j = h_{\al}$,
the diagonal part vanishes if $\ket{\al,N}$ and $\ket{\beta^*,N}$ have
the same conformal structure, i.e. the same conformal dimensions and null
 vectors at the same level, $\V_\al \sim \V_\beta$. For the diagonal
 representations, this is almost trivial. 

For $c_{p,1}$ models, there is a
 possibility for $\ket{\alpha}$ and $\ket{\beta}$ to be not in the same 
multiplet but in the same Jordan cell, the lowest weight space of JLWM.
Given that we have a
$rank$-$2$ JLWM with the upper JLWV $\ket{C}=\,\vert\,v^{(0)}\,\rangle$ and
the lower JLWV $\ket{D}=\,\vert\,{v^{(1)}}\,\rangle$, while
generators of the Virasoro algebra merely generate their descendants,
i.e. both submodules $\V_C$, $\V_D$ of JLWM do not contain any
other submodule. 
By forcing $\al, \beta$ to be in
the same cell, a vanishing diagonal part is assured and the conditions for boundary Ishibashi states
reduce to:
\bea
 \label{eq:nonW}
&&\hspace{-45pt}   ({\rm non{-}diag~part}) = 0  \qquad{\rm
   for~arbitrary~}j,k,n,N_1,N_2 ,  
\nn
   lhs &=& \delta_{N_1,N_2-n} \Bigg[
 \delta_{j,\al} \delta_{k,C} \delta_{\beta,D}
    \Big( \bra{\al,N_1} L_n \ket{\al,N_2} \bra{C,N_2} -
    \bra{C,N_1}L_n \Big) \ket{D,N_2} 
\nn
&&{}+ \delta_{\beta,k} \delta_{j,C} \delta_{\al,D}
    \bra{C,N_1} \Big( L_n \ket{D,N_2}  -
     \ket{D,N_1} \bra{\beta^*,N_1} L_n \ket{\beta^*,N_2} \Big) 
\nn
&&{}+ \delta_{j,k,C} \delta_{\al,\beta,D}
    \braket{C,N_1|L_n|D,N_2} \Big( \braket{C,N_2|D,N_2} -
     \braket{C,N_1|D,N_1} \Big) \Bigg]
. 
\eea
Then, finally one reaches the following conditions, 
\ben
 \label{eq:nonW-bc}
\delta_{\al,D} = \delta_{\beta,D} = 0.
\een
One can easily check every detail of this derivation by substituting three sets of values,
$j=k=C;j=C,k\not=C;j\not=C,k=C$.
Hence, we obtain the only allowed Ishibashi state of the form (\ref{def:B}) in the Jordan cell,
\ben
 \label{def:bs-0}
   \ket{B} = \sum_{\{N\}} \ket{C, N} \otimes 
   \ol{\ket{C, N}} . 
\een
This result in \cite{ishimoto1} is
valid for all $rank$-$2$ indecomposable representations of the type (\ref{def:B}), 
as long as $\V_C\sim\V_D$ is satisfied.
Unless the assumption is violated, one can extend the chiral algebra to any extent.

This calculation shows that there is only one Ishibashi state of the
restricted form (\ref{def:B}) belonging to the JLWM and that this state
is built on $C$ and not $D$. However the form (\ref{def:B}) is not 
the most general possible form which was investigated by Bredthauer and
Flohr in \cite{bredthauer} and which we now turn to.

\subsection{Generalised Ishibashi States at $c=-2$}
\label{sec:ishibashiBF}

We are now going to review and confirm the work by Bredthauer and Flohr at $c=-2$ in this subsection \cite{bredthauer}.

For the initial states defined on the Hilbert space (\ref{def:hilbert space}),
there is another and more general form for the solutions of
eq. (\ref{def:bc1}). Relax the form (\ref{def:B}) such that it admits linear combinations of
different $(\alpha, \beta)$ combinations by introducing unknown
coefficients $c_{m\,n}^{\alpha \beta\,N N^\prime}$:
\begin{eqnarray}
\label{def:BFB}
 \ket{B} = \sum_{\tiny \begin{array}{c}\alpha,\beta,N,N^\prime\\ m,n\end{array}}
 c_{m\,n}^{\alpha \beta\,N N^\prime} \ket{\alpha,N,m}\otimes\ol{\ket{\beta,N^\prime,n}}.
\end{eqnarray}
$\ket{\alpha}$ and $\ket{\beta}$ are in the Jordan cell in question. 
$N$ ($N^\prime$) is the level of the descendants. $m$ and $n$ label
suitable basis states at the level $N$ and $N^\prime$ respectively.
As the Ishibashi state
is originally defined such that $\alpha$ and $\beta$ be the same for all
levels, the states defined above are called the `generalised
Ishibashi state' when $\alpha$ and $\beta$ differ from such a definition.
Our notation for the coefficients is slightly changed from
\cite{bredthauer}, in order to manifest the nature of their 
$c^{\;l\; l^\prime}_{m n}$.

Obviously, the coefficients $c_{m\,n}^{\alpha \beta\,N N^\prime}$ are
the key ingredients which determine the explicit forms of boundary states. 
In order to determine them, we apply the conditions (\ref{def:minimal bc}).
Extracting the zero mode part from the conditions (\ref{def:minimal bc}), one finds:
\pagebreak
\begin{eqnarray}
 \label{cardy0}
  0 &=& \left( L_0 - \ol L_0 \right) \ket{B}
\nn
  &=& \sum_{\tiny \begin{array}{c}\alpha,\beta,N,N^\prime\\ m,n\end{array}}
 c_{m\,n}^{\alpha\beta\,N N^\prime} 
 \left\{ N- N^\prime + \left( L_0^{nil} - \ol L_0^{nil} \right) \right\} \ket{\alpha,N,m; \beta,N^\prime,n} .
\end{eqnarray}
The states $\ket{\alpha,N,m}\otimes\ket{\beta,N^\prime,n}$ are relabeled
by $\ket{\alpha,N,m;\beta,N^\prime,n}$ 
for convenience. These states are linearly
independent and thus the above identity must hold for each basis state. 
For instance from $(\alpha,\beta)=(\omega,\omega)$ for all levels, one can read $N=N^\prime$.
Moreover, one finds for any $N$:
\begin{eqnarray}
 \label{cond:L0}
  0= \sum_{\alpha,\beta,m,n} c_{m\,n}^{\alpha\beta\,N N} \left( L_0^{nil} - \ol L_0^{nil} \right) \ket{\alpha,N,m;\beta,N,n} . 
\end{eqnarray}
This condition singles out possible combinations of $(\alpha,\beta)$ at each level, for
example for $c^{\alpha\beta}\equiv c_{1\,1}^{\alpha\beta\,0 0}$:
\begin{eqnarray}
 c^{\omega\omega} = 0,\quad c^{\Omega\omega} = c^{\omega\Omega}.
\end{eqnarray}

Next, we expand the condition for $n=1$ in (\ref{def:bc1}, \ref{def:minimal bc}) in
the following way:
\begin{eqnarray}
 \label{cardy1}
  0 &=& \left( L_1 - \ol L_{-1} \right) \ket{B}
\nn
  &=&\!\!\!\!
 \sum_{\tiny \begin{array}{c}\alpha,\beta,N\\ m,n\end{array}}\!\!\!\!
 c_{m\,n}^{\alpha\beta\,N N}  \Biggl( \sum_{\gamma, a} {E_{\alpha}^{\gamma N m}}_{a} \!\ket{\gamma, N\!-\!1, a; \beta,N,n}
  - \sum_{b} {F^{\beta N n}}_{b} \!\ket{\alpha, N, m; \beta,N\!+\!1,b} \Biggr) 
\nn
  &=&\!\!\!\!
 \sum_{\tiny \begin{array}{c}\alpha,\beta,N\\ m,n\end{array}} \!\!\!\!
 \left( c_{a\,n}^{\gamma\beta\,N N} {E_{\gamma}^{\alpha N a}}_{m} 
  - c_{m\,b}^{\alpha\beta\,N-1\,N-1} {F^{\beta N-1\,b}}_{n} \right)
 \ket{\alpha, N\!-\!1, m; \beta,N,n} .
\end{eqnarray}
Einstein's notation for contraction is used for summations over $\gamma, a, b$.
The substituted expansions, and thus the coefficients ${E_{\alpha}^{\beta N m}}_{a}$ and ${F^{\alpha N m}}_{b}$, are
defined as follows:
\begin{eqnarray}
 L_1 \ket{\alpha, N, m} &=& \sum_{\beta, a} {E_{\alpha}^{\beta N m}}_{a} \ket{\beta, N-1, a} , 
\nn
 L_{-1} \ket{\alpha, N, m} &=& \sum_{b} {F^{\alpha N m}}_{b} \ket{\alpha, N+1, b} .
\end{eqnarray}
Consequently, the conditions for $n=\pm 1$ amount to the conditions for all
possible $(\alpha,\beta,N\geq 1,m,n)$:
\begin{eqnarray}
\label{cond:L1}
 c_{a\,n}^{\gamma\beta\,N N} {E_{\gamma}^{\alpha N a}}_{m} 
  - c_{m\,b}^{\alpha\beta\,N-1\,N-1} {F^{\beta N-1\,b}}_{n} &=& 0 ,
\nn
 c_{a\,n}^{\alpha\gamma\,N N} {E_{\gamma}^{\beta N a}}_{m} 
  - c_{m\,b}^{\alpha\beta\,N-1\,N-1} {F^{\alpha N-1\,b}}_{n} &=& 0 .
\end{eqnarray}
They are symmetric under 
$c_{m\,n}^{\alpha\beta\,N N} \longleftrightarrow c_{m\,n}^{\beta\alpha\,N N}$, 
and therefore 
$c_{m\,n}^{\alpha\beta\,N N} = c_{m\,n}^{\beta\alpha\,N N}$.
The conditions for $n=\pm 2$ in (\ref{def:bc1}, \ref{def:minimal bc})
can similarly be expressed in terms of the coefficients.

Working with these condition
(\ref{def:minimal bc}, \ref{cond:L0}, \ref{cond:L1}) in the
$c=-2$ triplet model,
Bredthauer and Flohr showed the following two solutions at $h=0$ \cite{bredthauer}:
\begin{eqnarray}
 \label{flohr's sol on bs}
  \ket{R_0} &\equiv& \ket{h=0; c^{\Omega\Omega} = -d, c^{\Omega\omega} = c^{\omega\Omega} =1},
\nn
  \ket{\V_0} &\equiv& \ket{h=0; c^{\Omega\Omega} = 1, c^{\Omega\omega} = c^{\omega\Omega} =0 } .
\end{eqnarray}
$d$ is defined as minus the normalisation constant of the upper JLWV:
\begin{eqnarray}
\label{BF normalisation}
 \braket{\omega|\omega} &=& -d , 
 \nn
 \braket{\Omega|\omega} &=& \braket{\omega|\Omega} = 1 . 
 \nn
 \braket{\Omega|\Omega} &=& 0.
\end{eqnarray}
The anti-holomorphic counterpart is given similarly.
Note that their notation of the bra-states is contrary to ours in the preceding sections. 
By calculating the conditions (\ref{cond:L0}, \ref{cond:L1}) 
up to level three, 
one can confirm that the coefficients $c_{m\,n}^{\alpha\beta\,N N}$ for
$N\leq 3$ are determined by $c^{\alpha\beta}$ of ground states, such
that two coefficients $c^{\Omega\Omega}$ and $c^{\Omega\omega}$ are
independent, and that the solutions are of the form
(\ref{flohr's sol on bs}). 
We see that $\ket{\V_0}$ is just the Ishibashi state we considered in
the previous section.

It should be mentioned here that, for the solutions to satisfy the $\W$
boundary conditions in (\ref{def:minimal bc}), the anti-unitary operator
$U$ must be inserted in front of each anti-holomorphic sector:
\begin{eqnarray}
 \ket{\alpha,N,m;\beta,N,n} \equiv \ket{\alpha,N,m}\otimes U \ol{\ket{\beta,N,n}} ,
\end{eqnarray}
This operator satisfies the identities: $U^\dag W^a_m U = - W^a_m$, $U^\dag L_m U = L_m$.

In addition, there is the fermionic ground state $\ket{\xi^\alpha}$ that
appeared in Fig.\ref{fig:triplet_c-2} in chapter \ref{ch:LCFT}.
This gives rise to more boundary Ishibashi states at $h=0$:
\begin{eqnarray}
 \ket{R_{01}^\pm} = \ket{h=0; c^{\Omega \xi^\pm}=1 } ,\,\,
 \ket{R_{10}^\pm} = \ket{h=0; c^{\xi^\pm \Omega}=1 } .
\end{eqnarray}
As for $\phi_i^\alpha$ in the same figure, one can see from the conditions
(\ref{cond:L0}, \ref{cond:L1}) up to level two that the
conformal family of $\phi_i^\alpha$ is contained in $\ket{R_0}$.

The same argument can be applied to the Jordan cell at $h=1$ and
two boundary Ishibashi states at $h=1$ were also obtained. We list their ten solutions below:
\begin{eqnarray}
\label{BFBS}
 \ket{R_0}, \ket{\V_0}, \ket{R_1}, \ket{\V_1}, \ket{R_{01}^\pm}, \ket{R_{10}^\pm}, \ket{\V_{-\frac18}}, \ket{\V_\frac38} .
\end{eqnarray}
The basis of bra-boundary states can be obtained similarly. With them, 
one can find that
among these (generalised) Ishibashi states only four states can give
nontrivial inner products between bra- and ket-states. The other six states
$\{\ket{\V_0}, \ket{\V_1}, \ket{R_{01}^\pm}, \ket{R_{10}^\pm}\}$ and 
their corresponding bra-states are decoupled from the theory, 
owing to the normalisations (\ref{BF normalisation}): 
for any bra-state $\alpha$ corresponding to (\ref{BFBS}),
\begin{eqnarray}
\label{marginal states}
 \braket{\alpha|\V_0}= \braket{\alpha|\V_1}= \braket{\alpha|R_{01}^\pm}= \braket{\alpha|R_{10}^\pm}=0.
\end{eqnarray}
Therefore, the general boundary state which satisfies the condition
(\ref{def:bc1}) can be written down with four constants by:
\begin{eqnarray}
\label{sol:general BF}
 \ket{B} = a \ket{R_0} + b \ket{R_1} + c \ket{\V_{-\frac18}} + d \ket{\V_\frac38}, 
\end{eqnarray}
up to those marginal states in (\ref{marginal states}).

They also computed the cylinder amplitudes
out of their solutions: 
\begin{eqnarray}
\label{character BF}
 \bra{R_0} \wt q^{\frac12(L_0+\ol L_0)+\frac1{12}}\ket{R_0} &=&
 \bra{R_1} \wt q^{\frac12(L_0+\ol L_0)+\frac1{12}}\ket{R_1}  = \chi_{\cal R} (\wt q) ,
\nn
 {\tiny \bra{\V_{-\frac18}} \wt q^{\frac12(L_0+\ol L_0)+\frac1{12}}\ket{\V_{-\frac18}}} &=& \chi_{\V_{-1/8}} (\wt q) ,
\nn
 \bra{\V_\frac38} \wt q^{L_0+\frac1{12}}\ket{\V_\frac38} &=& \chi_{\V_{3/8}} (\wt q) ,
\end{eqnarray}
where $\wt q=e^{2\pi i \wt q}$ appeared in eq. (\ref{def:Zclose}), and
the $\chi_{}(\wt q)$ are the characters of the $c=-2$ triplet model
which appeared in chapter \ref{ch:LCFT} \cite{roh1, gab4}.
The naive investigation on the generalised Ishibashi states 
results in the three-dimensional representation of the characters. 
In order to produce a higher-dimensional representation, one must
introduce { weak} boundary states \cite{bredthauer}. 
We do not deal with them  in this thesis.
The physical boundary states which correspond to the boundary
conditions can be obtained by solving Cardy's equation
(\ref{def:modified Cardy eq}) with an appropriate $S$-matrix.

\section{Coherent States with Symplectic Fermions}
\label{sec:coherent}

In this section, we focus on the boundary coherent states found by Kawai
and Wheater \cite{kaw}.
To begin with, we introduce the action of the $\xi$-$\eta$ fermionic ghost system:
\begin{eqnarray}
\label{ghost action}
  S_{ghost} [\xi, \eta] = \frac{1}{\pi} \int_\Sigma d^2 z \left(\eta(z)
  \bar \pa \xi(z) + \ol \eta(\bz) \pa \ol \xi(\bz) \right) . 
\end{eqnarray}
$\Sigma$ is a two-dimensional flat surface with one or more boundaries,
and $\eta$ and $\xi$ fermions are conformal dimension one and zero respectively.
Their propagators reflect their Fermi statistics:
\begin{eqnarray}
  \xi(z)\eta(w) \sim \eta(z)\xi(w) \sim \frac{1}{z-w}.
\end{eqnarray}
The anti-holomorphic sectors are given similarly. 
Symplectic fermions of conformal dimension one are defined by $\eta$ and
$\pa \xi$ such that \cite{kau2}:
\begin{eqnarray}
  \left( \begin{array}{c}
         \chi^+ \\ \chi^-
         \end{array}\right) \equiv \left( \begin{array}{c}
                                      \eta \\ \pa \xi
                                    \end{array} \right) .
\end{eqnarray}
Their mode expansions with its commutation relations are:
\begin{eqnarray}
  \chi^\alpha(z) \equiv \sum_{n\in\Z+\lambda} \chi_n^\alpha z^{-n-1} \,, \,\, 
  \{\chi_m^\alpha \,,\, \chi_n^\beta \} = m d^{\alpha\beta} \delta_{m+n,0} 
~~~~{\rm for~} m,n \in\Z+\lambda . 
\end{eqnarray}
$\alpha$ and $\beta$ take either $+$ or $-$, and $d^{\alpha\beta}$ is
the anti-symmetric tensor: $d^{+-}=-d^{-+}=1$. We impose the $Z_2$
twisting on the theory while $\lambda=0,\frac12$ denotes its untwisted and
twisted sectors.
The stress tensor is given by
\begin{eqnarray}
\label{def:symp T}
  T= :\chi^-\chi^+:_\lambda + \frac{\lambda(\lambda-1)}{2}\frac1{z^2} ,
\end{eqnarray}
where the operator normal ordering $:{\cal O}:_\lambda$ is taken and the
constant shift of the zero mode arises from the second term subject to the $Z_2$ twisting.
The central charge $c$ is $-2$ and the $\W$ algebra, or the triplet algebra, is realised by the symplectic fermions \cite{kaw,kau2,kau3}:
\begin{eqnarray}
  W^0 &=& -\frac12 \left(:\pa\chi^-\chi^+ + \pa\chi^+\chi^-: \right) , 
\nn
  W^\pm &=& \pa\chi^\pm \chi^\pm . 
\end{eqnarray}

The boundary conditions that come from the variation of the action
(\ref{ghost action}),
take the following simple form:
\begin{eqnarray}
  \delta\left( \eta\xi + \ol \xi \ol \eta \right)_{boundary} = 0 .
\end{eqnarray}
Without loss of generality, this can be translated into gluing
conditions for the ghost fields as 
$\eta=e^{i\phi}\ol \eta,\; \xi=e^{-i\phi} \ol \xi$ with 
$0\leq\phi<2 \pi$ at the boundary.
For the symplectic fermions, these are :
\begin{eqnarray}
  \left( \chi^\pm(z) - e^{\pm i\phi}\ol \chi^\pm(\bz) \right)_{boundary} =0 .
\end{eqnarray}
By the same procedure for eq. (\ref{def:bc1}) appeared in chapter \ref{ch:BCFT},
these conditions can be interpreted on an annulus, whence the boundary
conditions become those on {boundary states}.
In terms of modes, they are \cite{kaw}:
\begin{eqnarray}
\label{def:symp bc}
 \left( \chi_m^\pm - e^{\pm i\phi}\ol \chi_{-m}^\pm \right) \ket{B_\phi} =0 .
\end{eqnarray}
$m$ is integer for the untwisted sector and half-integer for the twisted sector.
$\ket{B_\phi}$ is a state at the initial boundary. The conditions for bra-states are given similarly. 
From the expression (\ref{def:symp T}) of the Virasoro algebra, it turns
out that the
solutions of the above fermionic conditions always satisfy the Virasoro
conditions (\ref{def:bc1}). 
Imposing the $\W$-symmetry on the boundary gives the constraint $e^{2i\phi}=1$, 
that is, $\phi=0 ~mod~ \pi$. Since $0\leq\phi<2\pi$, this finally amounts to:
\begin{eqnarray}
 \phi=0 {\rm ~or~} \pi ,
\end{eqnarray}
which are conventionally labeled by $+,-$ respectively.

Kawai and Wheater showed that coherent states are solutions for the above equations \cite{kaw}:
\begin{eqnarray}
 \ket{B_{\lambda \phi}} \equiv N\, \exp\left( \sum_{k>0}  \frac{e^{i\phi}}{k} \chi^-_{-k} \ol \chi^+_{-k} + \frac{e^{-i\phi}}{k} \ol \chi^{\,-}_{-k} \chi^+_{-k} \right) \ket{\lambda_{\phi}} ,
\end{eqnarray}
where $\lambda=0$ or $\frac12$, and the normalisation constant $N$ is added for generality.
By substituting the above into the conditions (\ref{def:symp bc}), 
it is easy to see that the state $\ket{0_\phi}$ must satisfy the following conditions 
\begin{eqnarray}
\label{def:symp bc 0}
  \chi^a_{m>0} \ket{0_\phi} =  \ol \chi^a_{m>0} \ket{0_\phi} = 0 , 
\nn
  \left( \chi^\pm_0 - e^{\pm i \phi} \ol \chi^\pm_0 \right) \ket{0_\phi} = 0 .
\end{eqnarray}
The first line expresses nothing but the lowest weight state conditions.
The latter condition on the zero modes is purely a boundary effect for the untwisted sector.
As the twisted sector has a unique vacuum and does not contain zero modes, 
the resulting coherent states turn out to be those of diagonal representations. Thus, we concentrate on the ket-states of the untwisted sector in what follows.

From the lesson in the chiral $c=-2$ triplet model
\cite{gab4}, it was assumed that such non-chiral states 
$\ket{0_\phi}$ are
in $\{ \ket{\Omega}\otimes\ol{\ket{\Omega}}, \ket{\omega}\otimes\ol{\ket{\Omega}}, \ket{\Omega}\otimes\ol{\ket{\omega}}, \ket{\omega}\otimes\ol{\ket{\omega}} \}$. 
Since $\ket{\omega}\otimes\ol{\ket{\Omega}}$ and 
$\ket{\Omega}\otimes\ol{\ket{\omega}}$ can never satisfy the condition, 
only two states $\ket{\Omega}\otimes\ol{\ket{\Omega}}$ and 
$\ket{\omega}\otimes\ol{\ket{\omega}}$ were selected in
\cite{kaw}.
The $sl2\times sl2$-invariant state $\ket{\Omega}\otimes\ol{\ket{\Omega}}$
naturally obeys (\ref{def:symp bc 0}), because $\ket{\Omega}$ 
can be defined by
$\ket{\omega}$ and the zero modes \cite{gab4}:
\begin{eqnarray}
\label{def:symp h=0 ground}
 \ket{\Omega} \equiv \chi^-_0 \chi^+_0 \ket{\omega} ,\,\,
 \ol{\ket{\Omega}} \equiv \ol \chi^{\,-}_0 \ol \chi^+_0 \ol{\ket{\omega}} .
\end{eqnarray}
On the other hand, as $\ket{\omega}\otimes\ol{\ket{\omega}}$ does not naturally obey (\ref{def:symp bc 0}), the following constraint was imposed:
\begin{eqnarray}
\label{symp const}
  \left( \chi^\pm_0 - e^{\pm i \phi} \ol \chi^\pm_0 \right)\ket{\omega}\otimes\ol{\ket{\omega}} =0 .
\end{eqnarray}
In fact, the above constraints drastically change the vacuum structure
and lead to the following equivalence on 
$\ket{\omega, \omega}\equiv\ket{\omega}\otimes\ol{\ket{\omega}}$.
Then the former definition of $\ket{\Omega, \Omega}\equiv\ket{\Omega}\otimes\ol{\ket{\Omega}}$ is
replaced by another definition with the $sl2$-invariance:
\begin{eqnarray}
\label{non-chiral symp states}
 \chi^-_0 \chi^+_0 \ket{\omega,\omega} 
 &\sim& \ol \chi^{\,-}_0 \ol \chi^+_0 \ket{\omega,\omega} 
 \,\sim\, \ket{\Omega,\Omega}, 
\nn
 L_0 \ket{\omega,\omega} &\sim& \ol L_0 \ket{\omega,\omega} \,\sim\, \ket{\Omega,\Omega}, 
\nn
 L_0 \ket{\Omega,\Omega} &\sim& \ol L_0 \ket{\Omega,\Omega} \,\sim\, 0 .
\end{eqnarray}
The equivalence ($\sim$) means equality ($=$) when the constraint
(\ref{symp const}) strictly holds.
We will discuss this point later.

With these two degenerate vacua with two choices of $\phi=\pm$, 
there are four possible states in the untwisted sector.
Together with $\ket{B_{\mu\pm}}\equiv\ket{B_{\lambda=\frac12, \phi=\pm}}$,
this gives six coherent states which form a basis of the boundary states:
\begin{eqnarray}
\label{six coherent}
 \ket{B_{\Omega\pm}}, \ket{B_{\omega\pm}}, \ket{B_{\mu\pm}} .
\end{eqnarray} 
Setting the normalisations of the vacua as:
\begin{eqnarray}
 \label{metric KW}
  \braket{\omega, \omega\,|\,\omega, \omega} &=& d , 
\nn
  \braket{\omega, \omega\,|\,\Omega, \Omega} &=&  \braket{\Omega, \Omega\,|\,\omega, \omega} \,=\, -1 \, 
\nn
  \braket{\Omega, \Omega\,|\,\Omega, \Omega} &=& 0 , 
\end{eqnarray}
one can calculate the cylinder amplitudes:
\begin{eqnarray}
\label{character KW1}
&& \bra{B_{\alpha\phi}} \wt q^{L_0 + \frac1{12}}
   \ket{B_{\beta\phi^\prime}}
\nn
 &=& N^2 \left\{ \braket{\alpha|\beta}+\bra{\alpha}L_0^{nil}\ket{\beta}\log \wt q\right\}
     \wt q^{\frac1{12}}\prod_{i=1}^{\infty} \left( 1 - e^{i(\phi+\phi^\prime)} \wt q^n\right)^2
\nn
 &=&
 \begin{array}{rl}
  &\quad\; \ket{B_{\Omega+}} \quad\;\ket{B_{\Omega-}} \qquad\qquad\ket{B_{\omega+}} \qquad\qquad\ket{B_{\omega-}} \\
  \begin{array}{r} \bra{B_{\Omega+}} \\\bra{B_{\Omega-}} \\\bra{B_{\omega+}} \\\bra{B_{\omega-}} \end{array}
  &\left(  \begin{array}{cccc}
   0 & 0 & -\eta^2(\wt q) & -\Lambda_{1,2} (\wt q) \\
   0 & 0 & -\Lambda_{1,2} (\wt q) & -\eta^2(\wt q) \\
   -\eta^2(\wt q) & -\Lambda_{1,2} (\wt q) & (d - \log \wt q)\eta^2(\wt q) & (d - \log \wt q)\Lambda_{1,2} (\wt q) \\
   -\Lambda_{1,2} (\wt q) & -\eta^2(\wt q) & (d - \log \wt q)\Lambda_{1,2} (\wt q) & (d - \log \wt q)\eta^2(\wt q) \\
                     \end{array}  \right)
                             \end{array} .
\end{eqnarray}
where $\phi,\phi^\prime=0~{\rm or}~\pi$, the normalisation $N$ is set to be $1$, 
$\Lambda_{r,k}(q) = \Theta_{r,k}(q)/{\eta(q)}$ and $\eta(q)$
is the Dedekind eta function [see chapter \ref{ch:LCFT} for their definitions].
The functions of the twisted sector are given similarly:
\begin{eqnarray}
\label{character KW2}
  \bra{B_{\mu\phi}} q^{L_0 + \frac1{12}} \ket{B_{\mu\phi^\prime}}
 = \Lambda_{0,2}(\wt q) - e^{i(\phi+\phi^\prime)} \Lambda_{2,2} (\wt q) ,
\end{eqnarray}
where the relevant ground state of conformal dimension $-\frac18$ is
unique and normalised as $\braket{\mu|\mu}=1$.
In both derivations, the famous Jacobi triple product identity is used.

Making a modular transformation, one finds that this set of cylinder amplitudes is not closed unless $(\log \wt q) \Lambda_{1,2}(\wt q)$ is deleted.
It follows that either $\ket{B_{\omega +}}$ or $\ket{B_{\omega -}}$ must drop out from the physical boundary states. 
Consequently, only five distinct coherent states become physical: 
\begin{eqnarray}
\label{sol:general KW}
 \ket{B} = a_\pm \ket{B_{\Omega\pm}} +b_\pm \ket{B_{\mu\pm}} +c \left( \ket{B_{\omega +}} {\rm or} \ket{B_{\omega -}}\right) .
\end{eqnarray}

After one of $\ket{B_{\omega\pm}}$ drops out, the cylinder amplitudes of the untwisted sector are realised by a 3 by 3 matrix. This matrix has one
zero-eigenvalue, and therefore a degree of freedom out of five in
(\ref{sol:general KW}) become null in its cylinder amplitude. For
example, dropping $\ket{B_{\omega-}}$, one finds:
\begin{eqnarray}
 \bra{\alpha} \wt q^{L_0 + \frac1{12}}
 \Bigl\{ \Lambda_{1,2}(\wt q) \ket{B_{\Omega+}}
   - \eta^2(\wt q) \ket{B_{\Omega-}} \Bigr\} = 0,
\end{eqnarray}
where $\bra{\alpha}$ is any bra-boundary state of the type (\ref{sol:general KW}).
Then the expression (\ref{sol:general KW}) turns out to be four-dimensional.
The physical boundary states can be obtained by solving Cardy's equation
(\ref{def:modified Cardy eq}) in chapter \ref{ch:BCFT} with an $S$-matrix.

\section{Comparison of the Constructions at $c=-2$}
\label{sec:comparison}

It is now clear that the obtained basis of the boundary states 
in the preceding two sections are different. 
More precisely, the character functions which the basis states generate 
are different for each construction: 
the (generalised) Ishibashi state constructions in section \ref{sec:ishibashi} contain a smaller set of functions as in
(\ref{character BF}) than 
the coherent state construction does in
(\ref{character KW1}, \ref{character KW2}) in section \ref{sec:coherent}.
It seems to suggest that the two basis do not correspond to the same set
of physical boundary states. 

Our primary concern will be addressed on whether this practical difference is
admissible and/or possible in the $c=-2$ triplet model. In addition, if
so, it will be followed by a question of whether it is possible to
determine which case is being examined from physical quantities. 
Our answer to the primary concern is that the mentioned difference is
admissible and possible. 
The reason is because it comes from some fundamental differences between two
constructions, which make the reference theory quite different.
In other words, the theory is different.
Before moving onto the fundamental differences,
we make a brief remark on the latter question.

In some cases, it is possible to point out the difference between the
two constructions from physical quantities, 
but there are other cases where it may not be possible.
Physical observation must be single-valued so that, 
if one $c=-2$ theory emerges in reality, nature will select only one set of
solutions for one particular setting. 
From such results, one may easily spot the difference on the 
partition functions in the closed string picture. The Ishibashi state
construction contains only $\Lambda_{i,2}(\wt q)$ whereas the coherent
state construction contains in addition:
\begin{eqnarray}
 \eta^2({\wt q}) {\rm ~~and~~}(\log \wt q) \eta^2(\wt q).
\end{eqnarray}
Therefore, if a logarithmic quantity emerges in
the cylindrical amplitude, the latter is the case. 
However, if such a quantity is absent, it is in general not possible 
to determine which case is being examined. 
There may appear only a smaller set of the coherent states. 
Another probe is necessary in order to distinguish them.

Roughly speaking, there are four fundamental differences.
Firstly, the Ishibashi state construction has the boundary conditions
(\ref{def:bc1}) for the Virasoro symmetry only, 
or (\ref{def:minimal bc}) for the extended case, 
while the coherent state construction has the fermionic boundary
conditions (\ref{def:symp bc}).
Secondly, the former construction is the superposition of the 
holomorphic/anti-holomorphic decomposition of the non-chiral states,
while the latter is intrinsically non-chiral and indecomposable to
holomorphic sectors.
This is related to the treatment of the zero-mode conditions
in (\ref{def:symp bc 0}).
Thirdly, 
the generalised Ishibashi state
construction deals with the mixed lowest weight representation whereas
the coherent state construction does not. \alert{One might think that} this may \alert{lead to
 some additional boundary states in
the latter.}
Finally, the appearance of each construction is of course 
different from each other, 
since for example the descendants are constructed in distinctive ways.
In ordinary CFT, such as the unitary minimal models, both constructions
may provide the same basis, since operators in the operator formalism
provide the generators of the Virasoro algebra, or the chiral algebra in
general \cite{kawaic1,kawaic2}.
However, the action of the Virasoro algebra and that of the symplectic
fermion modes are essentially different.
This would be related to the difference between Hilbert space generated by
the Virasoro algebra and Hilbert {\it Fock} space generated by the
fermionic modes.

Among those differences,
the most critical difference is the second one on the definitions of the
lowest weight states. 
In \cite{bredthauer2}, Bredthauer gave a brief comparison between two constructions and
showed that even with the fermionic boundary 
conditions (\ref{def:symp bc}) his and Flohr's way leads to the same number of
basis states as in the case of (\ref{def:minimal bc}) with no fermionic
conditions. 
It was also indicated that the difference would come from the imposition
of the constraint: 
\begin{eqnarray}
  \left(L_0 - \ol L_0 \right) \ket{LWS} =0 
\end{eqnarray}
on the lowest weight states, and from the treatment of the mixed states.
Roughly speaking, this indication is true, but precisely the above
condition must be replaced by the zero-mode conditions in (\ref{def:symp bc 0}): 
\begin{eqnarray*}
  \left( \chi^\pm_0 - e^{i\phi} \ol \chi^\pm_0 \right) \ket{LWS} = 0  
~~~for~~ \phi=0,\pi,
\end{eqnarray*}
in the untwisted sector.
Along the similar line, we state that
the two different constructions deal with different theories 
in a sense that they are based
on the different LWS spaces and hence the different Hilbert spaces.

In the rest of this section, we begin with the LWS space of $h=0$ in the
triplet model and clarify such a critical difference in the symplectic
fermion model. 
In addition,
other possibilities will be presented in due course beyond the cases in section \ref{sec:ishibashi} and \ref{sec:coherent} 
\cite{ishimoto1, kaw, bredthauer, bredthauer2, YI2}.
We are not going to reconfirm the approach 3 in \cite{bredthauer} 
but will also investigate his indication as well.

The chiral $c=-2$ triplet model is known to possess
the logarithmic pair at $h=0$ and, moreover, the fermionic ground
states of the same conformal dimension. They are expressed as:
\begin{eqnarray}
 \label{def:4 ground states}
  \ket{\Omega}, \,\, \ket{\xi^\pm}, \,\,  \ket{\omega} , 
\end{eqnarray}
for the holomorphic sector [see Fig.\ref{fig:triplet_c-2}].
With the triplet algebra, two bosonic states are correlated as $L_0\ket{\omega}=\ket{\Omega}$ whereas the
doublet state $\ket{\xi^\pm}$ is not correlated to other two. 
On the other hand, in the language of symplectic fermions, 
all the states in (\ref{def:4 ground states}) 
are related by the action of the zero modes as shown in (\ref{def:symp h=0 ground}):
\begin{eqnarray}
\label{rel:chiral LWS}
  \chi_0^\pm : \ket{\omega} \to \ket{\xi^\pm} \to \ket{\Omega}
\end{eqnarray}
The arrow stands for the action of either $\chi_0^+$ or $\chi_0^-$.

Let us define two projection operators ${\cal B}^\pm$ for later
convenience:
\begin{eqnarray}
  {\cal B}^\pm_\phi \equiv \left( \chi_0^\pm - e^{i\phi} \ol \chi_0^\pm \right) .
\end{eqnarray}
The zero-mode conditions can now be written as ${\cal B}^\pm_\phi \ket{LWS}=0$.
If one treats ${\cal B}^\pm_\phi \ket{LWS}=0$ as the constraints, which is
the case in section \ref{sec:coherent}, it
immediately leads to the following equivalence of the zero modes on the states at
the boundary:
\begin{eqnarray}
 \chi_0^\pm \sim \ol \chi_0^\pm, 
\end{eqnarray}
up to signs depending on the phase $\phi$. This is essentially same as
in section 4 of \cite{kau2}, and the non-chiral LWS space turns out to be four-dimensional:
\begin{eqnarray}
\label{def:4 nc states}
&& \ket{\mathbf \omega}_{boundary}, 
\nn
&& \ket{\mathbf \xi^\pm}_{boundary} \equiv - \chi_0^\pm \ket{\mathbf \omega}_{boundary},
\nn
&& \ket{\mathbf \Omega}_{boundary} \equiv \chi_0^- \chi_0^+ \ket{\mathbf \omega}_{boundary}.
\end{eqnarray}
This is partially equivalent to (\ref{non-chiral symp states}) with the
replacement of 
$\ket{\omega,\omega}\to \ket{\mathbf \omega}_{boundary}$ and 
$\ket{\Omega,\Omega}\to \ket{\mathbf \Omega}_{boundary}$.
This space is also quite similar to that of the non-chiral theory in \cite{gab5}, 
where our $\mathbf{\xi^\pm}$ should be interpreted as $\mathbf{\theta^\pm}$.
However, note that the definition of $\ket{\xi^\pm}$ by $\ol \chi_0^\pm$
depends on the phase $\phi$ in our case.

Since the states (\ref{def:4 nc states}) all satisfy the conditions
(\ref{def:symp bc 0}), we can establish eight coherent states on
them, four bosonic and four fermionic states:
\begin{eqnarray}
\label{sol:comparison1}
 \ket{B_{\Omega\pm}}, \ket{B_{\xi^\alpha \pm}}, \ket{B_{\omega\pm}}.
\end{eqnarray}
Together with $\ket{B_{\mu\pm}}$, they form a ten-dimensional space as was
suggested in \cite{bredthauer2}.
As the normalisation (\ref{metric KW}) implies
$\braket{\xi^\alpha|\xi^\beta}=-d^{\alpha \beta}$,
their character functions are given by (\ref{character KW1}, \ref{character KW2}) and 
\begin{eqnarray}
 \braket{B_{\xi^\alpha \phi}| \wt q^{L_0+\frac1{12}} |B_{\xi^\beta \phi^\prime}}
 = \left\{ \begin{array}{ll}
    -d^{\alpha \beta}\, \eta^2(\wt q)   &for~ \phi=\phi^\prime\\
    -d^{\alpha \beta}\, \Lambda_{1,2}(\wt q) &for~ \phi\not=\phi^\prime
           \end{array}  \right. , 
\end{eqnarray}
while other combinations vanish. This does not generate any new functions. 
Note that, for this LWS space, there is no need to consider the mixed states since
all LWS's satisfy the conditions (\ref{def:symp bc 0}).

Next, let us assume that the non-chiral states of $h=0$ at the boundary 
are the superpositions of the holomorphic and anti-holomorphic
sectors, leaving the boundary conditions intact.
Such LWS's are
linear combinations of sixteen different tensor products of
(\ref{def:4 ground states}). Four of them are
illustrated below:
\begin{eqnarray}
\label{16 diff states}
  \ket{\Omega}\otimes \ol{\ket{\Omega}} , \,\, 
  \ket{\Omega}\otimes\ol{\ket{\omega}} , \,\,
  \ket{\omega}\otimes\ol{\ket{\Omega}} , \,\,
  \ket{\Omega}\otimes\ol{\ket{\Omega}} .
\end{eqnarray}
As in (\ref{rel:chiral LWS}), the sixteen products of the type (\ref{16 diff states}) are
also related to each other in the same manner. 
In fact, the four non-chiral states (\ref{16 diff states}) can
be expressed by:
\bea
  \ket{\omega}\otimes\ol{\ket{\omega}} &\equiv& \ket{\omega, \omega}, 
\nn 
  \ket{\omega}\otimes\ol{\ket{\Omega}} &=&  \ol \chi_0^+ \ol \chi_0^- \ket{\omega, \omega}, \,\, 
\nn
  \ket{\Omega}\otimes\ol{\ket{\omega}} &=&  \chi_0^+ \chi_0^- \ket{\omega, \omega}, \,\, 
\nn
  \ket{\Omega}\otimes\ol{\ket{\Omega}} &=& 
  \chi_0^+ \chi_0^- \ol \chi_0^+ \ol \chi_0^- \ket{\omega, \omega} \equiv \ket{\Omega, \Omega}. \,\, 
\eea
With the lowest weight conditions $\chi_{n>0}^a\ket{\omega}=0$, the
last expression for $\ket{\Omega,\Omega}$ certainly exhibits the
$sl2\times sl2$ invariance:
\begin{eqnarray}
&&  \chi_{n>-1}^a \ket{\Omega, \Omega} = \ol \chi_{n>-1}^a \ket{\Omega, \Omega} = 0,
\nn
&&  L_{n\geq -1} \ket{\Omega, \Omega} = \ol L_{n\geq -1} \ket{\Omega, \Omega} =0 .
\end{eqnarray}
Twelve more states can be built on the state $\ket{\omega,\omega}$:
\bea
  \ket{\xi^\pm}\otimes\ol{\ket{\omega}} &=& 
  - \chi_{0}^\pm \ket{\omega,\omega} , 
\nn
  \ket{\omega}\otimes\ol{\ket{\xi^\pm}} &=& 
  - \ol \chi_{0}^\pm \ket{\omega,\omega} , 
\nn
  \ket{\xi^a}\otimes\ol{\ket{\xi^b}} &=& 
  \chi_{0}^a \ol \chi_{0}^b \ket{\omega,\omega} , 
\nn
  \ket{\Omega}\otimes\ol{\ket{\xi^\pm}} &=& 
  - \chi_{0}^- \chi_{0}^+ \ol \chi_{0}^\pm \ket{\omega,\omega} , 
\nn
  \ket{\xi^\pm}\otimes\ol{\ket{\Omega}} &=& 
  - \chi_{0}^\pm \ol \chi_{0}^{\,-} \ol \chi_{0}^+ \ket{\omega,\omega} .
\eea
As the actions of ${\cal B}^\pm_\phi$ generate the mapping from each
 state to a linear
combination of the above sixteen states, one can classify them into two
  sequences. Using the same notation
  $\ket{\alpha,\beta}\equiv\ket{\alpha}\otimes\ol{\ket{\beta}}$ in
  (\ref{metric KW}), they are:
\pagebreak
\begin{eqnarray}
  {\cal B}^\pm_\phi 
\,:\,\hspace{70pt}
 \ket{\omega, \omega}
  &\to& - \left(\ket{\xi^\pm, \omega} - e^{i\phi} \ket{\omega, \xi^\pm} \right)
\nn
 \ket{\xi^\alpha, \omega} - e^{i\phi} \ket{\omega, \xi^\alpha}
  &\to& d^{\alpha\beta} \left( \ket{\Omega,\omega} + \ket{\omega,\Omega} \right)
     + e^{i\phi}\left( \ket{\xi^\pm, \xi^\alpha} - \ket{\xi^\alpha, \xi^\pm} \right)
\nn
 \ket{\Omega,\omega} + \ket{\omega,\Omega} 
  &\to& 
  e^{i\phi} \left( \ket{\Omega,\xi^\pm} - e^{i\phi} \ket{\xi^\pm,\Omega} \right)
\nn
  \ket{\xi^+, \xi^-} - \ket{\xi^-, \xi^+}
  &\to& - \left(\ket{\Omega,\xi^\pm} - e^{i\phi} \ket{\xi^\pm,\Omega} \right)
\nn
  \ket{\Omega,\xi^\alpha} - e^{i\phi} \ket{\xi^\alpha,\Omega}
  &\to& - 2\,d^{\pm\alpha}\, e^{i\phi} \ket{\Omega,\Omega}
\nn
  \ket{\Omega,\Omega} 
  &\to& 0
\nn
\hline
\nn
  {\cal B}^\pm_\phi 
\,:\, 
  \ket{\xi^\alpha, \omega} + e^{i\phi} \ket{\omega, \xi^\alpha}
  &\to& d^{\pm\alpha} \left( \ket{\Omega, \omega} - \ket{\omega, \Omega} \right)
    - e^{i\phi} \left( \ket{\xi^\pm,\xi^\alpha}+\ket{\xi^\alpha,\xi^\pm} \right) 
\nn
 \ket{\Omega, \omega} - \ket{\omega, \Omega}
  &\to& e^{i\phi}\left( \ket{\Omega,\xi^\pm} + e^{i\phi}\ket{\xi^\pm,\Omega} \right)
\nn
 \ket{\xi^\alpha, \xi^\alpha}
  &\to& d^{\pm\alpha} \left( \ket{\Omega,\xi^\alpha} + e^{i\phi} \ket{\xi^\alpha, \Omega}\right)
\nn
 \ket{\xi^+, \xi^-} + \ket{\xi^-, \xi^+} 
  &\to& \pm \left( \ket{\Omega,\xi^\pm} + e^{i\phi}\ket{\xi^\pm,\Omega} \right)
\nn
 \ket{\Omega,\xi^\alpha} + e^{i\phi}\ket{\xi^\alpha,\Omega}
  &\to& 0
\end{eqnarray}
Schematically, they are:
\begin{eqnarray}
\label{symp chiral scheme}
 {\cal B}^\pm_\phi 
&:& \ket{\omega, \omega}
  \to \left(\ket{\xi^\alpha, \omega} - e^{i\phi} \ket{\omega, \xi^\alpha}\right)
  \to
\begin{array}[t]{l}
 \left( \ket{\Omega,\omega} + \ket{\omega,\Omega} \right)\\
 \left( \ket{\xi^+, \xi^-} - \ket{\xi^-, \xi^+} \right)
\end{array}
  \to\cdots
\nn&&\qquad\qquad
  \cdots\to \left( \ket{\Omega,\xi^\alpha} - e^{i\phi} \ket{\xi^\alpha, \Omega}\right)
  \to \ket{\Omega,\Omega}
  \to 0 \,,
\nn[10pt]
 {\cal B}^\pm_\phi 
&:&  \left(\ket{\xi^\alpha, \omega} + e^{i\phi} \ket{\omega, \xi^\alpha}\right)
\nn&&
  \to 
\begin{array}[t]{l}
  \left( \ket{\Omega, \omega} - \ket{\omega, \Omega} \right)\\
  \ket{\xi^\alpha, \xi^\alpha} \\
  \left( \ket{\xi^+, \xi^-} + \ket{\xi^-, \xi^+} \right) 
\end{array}
  \to \left( \ket{\Omega,\xi^\alpha} + e^{i\phi}\ket{\xi^\alpha,\Omega} \right)
  \to 0 \,\,.
\end{eqnarray}
From the above, we find that the states:
\begin{eqnarray}
\label{set symp chiral}
 \ket{\wt \Omega} &=& \ket{\Omega,\Omega} ,
\nn
 \ket{\wt \xi^\alpha ; \phi} &=& \ket{\Omega,\xi^\alpha} + e^{i\phi} \ket{\xi^\alpha,\Omega} ,
\nn
 \ket{\wt \omega; \phi} &=& 
  \ket{\Omega,\omega}+\ket{\omega,\Omega}
  + e^{i\phi} \left(\ket{\xi^+,\xi^-} - \ket{\xi^-,\xi^+}\right) 
\end{eqnarray}
satisfy the zero mode conditions (\ref{def:symp bc 0}).
Their corresponding coherent states are the following eight:
\begin{eqnarray}
\label{sol:comparison2}
 \ket{B_{\wt \Omega\pm}}, \ket{B_{\wt \xi^\alpha \pm}}, \ket{B_{\wt \omega \pm}} .
\end{eqnarray}
Since the normalisation $\braket{\Omega|\Omega}=0$ is given, the absence of
$\ket{\omega,\omega}$, $\ket{\omega,\xi^\alpha}$, and $\ket{\xi^\alpha,\omega}$ 
indicates that the first two entries
of the three pairs of states become null in their character functional form. 
Since those ground states give at most $\ket{\Omega,\Omega}$ when
$L_0$ acts on them, such null coherent states further shows that the resulting
character functions have no logarithms which require the ground state
pairing in $\bra{\alpha}L_0\ket{\beta}$ to be non-vanishing. 

When one defines $\braket{\xi^-,\xi^+|\xi^+,\xi^-}=-1$, 
the inner product of $\wt \omega$ becomes:
\linebreak
\mbox{
$
\braket{\wt \omega;\phi | \wt \omega;\phi^\prime} = 2(1+e^{i(\phi+\phi^\prime)}).
$}
Then, one obtains non-vanishing cylinder amplitudes from (\ref{set symp chiral}):
\begin{eqnarray}
 \bra{B_{\wt \omega \phi}} \wt q^{L_0+\frac1{12}} \ket{B_{\wt \omega \phi^\prime}}
 = 4 \,\delta_{\phi,\phi^\prime}\, \eta^2(\wt q) .
\end{eqnarray}
By a modular transformation, the $r.h.s.$ of (\ref{def:modified Cardy
eq}) cannot yield this function, and therefore this should be deleted.
Consequently, this result becomes similar to the case studied in section \ref{sec:ishibashiYI} \cite{ishimoto1,YI2}. On the other hand, when
$\braket{\xi^-,\xi^+|\xi^+,\xi^-}=1$, the inner product of $\wt \omega$
states amounts to $2(1-e^{i(\phi+\phi^\prime)})$. Then, one obtains non-vanishing cylinder amplitudes in the untwisted sector:
\begin{eqnarray}
 \bra{B_{\wt \omega\phi}} \wt q^{L_0+\frac1{12}} \ket{B_{\wt \omega\phi^\prime}}
 =  4 \left(1-\delta_{\phi,\phi^\prime}\right)\,\Lambda_{1,2}(\wt q)  .
\end{eqnarray}
This result is quite similar to the case 
in section \ref{sec:ishibashiBF} and is compatible with \cite{bredthauer2}.
The definitions of LWS spaces critically change the final results as such.

\chapter{Conclusions} 
\label{ch:conclusion}
\chaptermark{Conclusions}
 \fancyhead[LO]{\bfseries\leftmark} 

In this thesis, through chapters \ref{ch:intro} -- \ref{ch:BCFT},
we have presented a brief history and the basic facts on CFT, LCFT,
and BCFT.
Thereupon, 
our main works on the so-called boundary LCFTs have been presented in chapter \ref{ch:free} and \ref{ch:boundary states}.
We have explored the two-point functions and boundary states of BLCFTs,
particularly in its simplest cases, where the central charges are the
non-positive integers $-2$ or $0$ and the degenerate vacua are present.

In section \ref{sec:CG}, we have examined the $(p,q)$ model with the Neumann boundary condition in the Coulomb gas picture. 
In addition to the general case, a class of single solutions (\ref{sol:pq2}) for the logarithmic cases were shown for boundary two-point functions without the analytically continued operators. 
Thereafter, with the analytically continued screening currents, the general solutions (\ref{sol:pq double}) of the logarithmic cases
were shown. The cases are defined such that the 
differential equation for the function has a logarithmic solution at
$\xi=0$. 
In its final subsection, we turned into a specific case $c=-2$ and showed
the two-point functions $\braket{\mu(z_1,\bz_1)\mu(z_2,\bz_2)}$ and
$\braket{\nu(z_1,\bz_1)\mu(z_2,\bz_2)}$, from which the relations
between the normalisation constants (\ref{rel:constants}) were deduced.

In section \ref{sec:su2}, we have examined the free field realisation of
the $SU(2)_k$ AKM algebras motivated by the $SU(2)_k$ WZNW action.
From the action of the three free scalar bosons with the Neumann boundary
condition, we followed the calculation of \cite{Gerasimov:fi} and
confirmed the general solutions (\ref{sol:su2-1}, \ref{sol:su2-2}).
With the same technique in section \ref{sec:pq}, it was found that the
logarithmic cases for this class of doublet's solutions inevitably
becomes logarithmic at both $\xi=0$ and $\xi=1$. 
We then found the solutions regular at $\xi=1$
and restored the general solution (\ref{sol:su2log2}) of the differential
equation shown in \cite{KW}. Some relations between the normalisation 
constants and fusion rules are also listed. Our results also reveal that 
such relations may be different for different LCFTs.

We have not discussed the Dirichlet boundary condition since this can
only be examined easily when $p=q$ where no screening charges are
required.
This is an interesting problem for the next step, where one might see
the more natural emergence of logarithmic solutions or the restoration of the scale covariance, that is $B=0$ \cite{KW}.

The puncture-type operators in \cite{Kogan:1997fd} have not been dealt
with in this thesis.
However, as was stated
earlier, if one replaces our conjugate operators 
with appropriate puncture-type operators and defines 
$\braket{:\phi\,V_{2\alpha_0}:}\sim 1$, then one finds the same 
result. 
The boundary operators were discussed in section \ref{sec:c=-2 bop} whereas the boundary states for LCFTs were not.
Recently, it was shown that the boundary states of the minimal models can be realised in CG \cite{kawaic1, kawaic2}. 
Our charge unbalanced configurations with the chiral screening
charges may work well with it, for the precise determination of the
constants $\alpha$'s in eq. (\ref{sol:mumu numu}). 
Another possibility is to introduce new objects around the boundary as in \cite{schulze}. 
It may also give the logarithmic solution of eq. (\ref{gauss eq}) but our action should be modified as to be consistent with them.

There are several different models at $c=-2$. It is not
clear whether our free field realisation is equivalent to one of them, and whether the theory is (quasi-) rational or not.
Rigorous answers for these questions are beyond the scope of this thesis, 
instead we list the possible
candidates to compare with: the $c_{p,1}$ models in \cite{gab3}, which
close under fusion, and the singlet algebra in \cite{flohr1} which does
not have a nice modular property.
The techniques we used are quite common and, therefore, the same procedure
should be applicable to many other similar models with free boson fields.
In order to compare this formalism with the well-known
rational model of $c=-2$, one must have the vertex operator realisation
of the triplet algebra though \cite{kau1}. For example, the field $\nu$ of
dimension $3/8$ must be a doublet under the extended algebra \cite{gab4}. 
Recently, Nichols presented such a
sketch in \cite{Nichols:2003dj} along the line of
\cite{Kogan:1997fd,flohr4}.
This would be an interesting direction to study.

In section \ref{sec:ishibashi}, we have investigated two cases of the (generalised) Ishibashi states in the context of $rank$-2 Jordan lowest weight modules of boundary LCFTs.
In section \ref{sec:ishibashiYI}, we have shown that a naive form of the Ishibashi states (\ref{def:B}) 
leads to the results (\ref{def:bs-0}).
Next, by introducing linear combinations of the elements in (\ref{def:hilbert space}), we confirmed the results in \cite{bredthauer} that there are two generalised Ishibashi states in  the $c=-2$ triplet model, and that six states out of ten independent states become null in the partition function.
Apart from such generalised states, 
the general solution (\ref{sol:general BF}) is in agreement with the former case.
In all such cases, the logarithmic Ishibashi states of $\V_0$ and
$\V_1$ do not propagate in the closed string picture. 

In section \ref{sec:coherent}, the coherent state construction was
presented in the context of the $c=-2$ symplectic fermion system, 
which provides a free field realisation of the $c=-2$ 
triplet model \cite{kaw}. 
The fermionic boundary conditions (\ref{def:symp bc}) arise instead of (\ref{def:bc1}),
and the deduced conditions (\ref{def:symp bc 0}) prescribe the
number of possible boundary coherent states.
Six distinct states were found by Kawai and Wheater, out of which four 
form a basis of the physical boundary state space. 
The character functions (\ref{character KW1}) of those six states 
and the general solution (\ref{sol:general KW}) were also presented.
Their results indicate that the coherent states built on the Jordan cell do propagate but not diagonally.

Finally in section \ref{sec:comparison}, we have given a comparison 
between the above two constructions,
mainly from the partition function point of view. 
Their fundamental differences  
indicate that the critical point of difference is on the vacuum structure.
To compare with, we have focused on the coherent state construction in
the $c=-2$ triplet model, and have investigated two different vacuum
structures appearing in section \ref{sec:ishibashi} and \ref{sec:coherent}.
Then we have shown two different sets of boundary states, (\ref{sol:comparison1}) and (\ref{sol:comparison2}), in the same $c=-2$ triplet model.
The former solution simply gives the same set of the character
functions in section \ref{sec:coherent}, so that they are effectively same. 
The latter solution gives quite similar results
to the cases in section \ref{sec:ishibashiBF}, where there appear no
logarithms of $\wt q$ in the character functions.
In fact, the absence of such logarithms implies the absence of
$\chi_{\V_0}$ and $\chi_{\V_1}$ in the $r.h.s.$ of the modified
Cardy's eq. (\ref{def:modified Cardy eq}), and then leads to the
absence of $\eta^2(\wt q)$-term in the $l.h.s.$ 
Hence, the latter solution (\ref{sol:comparison2}) is practically
equivalent to the cases in section \ref{sec:ishibashiBF}.
This fact roughly resolves the construction dependence.

All the cases mentioned above possess ten boundary states. 
However, two solutions are apparently different, 
since the former contained $\log \wt q$-terms as in eq. (\ref{character KW1})  while the latter doesn't. 
Hence, one may conclude that the cases should be called different
theories or models
when their Hilbert spaces and physical partition functions are different.

In all of this, 
we have omitted the actual derivations of the physical boundary states. 
They are rather straightforward and can be confirmed in
[22-24, 55, 56, 83]. 
Note that the results may depend on the definitions of the
coefficients $n_{\al \beta}^{\;\; i}$ in eq. (\ref{def:modified Cardy eq}).
Because the actual forms of physical boundary states have not
been dealt with, it is not trivial to determine if the Verlinde formula holds. One may simply quote the literature 
[22-24, 55, 56, 83]. 
that the models involved in this thesis do not satisfy the
conventional expression of the formula
but lead instead to a block-diagonal form.

It would be intriguing to interpret the results in the D-brane context,
since the solutions would correspond to states on the D-branes.
On the other hand, we do not know 
the physical interpretation of boundary conditions in the way that we do with the free and fixed boundary conditions in the Ising model. 
So, this remains to be done.

Lastly, we would like to draw the reader's attention to the relations between
boundary conditions and boundary states. By deriving the boundary
states and their precise interpretation in free field
representations, one can fix the normalisation constants $\alpha$'s in
chapter \ref{ch:free}.
This means that one can also calculate the open-closed string interactions.
However, as was stated earlier in this chapter, the relation between
the model in chapter \ref{ch:free} and the triplet models in chapter
\ref{ch:boundary states} is still ambiguous.
Therefore, at the moment, we are not able to proceed further. 
Once such relations are established, 
the class of boundary LCFT become outstandingly tractable and
useful, not only for itself but for many branches of physics. 
We have just shown our very
first steps towards a total and deep understanding of boundary LCFT.

\clearpage 
\appendix
\renewcommand{\sectionmark}[1]{ \markright{#1}}
 \fancyhead[RE]{\bfseries\rightmark}
 \fancyhead[LO]{\bfseries\rightmark} 

\ni
{\huge\bf Appendix}
\addcontentsline{toc}{chapter}{Appendix}
\addcontentsline{toc}{section}{A \ \ The Radial Quantisation} 
\thispagestyle{plain}
\section*{A. The Radial Quantisation}
\label{sec:radial quantisation}

\begin{figure}[h]
\hspace*{4cm}
\epsfysize=4cm
\epsffile{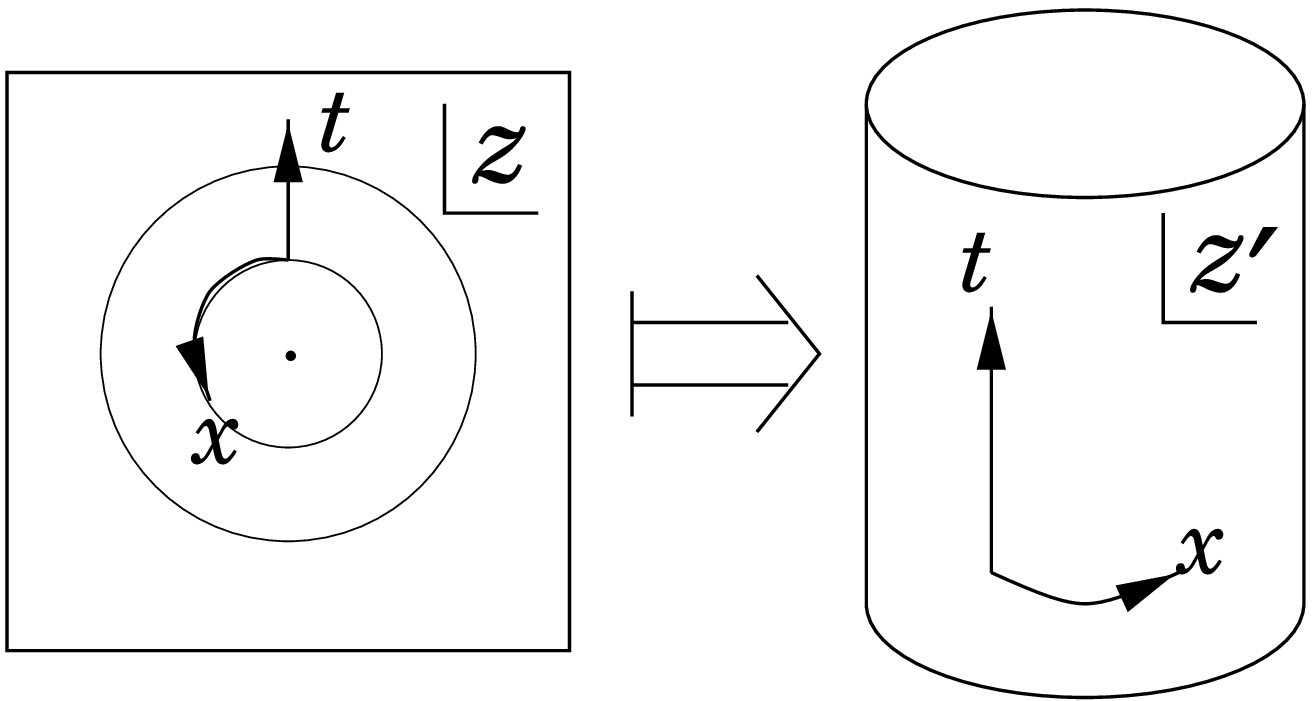}
 \caption{A conformal mapping from a complex plane to a cylinder.}
      \label{fig:PtoCyl}
\end{figure}

A map from the complex plane to a cylinder is given by
\bea
 \label{def:PtoCyl}
   z \too z^\prime = \ln (z) = t + i x,\quad
   {for~ t\in\R,\, 0\leq x < 2 \pi} .
\eea
Here, and in Fig.\ref{fig:PtoCyl}, $t$ is interpreted as a time coordinate and $x$ is a spatial coordinate.
The mapping transforms
any contour with a fixed radius around the origin on the plane, into an equal-time surface on the cylinder. 
Therefore, fields on the contour on the plane can be interpreted as
Hilbert states on the same equal-time surface. 
On the other hand, the $T(z)_{pl}$ is mapped into $T(z^\prime)_{cyl}$
\bea
 \label{eq:PtoCyl T}
   T(z^\prime)_{cyl} = z^2 T(z)_{pl} - \frac{c}{24},
\eea
where the suffices, $cyl$ and $pl$, refer to the geometries on which
the tensors are defined respectively.
The last extra term is the Schwartzian derivative of this case,
because of which the $T(z)$ is not a primary \cite{CFT,ketov}.
Note that $c$ appears as a parameter of the Casimir energy in the cylindrical picture.

On the cylinder, the system has a Hamiltonian and the equation of motion is given by
\bea
 \label{def:EOM on cyl}
   \frac{d \Phi}{d t} = \com{H}{\Phi}_{equal~ time},
\eea
or equivalently
\bea
 \label{def:time evolution on cyl}
   \delta_\tau \Phi = \com{\tau H}{\Phi}_{equal~ time},
\eea
where the Hamiltonian $H$ can be expressed by the dilatational generators.
\bea
 \label{def:H Vir}
   H = (L_0)_{cyl} + (\ol L_0)_{cyl}.
\eea
Under a general conformal transformation, the field on the cylinder is transformed as
\bea
 \label{def:general tf on cyl}
  \delta_{\epsilon, \bar \epsilon} \Phi(w,\bw) = \oint_w \frac{d z}{2\pi i} T(z) \Phi(w,\bw) + \oint_\bw \frac{d \bz}{2\pi i} \bar T(\bz) \Phi(w,\bw) .
\eea

In Minkowski space-time, the reality of $T(z)_{cyl}$ causes the
conjugation conditions on $\{L_{n\in\Z}\}$. 
\bea
 \label{def:unitary cond on Ln}
   L_n^+ = L_{-n} . 
\eea

\addcontentsline{toc}{section}{B \ \ Contiguous relations for the line integrals of the $SU(2)_k$ WZNW model} 
\sectionmark{Appendix}
\section*{B. Contiguous relations for the line integrals of the $SU(2)_k$ WZNW model}
\label{sec:contig}

For the relation (\ref{id:Fiab}) for $F_2$'s and the derivation of
(\ref{id:Fiab2}), we show the following identity:
\begin{eqnarray}
  F_i^{+-+-} = - F_i^{-++-} - F_i^{++--} ~~~for~i=1,2.
\end{eqnarray}
Let us introduce the following notation for the relevant chiral
four-point functions:
\begin{eqnarray}
  F_{ab}^i
  &=& - (z_{13} z_{24})^{-2h_{1/2}} \xi^{A/2} (1-\xi)^{A/2} \;
  I_{ab}^{~i} \,,
\nn
  I_{ab}^{~i} &=& \int_{w_a}^{w_b} dw\, \frac{(-w)^{-A}(\xi-w)^{-A}(1-w)^{-A}}{w_i -w} \,,
\end{eqnarray}
where $(w_1,w_2,w_3,w_4)=(0,\xi,1,\infty)$ and $2\alpha_\phi^2$ is replaced by $A=\frac{1}{k+2}$. They are
nothing but the functions (\ref{sol:su2-1}, \ref{sol:su2-2}) as
$F_1^{-++-}=F_{12}^1$. $i\in\{1,2,3\}$ denotes the position of $V_-$.

These integrals $I_{ab}^{~i}$ can be explicitly drawn as:
\begin{eqnarray}
 I_{12}^{~1}
  &=& - e^{A\pi i} \frac{\Gamma(-A)\Gamma(1-A)}{\Gamma(1-2A)} \;\xi^{-2A}\;
    {}_2 F_1 \left( A, -A; 1-2A; \xi \right) , 
\nn
 I_{12}^{~2}
  &=& e^{A\pi i} \frac{\Gamma(-A)\Gamma(1-A)}{\Gamma(1-2A)} \;\xi^{-2A}\;
    {}_2 F_1 \left( A, 1-A; 1-2A; \xi \right) ,
\nn
 I_{12}^{~3}
  &=& e^{A\pi i} \frac{\Gamma^2(1-A)}{\Gamma(2-2A)} \;\xi^{1-2A}\;
    {}_2 F_1 \left( 1+A, 1-A; 2-2A; \xi \right) ,
\nn
  &=& e^{A\pi i} \frac{\Gamma(-A)\Gamma(1-A)}{\Gamma(1-2A)} \left(\frac{-A}{1-2A}\right) \;\xi^{1-2A}\;
    {}_2 F_1 \left( 1+A, 1-A; 2-2A; \xi \right) ,
\nn
 I_{34}^{~1}
  &=& - e^{3 A\pi i} \frac{\Gamma(3A)\Gamma(1-A)}{\Gamma(1+2A)} 
    {}_2 F_1 \left( A, 3A; 1+2A; \xi \right) , 
\nn
 I_{34}^{~2}
  &=& - e^{3 A\pi i} \frac{\Gamma(3A)\Gamma(1-A)}{\Gamma(1+2A)} 
    {}_2 F_1 \left( 1+A, 3A; 1+2A; \xi \right) , 
\nn
 I_{34}^{~3}
  &=& - e^{3 A\pi i} \frac{\Gamma(3A)\Gamma(-A)}{\Gamma(2A)} 
    {}_2 F_1 \left( A, 3A; 2A; \xi \right) , 
\nn
  &=& - e^{3 A\pi i} \frac{\Gamma(3A)\Gamma(1-A)}{\Gamma(1+2A)} (-2)\; 
    {}_2 F_1 \left( A, 3A; 2A; \xi \right) .
\end{eqnarray}
Accordingly, one can find the ratios of them:
\begin{eqnarray}
 I_{12}^{~1}:I_{12}^{~2}:I_{12}^{~3} &=& 
  - {}_2 F_1 ( \alpha, \beta; \gamma; \xi ): {}_2 F_1 ( \alpha, \beta+1; \gamma; \xi ):
  - \left(\frac{\alpha}{\gamma}\right) \xi {}_2 F_1 ( \alpha+1, \beta+1; \gamma+1; \xi )
\nn&&\qquad {\rm ~~with~(\alpha,\beta,\gamma)=(A,-A,1-2A)},
\nn
 I_{34}^{~1}:I_{34}^{~2}:I_{34}^{~3} &=& 
    {}_2 F_1 ( \alpha, \beta; \gamma; \xi ): {}_2 F_1 ( \alpha+1, \beta; \gamma; \xi ):
 -2\,{}_2 F_1 ( \alpha, \beta; \gamma-1; \xi )
\nn&&\qquad {\rm ~~with~(\alpha,\beta,\gamma)=(A,3A,1+2A)} . 
\end{eqnarray}
From the contiguous relations of hypergeometric functions, (1.4.4) and
(1.4.16) of SL in \cite{slater}, with the permutation symmetry, one obtains:
\begin{eqnarray}
\label{eq:contig1}
 F(\beta+1) &=& F + \frac{\alpha}{\gamma}\, \xi\, F(\alpha+1,\beta+1,\gamma+1) , 
\end{eqnarray}
and from (1.4.5)
\begin{eqnarray}
\label{eq:contig2}
 F(\alpha+1) = \frac{\gamma-1}{\alpha} F(\gamma-1) - \frac{\gamma-\alpha-1}{\alpha} F .
\end{eqnarray}
The unchanged coefficients are omitted in the above such as
$F=F(\alpha,\beta;\gamma;\xi)$,
$F(\beta+1)=F(\alpha,\beta+1,\gamma;\xi)$, and so on.

Comparing (\ref{eq:contig1}) with $I_{12}^{~i}$, one further obtains:
\begin{eqnarray}
 I_{12}^{~2} = - I_{12}^{~1} - I_{12}^{~3} , 
\end{eqnarray}
and from the second one (\ref{eq:contig2}):
\begin{eqnarray}
 I_{34}^{~2} = - I_{34}^{~1} - I_{34}^{~3} .
\end{eqnarray}
To summarise, 
\begin{eqnarray}
 I_{ab}^{~2} = - I_{ab}^{~1} - I_{ab}^{~3}
 {~~~~~~for~~}(a,b)\in\{(1,2),(3,4)\} ,
\end{eqnarray}
and therefore,
\begin{eqnarray}
 F_{ab}^{2} = - F_{ab}^{1} - F_{ab}^{3} {~~~~~~for~~}(a,b)\in\{(1,2),(3,4)\}.
\end{eqnarray}
$(Q.E.D.)$

\end{document}